\documentclass[a4paper,fleqn,usenatbib]{mnras}

\usepackage{newtxtext,newtxmath}

\usepackage[T1]{fontenc}
\usepackage{ae,aecompl}

\usepackage{graphicx}	
\usepackage{amsmath}	
\usepackage{amssymb}	
\usepackage{graphicx}
\usepackage{rotating}
\usepackage{hyperref}
\hypersetup{draft}

\newcommand{\WISE}{{\it WISE}}

\title[A population of IR-bright GRB hosts]{Investigating a population of infrared-bright gamma-ray burst host galaxies}

\author[A. A. Chrimes et al.]{Ashley A. Chrimes,$^{1}$\thanks{E-mail: A.Chrimes@warwick.ac.uk}
Elizabeth. R. Stanway,$^{1}$
Andrew J. Levan,$^{1}$
Luke J. M. Davies,$^2$\newauthor
Charlotte R. Angus,$^3$
Stephanie M. L. Greis$^1$
\\
$^{1}$Department of Physics,  University of Warwick, Gibbet Hill Road, Coventry, CV4 7AL, UK\\
$^{2}$ICRAR, The University of Western Australia, 35 Stirling Highway, Crawley, WA 6009, Australia\\
$^3$School of Physics and Astronomy, University of Southampton, Southampton, SO17 1BJ, UK\\
}

\date{Accepted 2018 April 23. Received 2018 April 23; in original form 2017 November 28}

\pubyear{2017}

\begin{document}
\label{firstpage}
\pagerange{\pageref{firstpage}--\pageref{lastpage}}
\maketitle

\begin{abstract}
We identify and explore the properties of an infrared-bright gamma-ray burst (GRB) host population. Candidate hosts are selected by coincidence with sources in \WISE, with matching to random coordinates and a false alarm probability analysis showing that the contamination fraction is ${\sim}$ 0.5. This methodology has already identified the host galaxy of GRB\,080517. We combine survey photometry from Pan-STARRS, SDSS, APASS, 2MASS, GALEX and \WISE\ with our own WHT/ACAM and VLT/X-shooter observations to classify the candidates and identify interlopers. Galaxy SED fitting is performed using {\sc MAGPHYS}, in addition to stellar template fitting, yielding 13 possible IR-bright hosts. A further 7 candidates are identified from previously published work. We report a candidate host for GRB\,061002, previously unidentified as such. The remainder of the galaxies have already been noted as potential hosts. Comparing the IR-bright population properties including redshift $z$, stellar mass M$_{\star}$, star formation rate SFR and {\it V}-band attenuation A$_{V}$ to GRB host catalogues in the literature, we find that the infrared-bright population is biased toward low $z$, high M$_{\star}$ and high A$_{V}$. This naturally arises from their initial selection - local and dusty galaxies are more likely to have the required IR flux to be detected in \WISE. We conclude that while IR-bright GRB hosts are not a physically distinct class, they are useful for constraining existing GRB host populations, particularly for long GRBs.

\end{abstract}




\section{Introduction}
Gamma-ray bursts (GRBs) occur when relativistic jets are launched by a newly formed neutron star or black hole, along our line of sight, in the immediate aftermath of a cataclysmic event such as the collapse of a massive star or merger of two compact objects. The distribution of GRB $T_{90}$ durations (the time over which 90 per cent of the gamma-ray radiation arrives) is indicative of two main populations. Long bursts (T$_{90}$ $\gtrsim$ 2s) are thought to arise from the collapse of particularly massive, rapidly rotating stars \citep{Woolsey,Levan,Schady}, while short GRBs (SGRBs) are associated with the merger of compact objects \citep{NSNSGW1,NSNSmulti1,2017arXiv171005433T,2017ApJ...848L..14G,2017ApJ...848L..27T}. Throughout, we denote short and long bursts in the text as SGRBs and LGRBs respectively. Because the progenitors of LGRBs are massive stars, and these only exist for a short while after formation, we expect to find that their host galaxies are actively star-forming. Multi-wavelength observations of LGRB hosts have found this to be the case. Within their hosts, LGRBs appear to trace the regions of highest UV luminosity, and hence star formation \citep{Fruchter,Svensson,100316D,GRBHST,Lyman}. Similarly, we would expect the comoving rate density of LGRBs to scale with the star formation rate density throughout cosmic history \citep{HopkinsBeacom,Madau}. However, LGRBs have an apparent aversion to massive, luminous galaxies at fixed SFR, which has been interpreted as a host metallicity bias \citep{Fruchter,PerleyDust,TOUGH,Perley1}. The bias has been confirmed spectroscopically \citep{Kruhler}. For a massive star to produce a GRB, it is thought that the progenitor must have sufficient angular momentum (see \citet{BZmech}, and \citet{Penrose}). This is difficult if the stellar metallicity is high, because metals increase the opacity of the stellar envelope, driving mass and angular momentum loss through line-driven winds \citep{Vink,VinkBH}. Furthermore, if pre-burst mass loss produces a dense, high metallicity circumstellar medium, the opacity to gamma-rays is increased and may stifle the jet \citep{Hjorth}. 

Despite these theoretical considerations, the aversion to massive, dusty galaxies may also be due to observational biases \citep{PerleyDust}. The picture of LGRBs favouring low mass, low metallicity hosts is complicated by the existence of dark bursts. These are GRBs which have a steeper X-ray to optical slope, ${\beta}_{OX}$, than would be expected from extrapolation of their X-ray spectra \citep{Rol_darks}. The implication is that the optical emission is suppressed. Studies have shown that ${\sim}25$ per cent of all bursts fall into this category \citep{Jakobsson,GRBXS}. The proposed explanations for the high ${\beta}_{OX}$ values in this population include intrinsic faintness, the burst being at very high redshift, and dust obscuration within the host. The first of these is hard to explain on theoretical grounds, as the GRB and subsequent afterglow are though to arise from the same outflow of material interacting with the ISM; as such GRBs which are bright in X-rays ought to also be optically bright \citep[e.g. ][]{Dainotti}. If the burst is at very high redshift, observed optical bands correspond to rest-frame ultraviolet (UV), which is absorbed along the line of sight by neutral Hydrogen. However, such bursts are expected to be rare, and indeed only a small fraction of bursts have been confirmed at $z > 5$ \citep{Ruiz-Velasco,Tanvir2,Salvaterra,Greiner_hi-z,Cucchiara,ManyHiZ,Tanvir_hi-z,Bolmer}. The majority of dark bursts are thought to be optically faint due to dust extinction. Early studies of GRB host populations systematically missed dark burst hosts because localisation was performed with optical afterglows, particularly before the launch of \textit{Swift}. More recent studies have attempted to account for this by selecting hosts of bursts which have an undetected optical afterglow, using X-rays for localisation instead. However, \citet{Kruhler2011} and \citet{PerleyDust} found that even when considering dark burst hosts, which are typically more massive and dusty, the overall GRB population still shows a bias towards fainter, less massive systems than the typical star forming galaxy population at the same epoch. Moving to host identification in the infrared (IR) may provide a route to further reducing this bias, since ultraviolet light from young stellar populations is preferentially absorbed by dust, and re-emitted in the IR. This only aids the identification of dark GRB hosts, however, if they are dark due to extinction from galaxy-wide dust. If the extinction is local to the burst site, or exclusively along the line of sight, then a burst might be optically suppressed in an otherwise IR-faint galaxy with little dust re-emission. 

All-sky infrared surveys, such as the Wide-field Infrared Survey Explorer \citep[\WISE, ][]{Wright}, can be used for the purpose of host identification. Such surveys are shallow, favouring the identification of nearby \citep{Kovacs} or luminous and dusty hosts. A small number of GRB hosts have been confirmed at $z<0.1$, which we define as local. These include LGRBs\,051109B, 060218, 100316D and 111005A \citep{051109B,LLGRBs,100316D,Michalowski}, while the host of LGRB\,080517 was studied by \citet{Stanway} following initial selection through coincidence with a notably bright source in the \WISE\ bands. Subsequent follow-up resulted in characterisation of the stellar population and star formation rate in this galaxy through a number of indicators, including radio emission. It also secured the detection of molecular gas for only the third time in a GRB host, constraining the gas consumption timescale \citep{Stanway2}. 

In general, the benefits of identifying local GRB hosts are threefold. First, as discussed, proximity makes observation at radio, submillimetre and infrared wavelengths more feasible \citep[e.g.,][]{HIgrbhosts}. This is exemplified by the recent identification of infrared molecular hydrogen emission lines in low-redshift GRB host galaxies (Wiersema et al, in prep). Secondly, local galaxies will tend to have greater angular extent and thus be easier to spatially resolve for GRB environment studies, increasingly using IFUs \citep[e.g.][]{IFU1998,100316D}. Finally, a rare class of low-luminosity long GRB (LLGRB) has emerged thanks to their low redshift identification \citep[e.g.][]{1998discovery,Stanway}. Because the supernovae (SNe) associated with LLGRBs appear typical of GRB-SNe across the full range of LGRB energies, it seems unlikely that the progenitors of LLGRBs are different to `regular' LGRBs \citep{Schady}. The question then is, what factors can produce the wide range of inferred LGRB isotropic energies, while influencing the range of SN energies much less? Suggestions have included the effect of viewing angle, differences between central engine activity duration versus the shock breakout time, and progenitor metallicity having an impact on burst efficiency \citep{Hjorth,Levan,Schady}. Studies of a large sample of low redshift LGRB hosts will be invaluable in determining the conditions capable of producing LGRBs, including low-luminosity bursts, as well as for studying the evolution of LGRB hosts over cosmic time. 

In this paper we explore the properties of a population of IR-bright GRB host galaxies, detected in \WISE. The hosts are localised in X-rays and are selected from all bursts detected between 2005-2016 inclusive. Such galaxies may be nearby, or extremely luminous and dusty. The paper is structured as follows. Section \ref{sec:selection} describes the sample and selection criteria used. Section \ref{sec:obs} details observations of a sub-sample of these candidate hosts, including VLT/X-shooter and WHT imaging and spectroscopy, in addition to ATCA radio observations. We compile archival and survey data in section \ref{sec:archival}.  In section \ref{sec:prevreport}, previously studied hosts are identified. SED fitting is performed in section \ref{sec:sedfitting}. Section \ref{sec:results} presents our results and discussion, with the broader implications considered in section \ref{sec:interp}. Our conclusions follow in section \ref{sec:conclusions}. Where required, the standard ${\Lambda}$CDM cosmology is used, with $h$ = 0.7, ${\Omega}_\mathrm{M}$ = 0.3 and ${\Omega}_{\Lambda}$ = 0.7. Magnitudes are quoted in the AB system \citep{Oke}.

\section{Sample Selection}\label{sec:selection}
\subsection{Rationale}
The \textit{Swift} observatory \citep{Swift}, which has detected the bulk of GRBs since 2005, is mounted with an X-ray telescope \citep[XRT,][]{XRT}, a Gamma-ray Burst Alert Telescope \citep[BAT,][]{BAT} and an Ultraviolet and Optical Telescope \citep[UVOT,][]{UVOT}, as on-board instruments. UVOT provides the best localisations, however only one third of bursts with an X-ray detection have a UVOT determined position. At the other extreme, all detected GRBs have a BAT detection by definition, however the localisation is no better than a few arcminutes. The best balance between the number of detections and the ability to locate a host is therefore provided by X-rays, for which ${\sim}$ 98 per cent of bursts have a localisation. We have identified a sample of infrared-bright gamma-ray burst host galaxies by cross-matching the GRB X-ray afterglow coordinates with the ALLWISE IR all-sky catalogue from \WISE\ \citep[the {\it Widefield Infrared Survey Explorer, }][]{Wright}. This provides aperture matched photometry in four wavebands, $W1-4$, at 3.4, 4.6, 12 and 22\,$\mu$m. Any cross match procedure between these catalogues will identify both genuine matches and spurious matches to unassociated sources. Because the \WISE\ dataset is relatively shallow, we expect to see nearby or very luminous extragalactic sources, in addition to Galactic stellar contaminants. 

\subsection{Initial Cross-Matching and Cuts}\label{sec:xmatching}
Data for GRBs (detected by \textit{Swift}, {\it INTEGRAL}, Konus-{\it Wind} and the IPN) in the years 2005-2016 inclusive were downloaded from NASA's GRB catalogue\footnote{https://swift.gsfc.nasa.gov/archive/grb\_table/}. \textit{Swift} positions were checked against the \textit{Swift} XRT-GRB catalogue\footnote{http://www.swift.ac.uk/xrt\_live\_cat/}. The data include positions in the gamma-ray, X-ray and UV/optical bands, with their associated 90 per cent confidence error radii, in addition to the $T_{90}$ durations. We do not, at this stage, differentiate between long and short bursts. The total sample contains 1001 bursts, which are used for the following analysis. {\sc TOPCAT}\footnote{http://www.star.bris.ac.uk/~mbt/topcat/} \citep{TOPCAT} was used to cross-match the X-ray locations with sources in the ALLWISE catalogue. Matching is primarily to the W1 band, i.e. all of our sources have at least a W1 detection. No significance cut in W1 was made at this stage, since the quality of the sources are determined through flags and visual inspection, as described later.

We perform an initial cross-matching analysis with a fixed radius for all bursts. In order to determine the expected contamination fraction, we also match to a catalogue of positions created by shifting the 1001 GRB positions by ${\pm}$ 1 and 2\,deg in each of Right Ascension (RA) and Declination (Dec). Because the search radii used are of the order arcseconds, and the X-ray positional uncertainties are also on this scale, shifting by 1 or 2\,deg removes all physical correlations and creates a random sample of coordinates. Crucially however, the broad distribution of points in galactic latitude and hence surface density is preserved. Various trial radii from 1 to 10\,arcsec are tested. The difference between the number of matches to actual GRB coordinates, and to our $8008$ new pseudo-random coordinates, is used to estimate a significance through the Poisson cumulative distribution,
\begin{equation}
P({\geq}N_{A}|N_{R}) = \sum_{i = N_{A}}^{\infty}e^{-N_{R}}{\times}\frac{{N_{R}}^{i}}{i!},
\end{equation}
where the smallest $P({\geq}N_{A}|N_{R})$ corresponds to the best matching radius. We find that $r$ = 2.5\,arcsec minimizes the contamination fraction. The significance is further improved by employing cuts. This includes the removal of sources contaminated by diffraction spikes, optical ghosts and similar data artefacts, using the ALLWISE contamination and confusion flag (CCF). If one of the W1 or W2 bands is dominated by contaminating flux, the match is rejected, or if both of these bands are contaminated (but not dominated), the match is rejected. Matched sources which do not satisfy these criteria cannot be considered robust or reliable. The random matches include both brighter and fainter objects than the actual matches, therefore we limit the random matches to the same range in apparent magnitude to evaluate the probability of selecting the same population by chance. The cuts effectively act as a signal-to-noise filter, with the lowest W1 SNR after cuts of $\sim 4.8$. This gives a maximum centroiding error of $\sim$0.5\,arcsec (given ${\sigma}_{\mathrm{centroid}} = \mathrm{FWHM}/ 2.35*\mathrm{SNR}$). The X-ray positional uncertainties therefore dominate the cross-matching. A final cut was made by removing objects whose cross-matched counterparts were ambiguous or blended in W1 on visual inspection, but which otherwise satisfied the confusion flag cut. These are SGRB\,060801 and LGRB\,061007. Image cutouts of these (and all other) burst locations in the W1 band are given in the appendix. Given that visual inspection is both time consuming and subjective, this last cut was not applied to the randoms and as such all numerical comparisons between actual and random samples were made before this stage. 

With these cuts, and with a 2.5\,arcsec matching radius, we find 45 matches to actual GRB positions and 23 to random coordinates. The corresponding Poisson cumulative probability $P({\geq}N_{A}|N_{R})$ is ${\sim}10^{-5}$. Using this methodology, we estimate a contaminant fraction $f_{c}$ of 0.51, with a Poisson 95 per cent confidence interval covering the region 0.36 $\leq$ $f_{c}$ $\leq$ 0.67.

\subsection{Consideration of Burst Error Radii and Local Background Densities}\label{sec:fap}
In the previous section, we use the same matching radius for all bursts to give an estimate of the contamination fraction. However, this fails to consider two important factors. First, while 2.5\,arcsec is the best matching radius when averaging over the GRB sample, individual burst error radii vary and can be larger than this, so we may be missing genuine matches which lie further out. Second, because GRBs occur in galaxies which tend to exist as members of groups and clusters, our previous analysis considering only the effect of galactic latitude on chance alignment probability is incomplete. The true chance of random alignment may be greater than suggested by averaging over degree-scales because GRBs should preferentially occur in over-densities, which have angular scales much smaller than this.

To address the first issue concerning the tailoring of cross-matching radius to each GRB, we use a radius of 1.5${\times}$R$_{90}$. This is approximately the 99 per cent error radius, assuming a Gaussian profile for the X-ray probability function. From 1001 GRBs, this yields 60 GRBs with one IR match and 4 with two or more. Some \WISE\ sources are included in this count, and not in the 45 discussed previously, because their associated GRB has 1.5R$_{90}$>2.5\,arcsec and a match at $r$>2.5\,arcsec. Others are not included because the matched radius from the previous analysis is greater than 1.5R$_{90}$. These are GRBs\,050716, 060428B, 070208, 120119A, 120612A and 161108A. We add these 6 bursts back into the sample - a \textit{small} X-ray uncertainty is not used at this stage to reject an otherwise good match, because the source of the IR flux could plausibly be extended. 

\begin{figure}
\centering
\hspace*{0.02\textwidth}
\includegraphics[width=0.9\columnwidth]{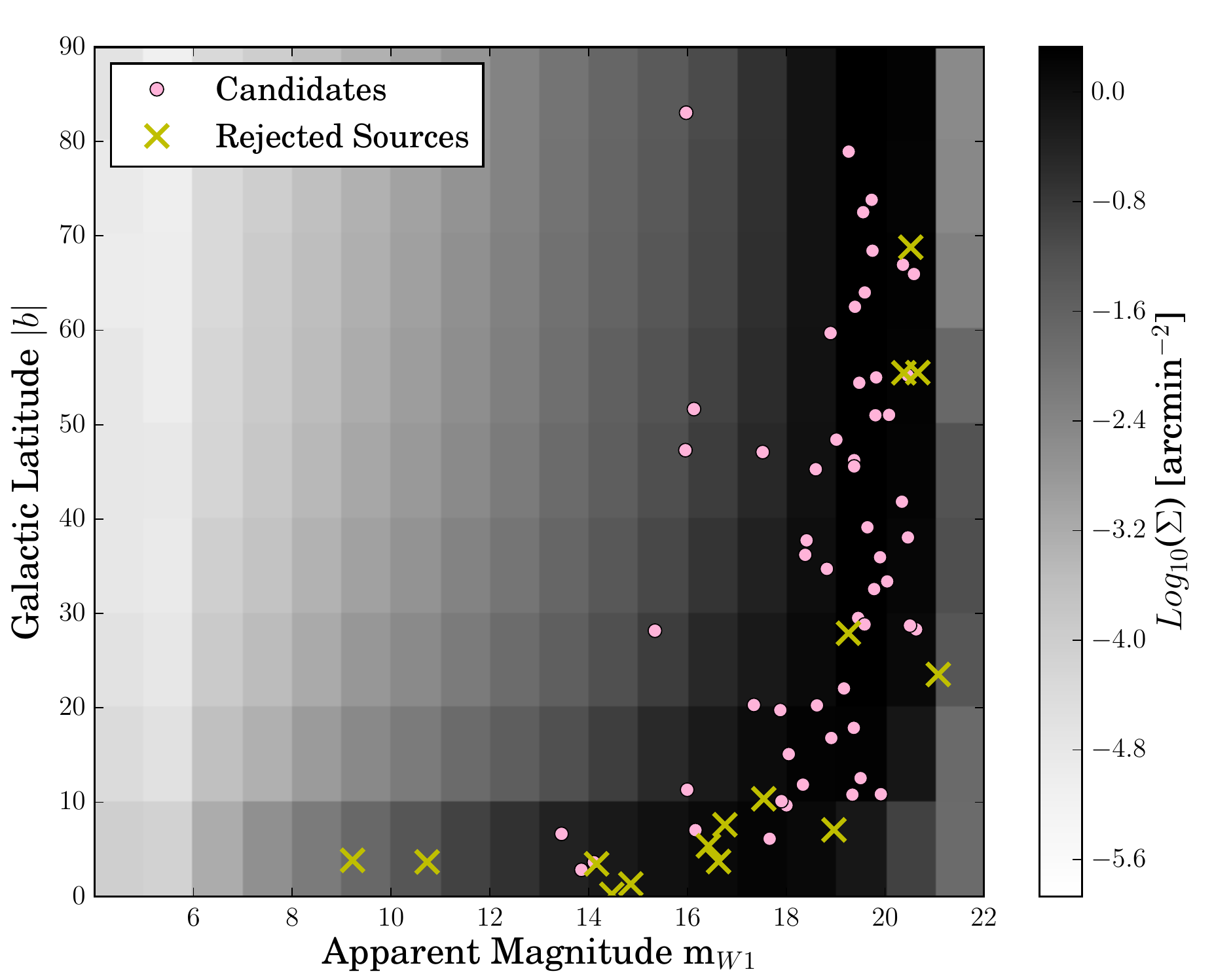}
\caption{The surface density of sources in the W1 band as a function of |b| and m$_{W1}$. The latitudes and magnitudes of the 55 final candidates described in section \ref{sec:sampsum} are indicated by dots, and the rejected sources by crosses.}
\label{fig:bg}
\end{figure}

\begin{figure}
\centering
\includegraphics[width=0.47\textwidth]{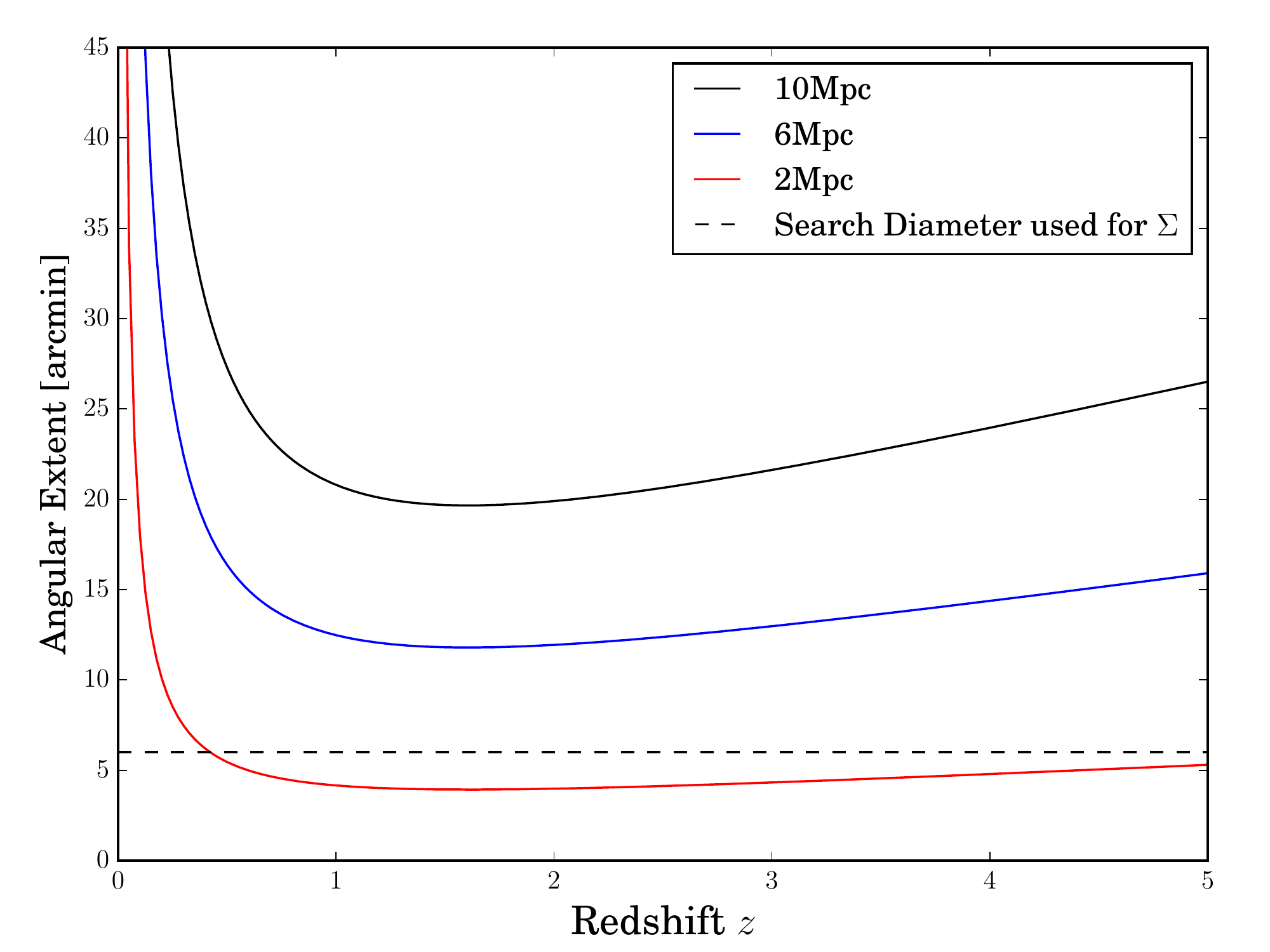}
\caption{The apparent angular extent of physical scales as a function of redshift, using the angular diameter distance. The chosen scale for calculating surface densities is sufficient to capture local variations, for all but the smallest cluster sizes at moderate redshift.}
\label{fig:angdiam}
\end{figure}

In order to estimate the chance that each association is genuine, we perform a false alarm probability (FAP) analysis. The surface density of sources in the entire ALLWISE catalogue is visualised in figure \ref{fig:bg}, as a function of latitude and W1 magnitude. The matched bursts are indicated. Clearly the apparent magnitude and galactic latitude both affect the probability that a match is spurious. However, sky object density also varies on small scales, in addition to the broad galactic latitude trend. To sample the local surface density ${\Sigma}$ around each burst, we cross-match the X-ray coordinates for each burst with a 3\,arcmin radius. Given that galaxy clusters have typical sizes of ${\sim}$10Mpc, a 6\,arcmin diameter is a sufficiently small angular scale to sample density variation due to clustering and cosmic variance. This is demonstrated in figure \ref{fig:angdiam}, which relates redshift to angular extent ${\theta}$ for physical scales $d$ of 2, 6 and 10Mpc, through ${\theta} = \frac{d}{D_{diam}}$, where ${D_{diam}}$ is the angular diameter distance.

\begin{table*}
\centering 
\caption{The 55 GRB X-ray positions for which a catalogued \WISE\ source is identified within 1.5R$_{90}$ of the X-ray error circle centroid, and is not rejected based on FAP, CCF or blend cuts. Our classification of the candidates is given in the final column, and is discussed throughout the paper in the appropriate sections. SGRB\,050724 is a disguised short burst, despite the long T$_{90}$.}
\begin{tabular}{p{1.0cm} c c c c c c c c c r}
\hline 
GRB & T$_{90}$ & Short/ & X-ray Ra & X-ray Dec & R$_{90}$ & \WISE\ Ra & \WISE\ Dec & Sep & FAP & Type  \\ [0.25ex] 
 & [s] & Long & [deg] & [deg] & [arcsec] & [deg] & [deg] & [arcsec] &   &     \\
\hline 
050219A & 23.7 & L & 166.4124 & -40.6842 & 1.9 & 166.4128 & -40.6847 & 2.13 & 0.013183 & NC/LG \\
050318 & 32 & L & 49.7129 & -46.3961 & 1.4 & 49.7129 & -46.3961 & 0.02 & 0.000002 & NC \\
050522 & 10.8 & L & 200.1458 & 24.7883 & 6 & 200.1440 & 24.7869 & 8.06 & 0.002003 & s \\
050716 & 69.1 & L & 338.5866 & 38.6843 & 1.4 & 338.5866 & 38.6850 & 2.42 & 0.014346 & ND/s \\
050721 & 98.4 & L & 253.4356 & -28.3811 & 1.7 & 253.4352 & -28.3814 & 1.90 & 0.018820 & s \\
050724 & 96 & S & 246.1847 & -27.5409 & 1.5 & 246.1849 & -27.5407 & 0.89 & 0.002065 & G \\
060428B & 57.9 & L & 235.3570 & 62.0248 & 1.4 & 235.3583 & 62.0249 & 2.26 & 0.002668 & CA \\
061002 & 17.6 & L & 220.3480 & 48.7414 & 2.6 & 220.3478 & 48.7413 & 0.65 & 0.000302 & PG \\
070208 & 47.7 & L & 197.8859 & 61.9651 & 1.5 & 197.8866 & 61.9656 & 2.37 & 0.011924 & PG/CA \\
070309 & ${\sim}$40 & L & 263.6658 & -37.9307 & 4.4 & 263.6647 & -37.9306 & 3.33 & 0.026938 & NC \\
070429B & 0.47 & S & 328.0159 & -38.8283 & 2.4 & 328.0156 & -38.8286 & 1.40 & 0.004958 & NC/LG \\
070724A & 0.4 & S & 27.8085 & -18.5944 & 1.7 & 27.8088 & -18.5944 & 1.02 & 0.001910 & G \\
071117 & 6.6 & L & 335.0439 & -63.4433 & 1.5 & 335.0444 & -63.4428 & 2.09 & 0.008035 & NC \\
080207 & 340 & L & 207.5122 & 7.5022 & 1.4 & 207.5124 & 7.5018 & 1.55 & 0.008959 & ND/LG \\
080307 & 125.9 & L & 136.6287 & 35.1388 & 1.4 & 136.6290 & 35.1392 & 1.98 & 0.011835 & ND/CA \\
080405 & 40 & L & 162.5996 & -4.2888 & 2.5 & 162.5988 & -4.2888 & 2.71 & 0.001357 & s \\
080517 & 64.6 & L & 102.2420 & 50.7352 & 1.6 & 102.2415 & 50.7353 & 1.06 & 0.000455 & G \\
080605 & 20 & L & 262.1252 & 4.0157 & 1.5 & 262.1254 & 4.0156 & 0.60 & 0.000889 & ND/LG \\
080623 & 15.2 & L & 237.6610 & -62.0491 & 1.4 & 237.6616 & -62.0487 & 1.56 & 0.012242 & NC \\
090904B & 47 & L & 264.1855 & -25.2132 & 1.4 & 264.1854 & -25.2129 & 1.11 & 0.005129 & s \\
091102 & 6.6 & L & 72.6155 & -72.5197 & 2 & 72.6149 & -72.5199 & 1.01 & 0.003337 & NC \\
100206A & 0.12 & S & 47.1626 & 13.1570 & 3.3 & 47.1631 & 13.1581 & 4.15 & 0.010560 & G \\
100316D & ${\geq}$1300 & L & 107.6276 & -56.2555 & 3.7 & 107.6255 & -56.2562 & 4.96 & 0.030711 & NC/LG \\
100816A & 2.9 &  L? & 351.7399 & 26.5784 & 1.4 & 351.7395 & 26.5780 & 1.97 & 0.009327 & ND/LG \\
110206A & ${\sim}$20 & L & 92.3343 & -58.8069 & 1.9 & 92.3331 & -58.8067 & 2.24 & 0.024210 & NC \\
110305A & 12 & L & 260.8806 & -15.8025 & 1.7 & 260.8810 & -15.8030 & 2.22 & 0.006510 & Ps \\
110918A & ${\sim}$22 & L & 32.5387 & -27.1061 & 1.5 & 32.5386 & -27.1057 & 1.24 & 0.003484 & ND/LG \\
111222A & ${\sim}$1 & S & 179.2197 & 69.0709 & 2.9 & 179.2208 & 69.0704 & 2.40 & 0.000177 & s \\
120119A & 253.8 & L & 120.0288 & -9.0817 & 1.4 & 120.0291 & -9.0824 & 2.49 & 0.021870 & Ps \\
120224A & 8.13 & L & 40.9422 & -17.7613 & 1.4 & 40.9424 & -17.7617 & 1.76 & 0.003069 & ND/g \\
120612A & 90 & L & 126.7217 & -17.5748 & 1.5 & 126.7212 & -17.5743 & 2.41 & 0.009252 & s \\
120819A & 71 & L & 235.9075 & -7.3091 & 1.7 & 235.9076 & -7.3093 & 0.92 & 0.002322 & ND \\
130515A & 0.29 & S & 283.4401 & -54.2791 & 2.4 & 283.4385 & -54.2792 & 3.44 & 0.040879 & NC/s \\
130527A & 44 & L & 309.2763 & -24.7250 & 1.4 & 309.2761 & -24.7247 & 1.31 & 0.005843 & ND/g \\
130528A & 59.4 & L & 139.5051 & 87.3012 & 1.9 & 139.4988 & 87.3015 & 1.48 & 0.008832 & ND/LG/CA \\
130603B & 0.18 & S & 172.2006 & 17.0714 & 1.4 & 172.2012 & 17.0714 & 1.85 & 0.006614 & G \\
130725A & 101.8 & L & 230.0324 & 0.6276 & 1.8 & 230.0318 & 0.6276 & 2.09 & 0.007389 & ND \\
130907A & >360 & L & 215.8922 & 45.6073 & 1.4 & 215.8921 & 45.6070 & 0.78 & 0.000969 & PG \\
131018B & ${\sim}$38 & L & 304.5369 & 23.1876 & 4.9 & 304.5361 & 23.1876 & 2.84 & 0.009425 & Ps \\
131122A & ${\sim}$70 & L & 152.5422 & 57.7277 & 4.8 & 152.5440 & 57.7292 & 6.50 & 0.044608 & PG \\
140331A & 209 & L & 134.8644 & 2.7173 & 1.7 & 134.8650 & 2.7175 & 2.05 & 0.007334 & G \\
140927A & 6.26 & L & 291.7916 & -65.3936 & 1.8 & 291.7922 & -65.3932 & 1.64 & 0.000745 & Ps \\
141212A & 0.3 & S & 39.1248 & 18.1470 & 2.6 & 39.1254 & 18.1468 & 2.23 & 0.018170 & G \\
150101B & 0.018 & S & 188.0205 & -10.9336 & 1.8 & 188.0207 & -10.9335 & 0.67 & 0.000070 & G \\
150120A & 1.2 & S & 10.3189 & 33.9949 & 1.8 & 10.3193 & 33.9952 & 1.49 & 0.003069 & PG \\
150323C & 159.4 & L & 192.6169 & 50.1912 & 1.6 & 192.6162 & 50.1909 & 1.93 & 0.010869 & ND/g \\
150626A & 144 & L & 111.3368 & -37.7808 & 1.8 & 111.3370 & -37.7813 & 1.97 & 0.005876 & NC \\
151111A & 76.93 & L & 56.8448 & -44.1615 & 1.5 & 56.8447 & -44.1619 & 1.53 & 0.004170 & NC \\
160703A & 44.4 & L & 287.4168 & 36.9175 & 3.9 & 287.4164 & 36.9174 & 1.16 & 0.005082 & Ps \\
161001A & 2.6 &  L? & 71.9200 & -57.2608 & 1.4 & 71.9195 & -57.2604 & 1.69 & 0.007076 & NC \\
161007A & 201.7 & L & 103.4090 & 23.3068 & 1.5 & 103.4087 & 23.3064 & 1.65 & 0.011553 & ND/g \\
161010A & ${\sim}$30 & L & 275.2143 & -28.7852 & 2.9 & 275.2144 & -28.7862 & 3.68 & 0.017848 & s \\
161104A & 0.1 & S & 77.8937 & -51.4601 & 3 & 77.8941 & -51.4613 & 4.47 & 0.024994 & NC \\
161108A & 105.1 & L & 180.7879 & 24.8682 & 1.5 & 180.7885 & 24.8678 & 2.44 & 0.006978 & PG \\
161214B & 24.8 & L & 3.8512 & 7.3524 & 1.5 & 3.8510 & 7.3524 & 0.73 & 0.000854 & s \\
\hline 
\end{tabular}
\newline
{s - star. Ps - photometric star. G - galaxy. PG - photometric galaxy. NC - no coverage. ND - optical non-detection. LG - identified as an IR-bright host galaxy by comparison to the published literature. CA - rejected due to possible or confirmed chance alignment.}
\label{tab:samp}
\end{table*}

\begin{table*}
\centering 
\caption{The 15 GRB X-ray positions which match to at least one \WISE\ source within 1.5R$_{90}$, but which have these matches rejected due to FAP, CCF or \WISE\ blending cuts.} 
\begin{tabular}{p{1.0cm} c c c c c c c c c r}
\hline 
GRB & T$_{90}$ & Short/ & X-ray Ra & X-ray Dec & R$_{90}$ & \WISE\ Ra & \WISE\ Dec & Sep & FAP & Type  \\ [0.25ex] 
 & [s] & Long & [deg] & [deg] & [arcsec] & [deg] & [deg] & [arcsec] & - & -   \\
\hline 
050117 & 166.6 & L & 358.4708 & 65.9389 & 15 & 358.4747 & 65.9404 & 7.78 & 0.787461 & FR \\
050306 & 158.3 & L & 282.3088 & -9.1531 & 6 & 282.3101 & -9.1545 & 6.87 & 0.004357 & CC \\
060223B & 10.3 & L & 254.2450 & -30.8128 & 10 & 254.2454 & -30.8141 & 4.94 & 0.197244 & FR \\
060502B & 0.131 & S & 278.9385 & 52.6315 & 15 & 278.9413 & 52.6328 & 7.70 & 0.275397 & FR \\
060801 & 0.49 & S & 213.0055 & 16.9818 & 1.5 & 213.0059 & 16.9818 & 1.40 & 0.007557 & WB \\
061007 & 75.3 & L & 46.3317 & -50.5007 & 1.4 & 46.3318 & -50.5007 & 0.23 & 0.000191 & WB \\
071001 & 58.5 & L & 149.7336 & -59.7818 & 6 & 149.7353 & -59.7822 & 3.44 & 0.004577 & CC \\
071109 & ${\sim}$30 & L & 289.9746 & 2.0465 & 9 & 289.9747 & 2.0463 & 0.96 & 0.022986 & CC \\
080212 & 123 & L & 231.1474 & -22.7417 & 1.4 & 231.1469 & -22.7415 & 1.68 & 0.007351 & CC \\
100909A & ${\sim}$70 & L & 73.9473 & 54.6594 & 5.4 & 73.9510 & 54.6594 & 7.75 & 0.161417 & CC \\
120419A & ${\sim}$20 & L & 187.3876 & -63.0079 & 4.5 & 187.3876 & -63.0095 & 5.62 & 0.075810 & CC \\
120811A & 166 & L & 257.1654 & -22.7106 & 2.8 & 257.1658 & -22.7114 & 3.31 & 0.047855 & WB \\
140103A & 17.3 & L & 232.0875 & 37.7592 & 3.6 & 232.0876 & 37.7577 & 5.31 & 0.127637 & FR \\
150301A & 0.48 & S & 244.3047 & -48.7131 & 5 & 244.3019 & -48.7136 & 6.81 & 0.188791 & FR \\
151004A & 128.4 & L & 213.6322 & -64.9391 & 7 &  213.6343 &  -64.9369 &  8.58 &  0.091087 &  FR \\
\hline 
\end{tabular}
\newline
{CC - flagged as confused in \WISE. WB - flagged as a blend in W1 band. FR - Rejected due to FAP>0.05}
\label{tab:samp_reject}
\end{table*}

For the region around each burst, the local density of sources of magnitude equal to the \WISE\ match or brighter is given by,
\begin{equation}
{\Sigma}(m{\leq}m_{g}) = \frac{N(m{\leq}m_{g})}{{\pi}r_{3}^{2}} ,
\end{equation}
where $N(N_{m{\leq}m_{g}})$ is the number of sources within 3\,arcmin of the burst of W1 magnitude $m_{g}$ or brighter, and $r_{3}$ = 3\,arcmin. The probability of a match at angular distance $r$ being genuine and not a false alarm can be written as,
\begin{equation}
P_{\mathrm{chance}} = e^{-{\Sigma}(m{\leq}m_{g}){\pi}r^{2}},
\end{equation}
which tends to 1 as $r$ tends to 0, and tends to 0 as $r$ tends to ${\infty}$, as required. Using this method, the false alarm probability (FAP) is given by $1 - P_{\mathrm{chance}}$. The CCF flag cuts from section \ref{sec:xmatching} are again used. After these are made, a cut of FAP<0.05 is chosen. This cut, when applied to the matching of random positions to ALLWISE, yields a theoretical maximum of 50 matches by chance. However, the distribution continues well below 0.05. The average FAP is therefore is much lower, and the number of false matches will also be lower. This is backed up by the addition of only 7 matches when going from a FAP cut of 0.025 to 0.05. Included in these 7 is LGRB\,100316D, which has a previously noted $z$=0.059 host galaxy. Using a FAP cut of 0.05 allows us to catch hosts which have larger projected sizes, such as that of LGRB\,100316D. In addition, it allows us to identify the hosts of bursts with large X-ray uncertainties, provided the field is not crowded and the \WISE\ source is sufficiently bright. 

As in section \ref{sec:xmatching}, SGRB\,060801 and LGRB\,061007 are removed due to possible blending in \WISE, in addition to LGRB\,120811A. Three GRBs with more than one IR match (GRBs\,060223B, 071007 and 071109) had all of their candidates rejected due to CCF or FAP cuts. The fourth example with more than one match, LGRB\,050117, lies in the galactic plane ($|b|$ = 3) and has two matches almost equidistant at ${\sim}$7\,arcsec, with similar false alarm probabilities. We cannot distinguish which IR source is more likely to be associated, and the line of sight extinction meant there was no optical afterglow reported for this burst, precluding a improved localisation. Therefore, we reject LGRB\,050117. This leaves us with a final sample of 55 bursts, each with one matched \WISE\ source. This differs from the sample derived in section \ref{sec:xmatching}, in that ten extra bursts are included: LGRB\,050522, LGRB\,070309, LGRB\,080405, SGRB\,100206A, LGRB\,100316D, LGRB\,130118B, SGRB\,130515A, LGRB\,131122A, LGRB\,161010A and SGRB\,161104A. 


\subsection{Sample Summary}\label{sec:sampsum}
The final sample includes candidate host galaxies for 55 GRBs. These are listed in table \ref{tab:samp}, with the candidates rejected for CCF flags, \WISE\ blending and high false alarm probabilities given in table \ref{tab:samp_reject}. The tables give the X-Ray coordinates, \WISE\ coordinates, T$_{90}$ estimates, the 90 per cent confidence interval on the X-ray position, the X-ray-\WISE\ separation and a false alarm probability for association with the \WISE\ source. Given the analysis in section \ref{sec:xmatching}, we are confident that around a third to two-thirds of the associations are spurious. However, some will be Galactic stars, and others may be galaxies with properties inconsistent with being a GRB host. These contaminants can be identified as such through their photometric and spectroscopic properties, as well as through better burst localisation. Observations and archival searches for these observations are thus the next objective of this analysis.

\begin{table}
\centering 
\caption{WHT/ACAM observations, taken on 2015 Jan 19/20. If a magnitude uncertainty is not given, the value corresponds to the 2\,${\sigma}$ limit at the position of the WISE source.} 
\begin{tabular}{l c c c c p{0.6cm}}
\hline 
Target & Filter & Int. [s] & Mag(AB) & 2\,${\sigma}$ depth & Seeing [arcsec] \\ [0.25ex] 
\hline 
061002       & g & 573 & 22.27 $\pm$ 0.06 & 24.6 & 1.86 \\
             & r & 573 & 21.84 $\pm$ 0.05 & 23.9 & 1.82 \\ 
             & i & 572 & 21.09 $\pm$ 0.05 & 23.8 & 1.88 \\ 
             & z & 573 & 20.45 $\pm$ 0.07 & 22.0 & 1.88\\ 
070208       & g & 573 & 19.81 $\pm$ 0.01 & 24.9 & 1.80 \\ 
             & r & 730 & 19.46 $\pm$ 0.01 & 24.1 & 2.25\\ 
             & i & 573 & 19.25 $\pm$ 0.02 & 22.8 & 1.75\\ 
             & z & 573 & 19.13 $\pm$ 0.03 & 20.8 & 1.66 \\ 
080307       & g & 897 & 24.1 $\pm$ 0.2 & 24.6 & 1.78 \\ 
             & r & 1653 & 23.0 $\pm$ 0.1 & 24.3 & 1.90\\ 
             & i & 213 & > 22.8 & 23.4 & 1.78\\ 
             & z & 731 & > 23.9 & 22.4 & 2.32 \\ 
100816A      & g & 573 & > 24.1 & 24.0 & 2.52 \\ 
             & r & 693 & 22.6 $\pm$ 0.1 & 23.6 & 1.67\\ 
             & i & 514 & > 21.3 & 21.8 & 1.49\\ 
111222A      & g & 261 & 19.11 $\pm$ 0.02 & 22.5 & 2.46\\ 
             & r & 81 & 18.14 $\pm$ 0.03 & 20.8 & 1.65 \\ 
             & i & 81 & 16.64 $\pm$ 0.01 & 20.5 & 1.73\\ 
             & z & 180 & 15.89 $\pm$ 0.02 & 22.2 & 1.79\\ 
140331A      & g & 491 & > 25.0 & 24.6 & 2.74\\ 
             & r & 933 & 22.59 $\pm$ 0.09 & 23.9 & 2.38\\ 
             & i & 573 & > 20.0 & 21.1 & 2.67\\ 
             & z & 371 & 20.80 $\pm$ 0.06 & 22.4 & 2.31 \\ 
141212A      & r & 573 & 22.8 $\pm$ 0.1 & 23.3 & 2.08\\ 
             & i &  573 & 22.7 $\pm$ 0.2 & 23.0 & 2.38 \\ 
             & z & 573 & > 21.9 & 22.4 & 1.46 \\  
\hline 
\end{tabular}
\label{tab:whtimg}
\end{table}

\section{Observations}\label{sec:obs}
In order to investigate the true hosts and determine which matches are spurious, we have observed subsets of the sample with WHT/ACAM, VLT/X-shooter and ATCA. 7 targets were observed with ACAM/WHT, 5 with VLT/X-shooter and 14 with ATCA.

\subsection{WHT Imaging}\label{sec:wht}
Observations of 7 candidate hosts were taken over two nights (2015 January 19 and 20) with the auxiliary-port camera (ACAM) on the William Herschel Telescope (WHT). These were associated with programme WHT/2015A/34. Both nights were severely affected by poor observing conditions. The object associated with SGRB\,111222A was observed on 2015 January 19, however observations were hampered by clouds and wind gusts in excess of 70\,km\,h$^{-1}$. The objects associated with LGRB\,100816A, SGRB\,141212A, LGRB\,140331A, LGRB\,070208, LGRB\,061002 and LGRB\,080307 were observed on 2015 January 20. Conditions were clearer but still windy, with poor seeing ($\sim$2\,arcsec). Sloan $g$, $r$, $i$ and $z$ filters were used. The images were reduced with standard {\sc IRAF} procedures and aperture photometry performed on the candidate hosts. Aperture sizes were chosen to be ${\sim}$2 times the seeing FWHM (of the largest band) if the target was a point source, or else ${\sim}$2 times the FWHM of the object of interest. Aperture sizes were kept constant for each source. Details of the observations and measured quantities are listed in table \ref{tab:whtimg}, along with 2\,${\sigma}$ depths and the seeing. The quoted magnitudes are in broad agreement with archival data where available.

\begin{table}
\centering 
\caption{Details of the VLT/X-shooter and WHT/ACAM spectroscopic observations.} 
\begin{tabular}{p{0.9cm}p{0.6cm}p{1.5cm}p{0.6cm}p{0.6cm}p{0.6cm}p{0.8cm}}
\hline 
Target & Obsv. & Date & Int.UV [s] & Int.Vis [s] & Int.IR [s] & Seeing [arcsec] \\ [0.25ex] 
\hline 
111222A & WHT &  2015 Jan 19 & 1260 & 1260 & - & 1.91 \\ 
140331A & WHT &  2015 Jan 20 & 1255 & 1255 & - & 2.53 \\  [1ex] 
091102  & VLT & 2015 Dec 07 & 1800 & 2220 & 2160 & < 0.8 \\ 
091102  & VLT &  2015 Dec 07 & 1800 & 2220 & 2160  & < 0.8 \\ 
120224A & VLT &  2015 Dec 13 & 1800 & 2220 & 2160 & < 0.8 \\ 
120224A & VLT &  2015 Dec 14 & 1800 & 2220 & 2160 & < 0.8 \\ 
120612A & VLT &  2015 Dec 13 & 1800 & 2220 & 2160 & < 0.8 \\ 
120612A & VLT &  2015 Dec 14 & 1800 & 2220 & 2160 & < 0.8 \\ 
140331A & VLT &  2015 Dec 14 & 1788 & 2190 & 2664 & < 0.8 \\ 
140331A & VLT &  2016 Jan 07 & 1788 & 2190 & 2664 & < 0.8 \\ 
141212A & VLT &  2015 Dec 08 & 1788 & 2190 & 2664 & < 0.8 \\ 
141212A & VLT &  2015 Dec 15 & 1788 & 2190 & 2664 & < 0.8 \\
\hline 
\end{tabular}
\label{tab:vltspc}
\end{table}

\begin{figure*}
\centering
\includegraphics[width=0.85\textwidth]{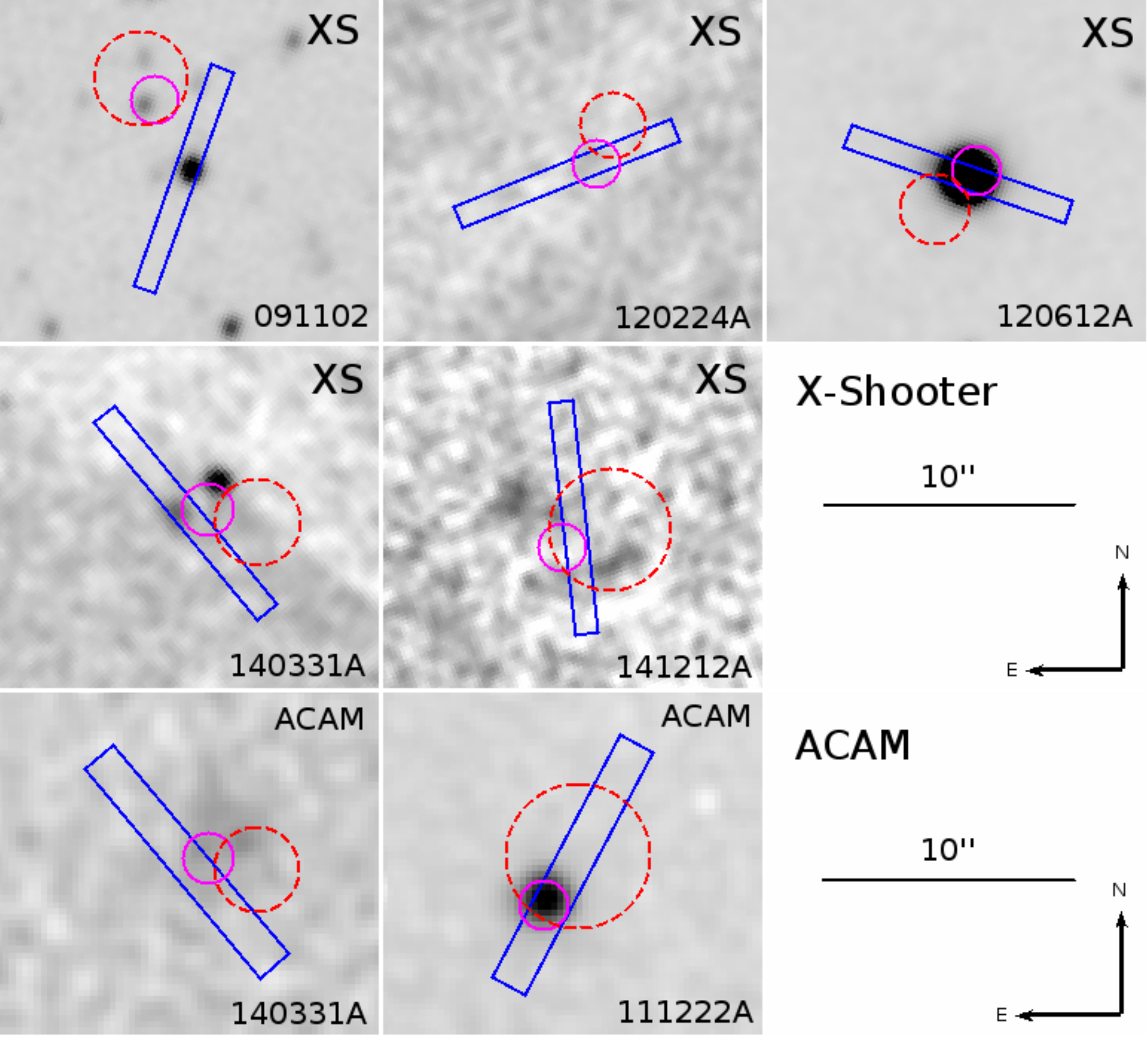}
\caption{Slit positions for the VLT X-shooter and WHT ACAM spectroscopic observations, with the enhanced 90 per cent X-ray error radii overlaid in red. Blue rectangles represent the slit positions. Solid magenta circles indicate the centroid of the \WISE\ sources. All images are in the {\it r}-band, and are stretched and smoothed with a Gaussian kernel of radius 3 pixels.}
\label{fig:slits}
\end{figure*}

\begin{figure}
\centering
\includegraphics[width=0.9\linewidth]{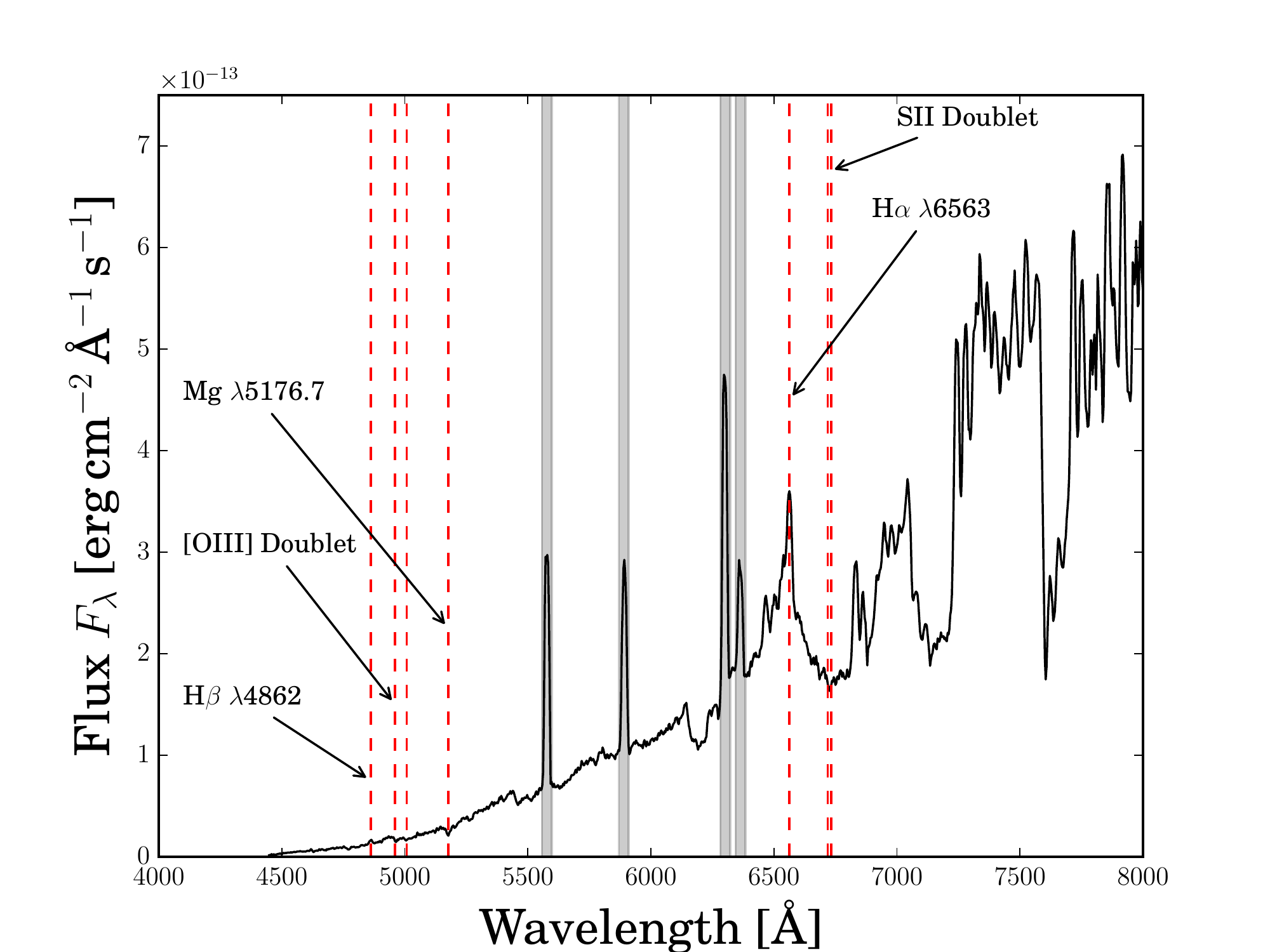}
\caption{The WHT optical spectrum of the object associated with SGRB\,111222A. The spectral shape and presence of absorption and emission lines at redshift ${\sim}$ 0 indicate that this is an M-star. OI, O$_{2}$ and NaD sky features are masked out.}
\label{fig:whtspec}
\end{figure}

\begin{figure}
\centering
\includegraphics[width=0.9\linewidth]{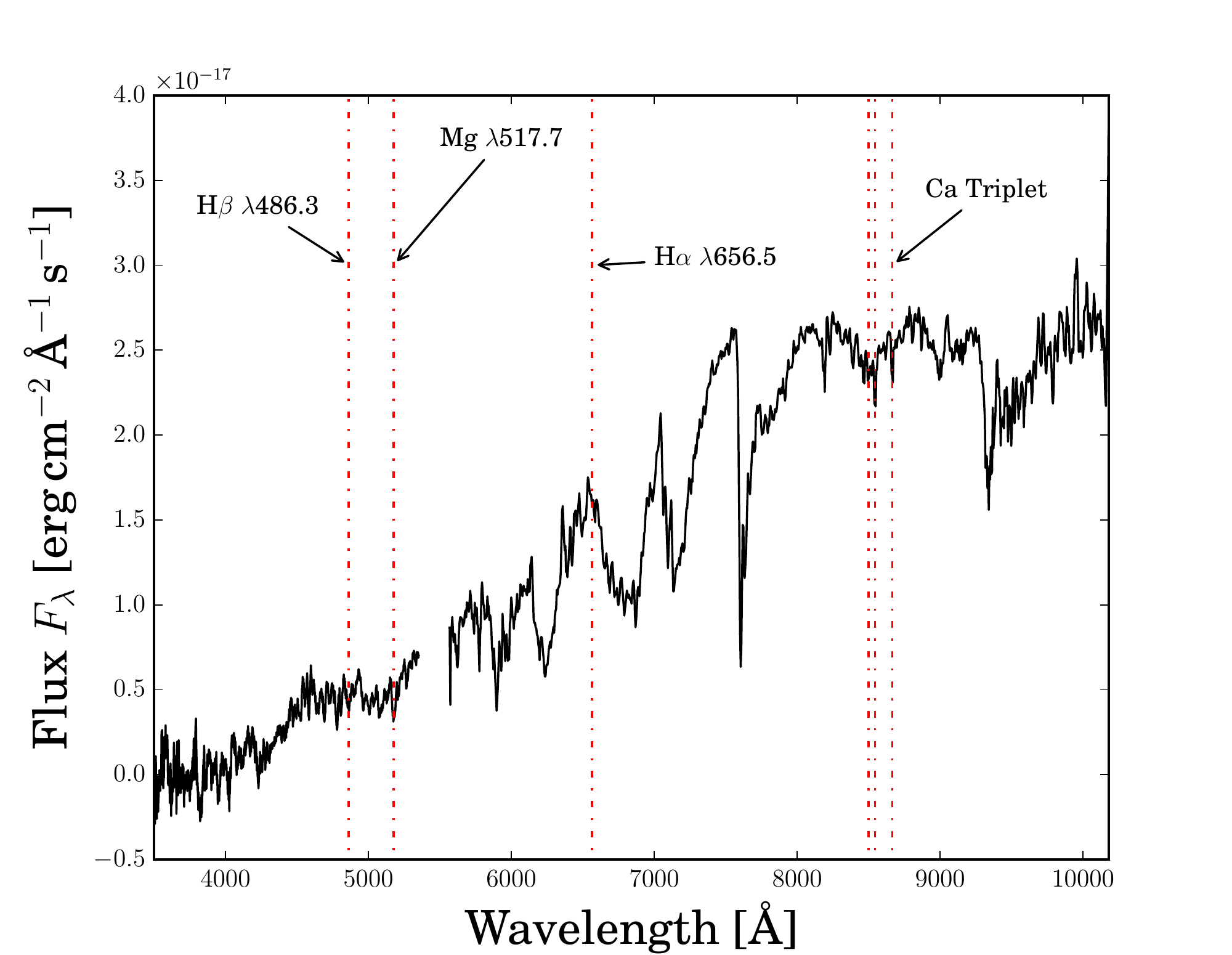}
\caption{The VLT/X-shooter spectrum of the object associated with LGRB\,091102 target. H${\beta}$, Mg, H${\alpha}$ and Ca absorption lines at negligible redshift confirm that this is an M-star. However, the slit was misaligned with the IR source, so we do not consider this identification any further.}
\label{fig:xs1}
\end{figure}

\begin{figure}
\centering
\includegraphics[width=0.9\linewidth]{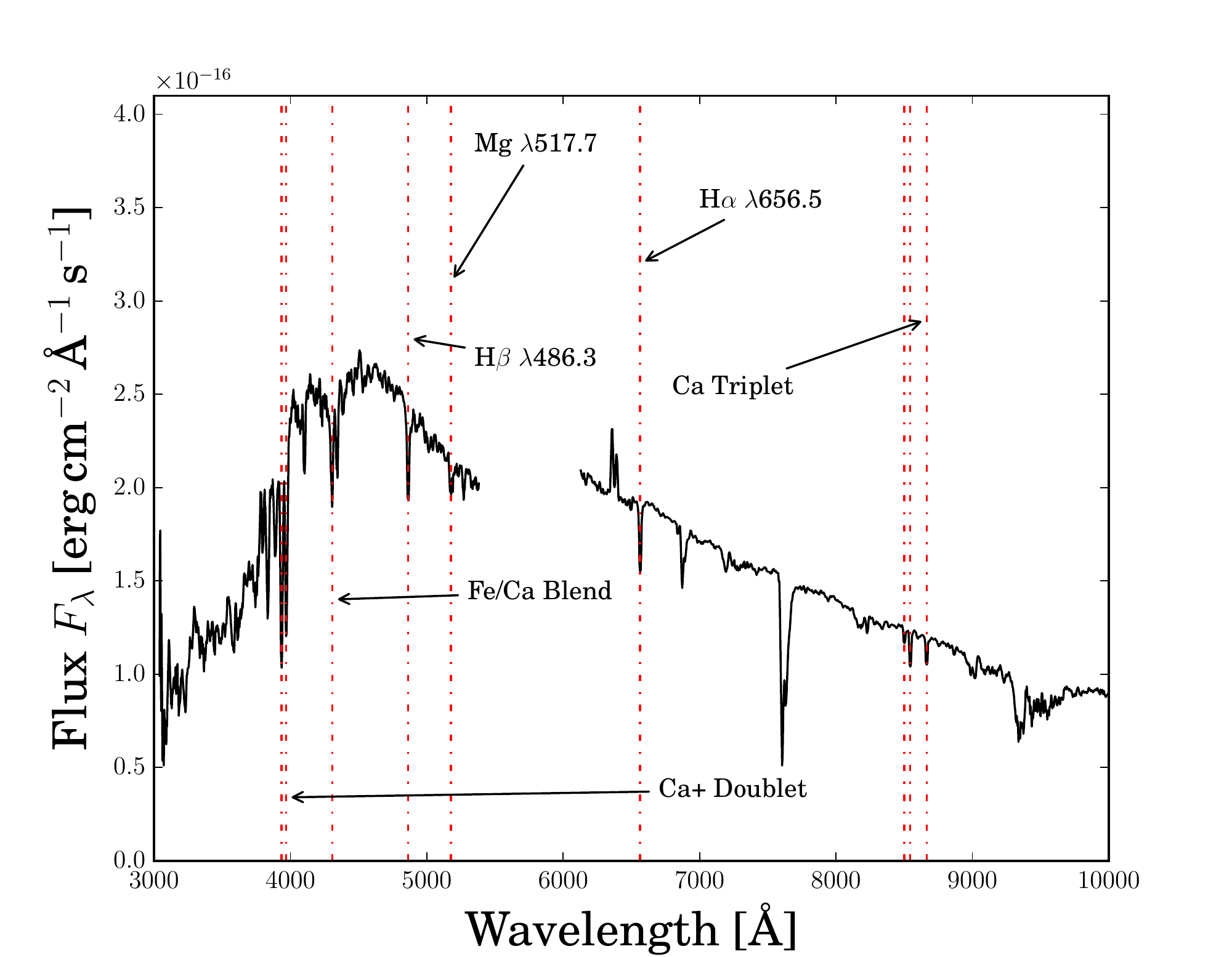}
\caption{The VLT/X-shooter optical spectrum of the LGRB\,120612A target. H${\beta}$, H${\alpha}$ and various metal absorption lines at $z {\approx}$ 0 indicate that this is a foreground star.}
\label{fig:xs2}
\end{figure}

\begin{figure*}
\centering
\includegraphics[width=1\textwidth]{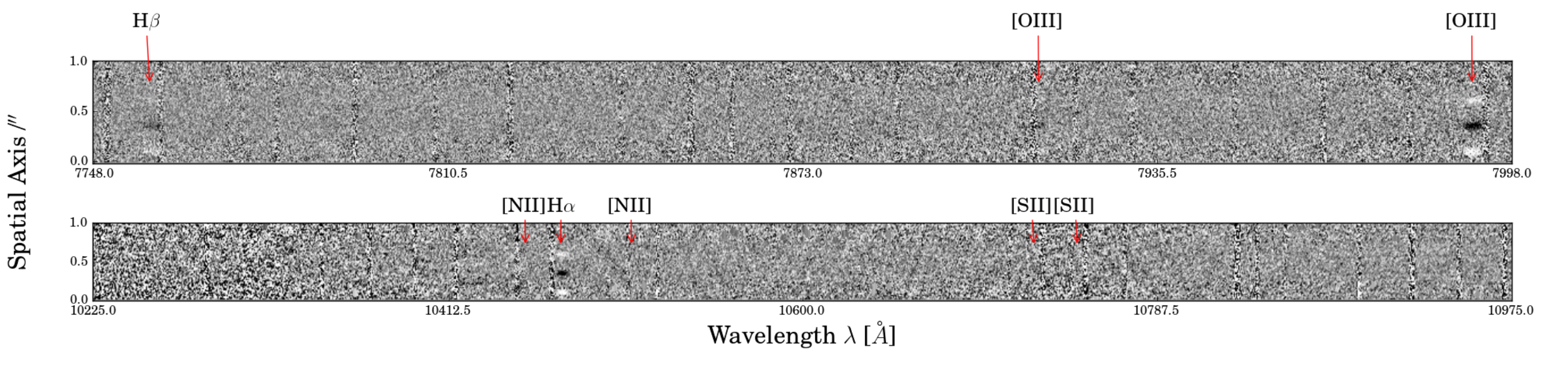}
\caption{Emission lines from the host galaxy of SGRB\,141212A. The upper nodded spectrum is from X-shooter's VIS arm, the lower from the NIR arm. The marked wavelengths correspond to, in order of increasing wavelength, H${\beta}$ and [OIII] on the VIS arm and [NII], H${\alpha}$, [NII] and the [SII] doublet in the NIR. The observed lines indicate a redshift of 0.596$\pm$0.001.}
\label{fig:emission}
\end{figure*}

\subsection{WHT Spectroscopy}
The candidate hosts of SGRB\,111222A and LGRB\,140331A were observed on 2015 January 19 and 20 respectively, using the V400 grism and a 1.5\,arcsec slit on ACAM. The position of the slit with respect to the 90 per cent XRT error circle and \WISE\ source are shown in figure \ref{fig:slits}, overlaid on {\it r}-band images. Given the poor seeing, slit losses were significant. The observations are listed in table \ref{tab:vltspc}. The LGRB\,140331A candidate counterpart was not detected. The SGRB\,111222A IR counterpart is identified as an M dwarf, as shown in figure \ref{fig:whtspec}. 

\subsection{VLT Spectroscopy}
We observed five GRB host candidates using the echelle spectrograph on VLT/X-shooter \citep{XS}. Observations were associated with programme 096.D-0260(A) (PI: Stanway) and are detailed in table \ref{tab:vltspc}. Images in the {\it r}-band and the position of the slit with respect to the XRT and \WISE\ positions are also shown in figure \ref{fig:slits}. The XS images are used primarily for visualising the slit placement. The 2\,${\sigma}$ depths of the XS images corresponding to GRBs\,091102, 120224A, 120612A, 140331A and 141212A are 23.5, 22.6, 24.1, 22.9 and 21.7 respectively. In each case, the slit placement was chosen to overlap with the \WISE\ source. 

The spectra were reduced using the standard ESO pipeline in {\sc Gasgano}. Of the five targets, two were marginally detected (LGRBs\,120224A and 140331A), one was detected with prominent emission lines (SGRB\,141212A), and two were found to be foreground stars (LGRBs\,091102 and 120612A). The two stellar spectra are shown in figures \ref{fig:xs1} and \ref{fig:xs2} respectively. We note that service-mode observations of the counterpart of LGRB\,091102 has been misaligned, likely due to the misidentification of a selected offset and alignment star. Thus the spectroscopic identification of this as a star is irrelevant to the GRB and is presented here to avoid confusion in future studies of archival data for these observations. This is the only case where we are required to attempt photometry on the XS imaging, measuring an {\it r}-band magnitude of the faint \WISE\ aligned source of ${\sim}$21.4 \citep[anchored to the APOP magnitude for the nearby bright star, ][]{APOP}. However, it is not detected in the other bands available, $u$ and $z$. Combined with GALEX and 2MASS non-detections, we deem there to be insufficient data for fitting the SED of this object. We classify it as having no coverage in table \ref{tab:samp}. 

The spectra of the LGRB\,120224A and 140331A targets are featureless, with only marginally detected continuum flux and no readily identifiable absorption or emission lines. LGRB\,120224A's candidate host has also been observed with X-shooter in a different programme, with similar results \citep{120224A}. If these are indeed the LGRB host galaxies, this requires them either to be mature stellar systems without nebular emission, or else heavily dust enshrouded. It should be noted that the WCS of the X-shooter images are misaligned with ALLWISE and the X-ray positions, leading to small offsets from their true positions. For example, the slit for LGRB\,140331A has been deliberately placed over the fainter object south-east of the error circle, because the centroid of the IR flux aligns with it, suggesting that it corresponds to the source of the IR emission. While the other, brighter object \textit{might} be the true host, it is likely not IR-bright, and would therefore be out of place in our sample. 

The potential host of SGRB\,141212A has a weak continuum with H${\alpha}$, H${\beta}$ and \ion{O}{iii} emission lines. The wavelength of these correspond to a redshift of 0.596$\pm$0.001. This is in agreement with \citet{Chornock}, who observed an object within the enhanced XRT error circle one day post burst with the Gemini-N spectrograph. They found that, out of two objects near the error circle, the likely host has a redshift $z$=0.596. Portions of the 2D spectrum covering key emission lines are shown in figure \ref{fig:emission}. Emission line measurements are listed in table \ref{label:141212A}. Owing to the low signal-to-noise ratio, meaningful constraints on the H$\alpha$/H$\beta$ ratio are not possible. However, we note that the presence of these lines is in qualitative agreement with the star-forming best-fitting SED as discussed later in section \ref{sec:sedfitting}. 

\begin{table}
\centering 
\caption{Emission line measurements from the galaxy associated with SGRB\,141212A.} 
\begin{tabular}{l c c}
\hline 
Line & $\lambda$  & Flux  \\
 & [\AA] & [$10^{-14}$\,ergs\,cm$^{-2}$\,s$^{-1}$]\\ [0.25ex] 
\hline 
H${\beta}$ & 7759$\pm$4 & 0.53$\pm$0.47 \\ 
$[$OIII$]$ & 7991$\pm$3 & 1.8$\pm$0.8 \\ 
H${\alpha}$ & 10474$\pm$5 & 2.0$\pm$0.9 \\  [1ex] 
\hline 
\end{tabular}
\label{label:141212A}
\end{table}

\begin{table}
\centering 
\caption{Observations and upper limits on the radio emission of the 14 ATCA targets.}
\begin{tabular}{l c c c r}
\hline 
GRB & z & Beam FWHM  & Image RMS  & 3\,${\sigma}$ SFR Limit  \\ 
 & & [arcsec] & [${\mu}$Jy] & [M$_{\odot}$yr$^{-1}$]\\
[0.25ex] 
\hline 
050219A &  0.212   &  9.7$\times$1.5    & 10.5  &      $<$10\\
050318  &   1.44   &  5.2$\times$1.9    & 10.2  &      $<$990\\
070429B &  0.902   &  8.9$\times$1.9    & 9.6   &      $<$290\\
070724A &  0.457   &   13$\times$1.8    & 10.7  &      $<$62\\
071117  &   1.33   &  6.4$\times$1.5    & 12.9  &      $<$1020\\
080623  &    --    &  5.5$\times$1.8    & 13.3  &   - \\
080702B &   2.09   &   51$\times$1.6    & 17.2  &      $<$4100\\
091102  &   --     &  4.0$\times$1.8    & 10.3  &  - \\
110206A &   --     &  3.4$\times$2.1    & 11.5  &  -\\
110918A &  0.984   &   12$\times$1.9    & 12.9  &      $<$480\\
120119A &   1.73   &   34$\times$1.8    & 10.0  &      $<$1500\\
120224A &   --     &   17$\times$1.8    & 9.0   &  -\\
120612A &   --     &   18$\times$1.7    & 11.1  &  -\\
120819A &   --     &   42$\times$1.7    & 10.8  &  -\\
\hline 
\end{tabular}
\label{label:radio}
\end{table}

\subsection{ATCA Radio Observations}\label{sec:radio}
Radio observations of 14 candidate hosts were made at central frequencies of 5.5 GHz and 9.0 GHz and a bandwidth of 2\,GHz per frequency, with the Australia Telescope Compact Array (ATCA). Science targets and secondary phase calibrators were observed during programme C3002 (PI: Stanway). Observations were taken on 2015 January 31 and 2015 February 1 and 2. The array was in its most extended, 6A, configuration with a maximum baseline of 6km and six antennae in use. Short observations were taken across a range of hour angles to secure reasonable $uv$-plane coverage. The data were reduced with the standard data reduction software {\sc Miriad}. Absolute flux calibration was performed using observations of PKS\,1934-638. 

None of the targets were detected. The observations are listed in table \ref{label:radio}. We also list the synthesized beam size, which varied significantly from source to source given their wide range of declinations, and the final image RMS noise level. Where a redshift for the source is known, we use the 1.4\,GHz flux to star formation rate (SFR) calibration of \citet{2012ARA&A..50..531K} to estimate a 3\,$\sigma$ upper limit on the star formation rate (assuming a radio spectral slope of -1).

\section{Archival and Survey Data}\label{sec:archival}

\begin{figure*}
\centering
\begin{minipage}[t]{0.49\linewidth}
\centering
\includegraphics[width=0.95\linewidth]{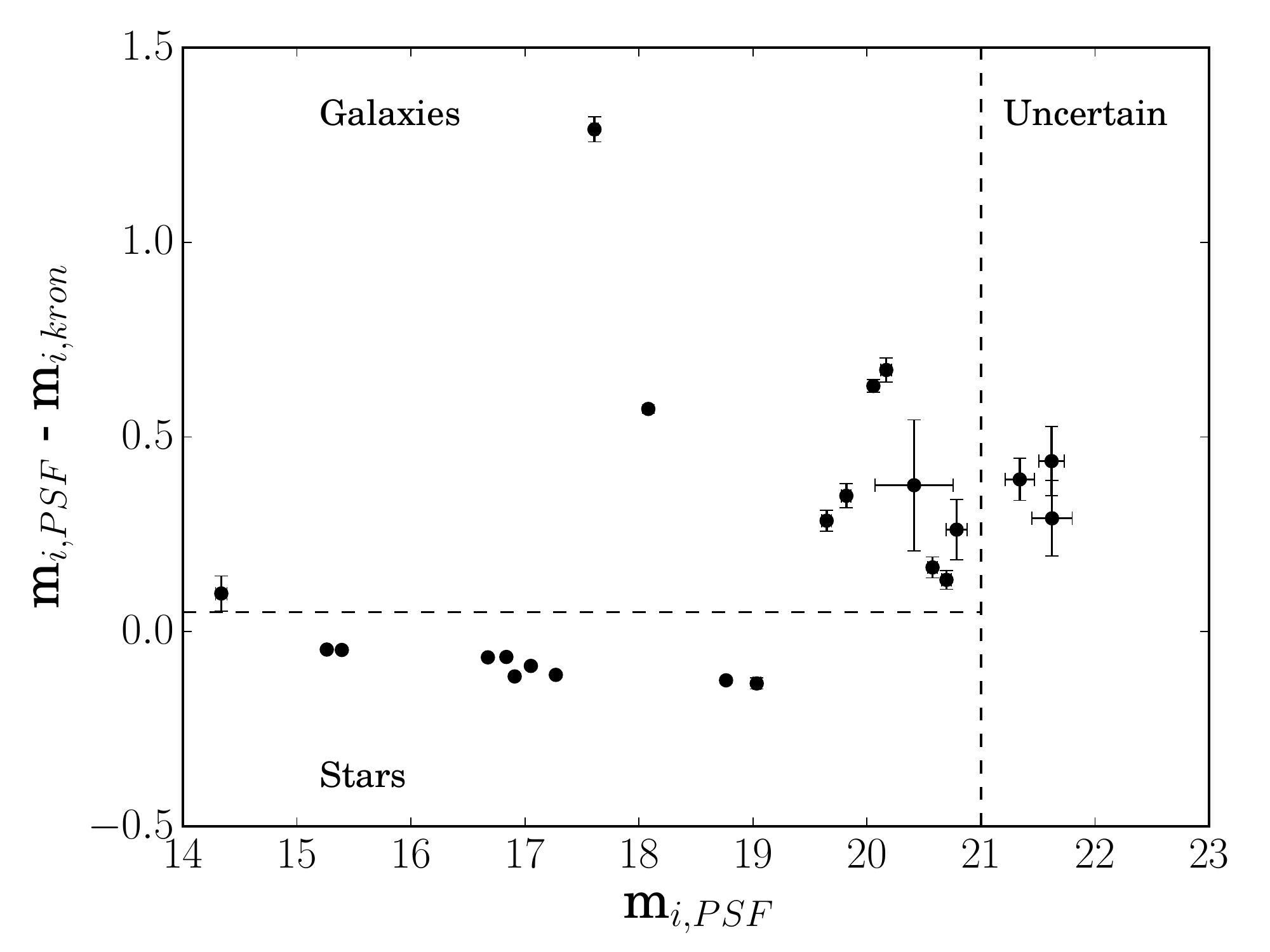}
\caption{Star-galaxy separation for our sources which have an {\it i}-band PSF and Kron magnitude. Faint sources cannot be reliably separated. }
\label{fig:psfkron}
\end{minipage}
\quad
\begin{minipage}[t]{0.49\linewidth}
\centering
\includegraphics[width=0.95\linewidth]{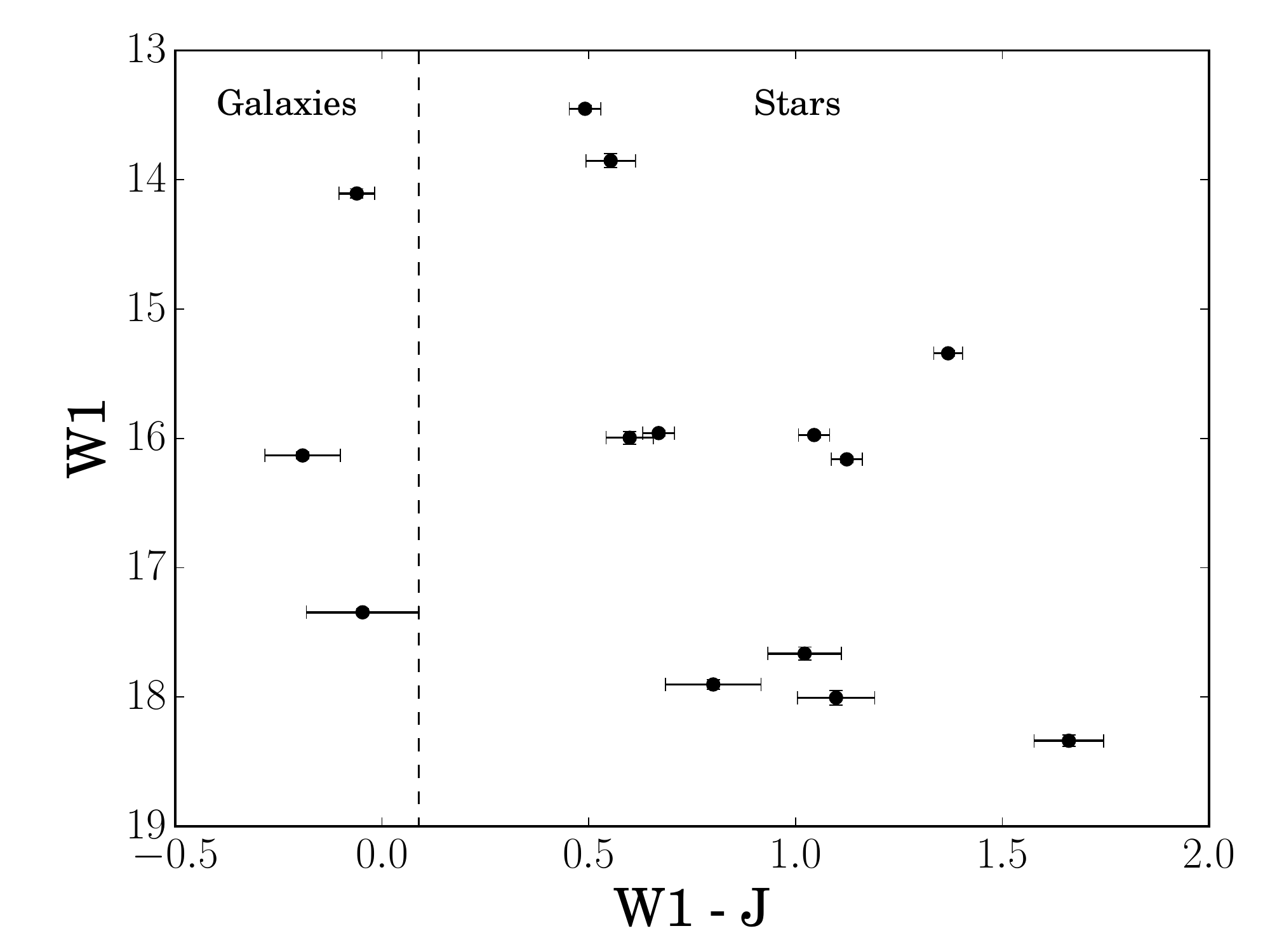}
\caption{Star-galaxy separation for our sources which have a W1 and {\it J}-band magnitude. The {\it J} magnitudes dominate the uncertainty here.}
\label{fig:wisesep}
\end{minipage}
\end{figure*} 

Additional information for the candidate GRB host counterparts was gathered from archival surveys and the literature. The main source of optical photometry is the Pan-STARRS survey \citep[DR1,][]{PSDR1}, which covers the whole sky north of -30\degr\ declination down to 3\,${\sigma}$ depths of $g,r,i,z,y$ $<$ 23.3,23.2,23.1,22.3,21.3. Cross-matching between the Pan-STARRS 1 science archive and the 55 \WISE\ counterparts was performed with a 2.5\,arcsec matching radius, producing 27 matches. Of these, a small subset have more than one possible optical counterpart, and these are carefully considered in section \ref{sec:ambiguity}. The candidate hosts for SGRB\,141212A are visible in Pan-STARRS but below the cataloging threshold, so we measure the magnitudes from image cutouts, complementing our WHT photometry presented in section \ref{sec:wht}. The Sloan Digital Sky Survey \citep[SDSS,][ we use DR12]{SDSS12}, VST/ATLAS \citep{ATLAS} and APASS \citep{APASS} were also searched by matching to within 2.5\,arcsec of the \WISE\ source. All of these surveys extend south of -30\degr\ declination. Two GRB locations have matches in VST/ATLAS, SGRBs\,070724A and 150101B, which are also covered by Pan-STARRS, and only one in APASS (LGRB\,140927A). There are 12 matches in SDSS, which provides the only optical photometry for LGRB\,161108A. In the remaining 11 matched cases, Pan-STARRS data also exists, and we use the best available combination of photometry. At the very least, $u$ band limits are used from the SDSS matches. Overall, we have a total of 29 optical survey detections.

The remaining 26 positions may fail to obtain a match because the source lies outside the Pan-STARRS survey region (14 objects), or because the galaxy is optically faint (12 instances). Where a GRB falls in a field covered by one of the surveys used, but no object is detected at that position, 2\,${\sigma}$ upper limits are used. Four host candidates lacking coverage have been studied in the literature (GRBs\,050219A, 070429B, 100316D and 130515A), so 10 are classified 'NC' (no coverage) in table \ref{tab:samp}. 

A total of 6 of the 12 optically undetected sources have been well studied in the literature. In these cases we use the results of those works. The total number of \WISE\ sources for which we have survey coverage but are lacking information on an optical counterpart, is therefore 6. The 12 undetected sources are discussed further in sections \ref{sec:prevreport} and \ref{sec:results}. 

The {\it GALEX} All Sky Imaging Survey \citep[AIS,][]{Martin} provides UV photometry or limits for all of our objects at near-constant depth \citep{GALEX}. {\it GALEX} has two photometric bands, the Far and Near UV (FUV and NUV), with effective wavelengths of 1528{\AA} and 2271{\AA} respectively. Ten of the sample have a NUV source within 5.3\,arcsec, the NUV PSF FWHM. Expanding the search radius to 10\,arcsec yields only one more match (at 9\,arcsec), suggesting that those matches identified are genuine. Of these, four (LGRB\,080405, LGRB\,080517, LGRB\,100316D and SGRB\,150101B) also have a FUV detection. Where we have no detection, we use the AIS mean 2\,${\sigma}$ upper limits, m$_{FUV}$ = 20.89 and m$_{NUV}$ = 21.79. 

The 2 Micron All Sky Survey (2MASS) was used to provide NIR data or limits for our sample \citep{2MASS}. A catalogued 2MASS source was identified for 14 of the 55 \WISE\ matches. Image cutouts for all the GRB positions were also inspected, and we measure $JHK$ 2\,${\sigma}$ upper limits for the remainder of the sample. We search the FIRST \citep{FIRST} and NRAO VLA Sky Survey \citep[NVSS,][]{NVSS} radio surveys, cross-matching to the \WISE\ coordinates, and find only one match. The host of SGRB\,150101B is detected in NVSS at 1.4\,GHz. The full table of FUV to W4 photometry derived from this compilation of observations and archival data is given in table \ref{tab:photresults} of the appendix. 

Initial checks were performed with the available data to discern the physical nature of the sources. The first method employed uses the difference between PSF and Kron magnitudes \citep[][ and references therein]{Farrow}, which is a recommended technique for star-galaxy separation in Pan-STARRS. Because the Kron radii vary with the light distribution of the object in question, and the PSF does not, extended sources such as galaxies (or saturated stars) show discrepancies between the two magnitudes. A plot showing star-galaxy separation in our sample using this method is given in figure \ref{fig:psfkron}. Beyond an {\it i}-band apparent magnitude of $\sim$21, the separation becomes unreliable. Additionally, PSF-Kron positions towards the lower right of the galaxy region might be contaminants, and we do not use positioning in this region as grounds for galaxy classification. For all sources, a visual check for extension was also made. It should be noted that while the PSF-Kron method can confirm an object as a galaxy, it cannot definitively classify stars. In particular, compact, dwarf or distant galaxies may be unresolved at Pan-STARRS resolution and appear 'starlike' by this classifier. Where the only information we have is an insecure galaxy or star classification using PSF-Kron, and the object is not obviously extended by eye, we use SED fitting to distinguish the possibilities (see section \ref{sec:sedfitting}). 

An alternative method, proposed by \citet{Kovacs}, separates stars and galaxies with the aid of the {\it J}-band. This is demonstrated in figure \ref{fig:wisesep}. At fixed W1 magnitude, objects that are bluer in W1-{\it J} are more likely to be galaxies. \citet{Kovacs} found that a cut at W1-{\it J} = 0.09 (AB magnitudes) is an effective star-galaxy separator, with a stellar contamination on the galaxy side of only ${\sim}$1.8 per cent. However, we caution that this technique was applied to brighter sources than we are dealing with, with the galaxies lying at a median redshift of 0.14, much lower than our in our sample. Because it is unknown how the W1-{\it J} colours vary for lower mass, star forming galaxies across a range of redshift, the use of the cut here is a suggestive, but not decisive, diagnostic. 

We also check for proper motion in \WISE\, classifying those with notable proper motion (we require a total proper motion of at least 2\,${\sigma}$ significance) as stars. The HSOY catalogue \citep{HSOY} is also searched. This is a precursor to the full Gaia DR2. We identify several optical sources, all within 1\,arcsec of their respective \WISE\ match, which have proper motions allowing us to rule them out as stars. A small number of sources have notable PM in HSOY but not \WISE\. These might be chance alignments between a foreground star and a background IR source. However, due to large uncertainties on the \WISE\ proper motions, we are unable to tell. Where this scenario arises, we assume that the optical counterparts are associated with the IR sources, given that this is most likely. Table \ref{tab:stargalaxy} gives the results for those sources that have data available for at least one of the star-galaxy checks discussed.  

\begin{table}
\centering 
\caption{Initial star galaxy separation results using archival photometry and catalogued data products, for those sources that had sufficient data for at least one of the separation tests. Note that sources with PSF-Kron positions towards the lower right of the galaxy region may actually be stellar contaminants. The 5 objects listed below the line lack coverage in Pan-STARRS and SDSS, but are classified in other ways.}
\begin{tabular}{p{1.0cm} c c c c c}
\hline 
GRB & PM WISE & PM HSOY & PSF-Kron & W1-{\it J} & Type\\ [0.25ex] 
\hline 
050522 & - & S & S & S & S \\
050716 & S & - & S & - & S \\
050721 & - & S & S & S & S \\
050724 & - & - & G & - & G \\
060428B & - & - & G & - & G \\
061002 & - & - & U & - & U \\
070208 & - & - & G & - & G \\
070724A & - & - & G & - & G \\
080405 & - & S & G & - & S \\
080517 & - & - & G & G & G \\
090904B & S & S & S & G & S \\
100206A & - & - & U & - & U \\
110305A & - & - & S & S & S \\
111222A & - & S & S & S & S \\
120119A & - & - & G & - & G \\
120612A & - & - & S & S & S \\
130603B & - & - & G & - & G \\
131018B & - & - & S & S & S \\
140927A & - & - & - & S & S \\
150101B & - & - & G & G & G \\
150120A & - & - & U & - & U \\
160703A & - & S & S & - & S \\
161010A & S & - & G & S & S \\
161214B & - & S & S & - & S \\
\hline 
070309 & - & - & - & S & S \\
080623 & S & S & - & S & S \\
150626A & - & S & - & S & S \\
161001A & - & S & - & - & S \\
161104A & S & - & - & - & S \\ 
\hline 
\end{tabular}
\newline
{G - galaxy. S - star. U - uncertain.}
\label{tab:stargalaxy}
\end{table}


\section{Previously Reported GRB Hosts and Observations}\label{sec:prevreport}
Many of the GRBs in our sample of 55 have previously been studied, yielding useful information for our analysis. In this section, we compile reported afterglow positions, detailed host studies and other noteworthy information. This allows us to rule out some IR sources as chance alignments, and add some hosts to our sample for which we lacked the required observations. We split these into 3 categories: matched IR sources with an optical detection, those without, and those lying outside the optical surveys searched in this paper.

\subsection{Optical Survey Detections}
\noindent{\it LGRB\,060428B:}\ The candidate host represents a single IR source corresponding to what appears to be a single optical source.  However, \citet{lensedgrb} has suggested that a compact blue galaxy lies underneath the foreground elliptical's light, at an offset of 2.6\,arcsec. This corresponds to the foreground object's Einstein radius, as such they claim that LGRB\,060428B is likely a gravitationally lensed event originating from the higher redshift, bluer galaxy. While this work has not been fully published, it seems a plausible explanation. We therefore take a conservative approach and exclude this source from later analysis.

\smallskip
\noindent{\it LGRB\,160703A:}\ \cite{160703A} observed the afterglow with Keck-I in the {\it g} and {\it r} bands. The improved positional certainty over XRT suggests that the IR source is not the host. The object has proper motion (see table \ref{tab:stargalaxy}), confirming this interpretation. Therefore we remove the source from further analysis.

\smallskip
\noindent{\it LGRB\,161214A:}\ The object associated with the matched \WISE\ source was observed by \citet{161214A}, who obtained spectroscopy identifying it as a K or M star. 

\smallskip
\noindent There are cases where an improved burst localisation (e.g. in the optical) strengthens the IR source association, rather than ruling it out. There are also hosts where we have sufficient photometry for SED fitting, in addition to previously reported host parameters. We use this information to compare to our SED fitting results. These studies and observations are referenced for each burst in table \ref{table:results} of section \ref{sec:sedfitting}. 
 
\subsection{Optical Survey Non-Detections}
\noindent{\it LGRB\,050716:}\ No object is detected in optical imaging. \citet{050716_rol} deduced a burst redshift $>2$ based on the optical to X-ray afterglow SED, however the matched \WISE\ source has proper motion - assuming that the flux is entirely from this star, we reject this association as a chance alignment.

\smallskip
\noindent{\it LGRBs\,070208 and 120119A:}\ \citet{GRBHST} report {\it HST} imaging of the burst locations in these cases, with the optical afterglow positions indicating that these may be chance alignments with the IR-bright sources. This is backed up by our analysis in section \ref{sec:sedfitting}. 

\smallskip
\noindent{\it LGRB\,080207:}\ We successfully identify this host, which has been extensively studied in the literature \citep{080207a,080207,TOUGH,TOUGH2,2017arXiv170900424A} as an example of a red, dusty luminous infrared galaxy. There is a 1\,arcsec separation between the \WISE\ centroid and the Chandra position provided by \citet{080207}, which itself is clearly placed over the galaxy in question. Therefore, we include the reported parameters for this galaxy in our analysis. The optical faintness of this galaxy shows that such sources can have steep optical to IR spectral slopes, and demonstrates that other optical non-detections could be similar in nature. 

\smallskip
\noindent{\it LGRB\,080307:}\ This burst has an X-ray detected AGN a few arcseconds away \citep{Page}. Both the \WISE\ IR emission and X-ray flux have levels consistent with expectations for local AGN \citep{Eckart}. Therefore, we suggest that the AGN is the most likely source of the IR flux. While it is possible that the AGN resides in the host galaxy (as is the case with SGRB\,150101B, see \citet{Fong}), we use {\sc MAGPHYS} for our SED fitting in the next section, which does not have a prescription for AGN. Because the IR flux is consistent with being AGN dominated, this would lead to incorrect parameters when the SED is fitted by {\sc MAGPHYS}. Therefore, we take a cautious approach and remove LGRB\,080307 from further analysis. 

\smallskip
\noindent{\it LGRB\,080605:}\ In the imaging provided by \citet{080605} and \citet{GRBHST}, we can see that the \WISE\ flux centroid is centered on a $z$ = 1.64 galaxy undetected in Pan-STARRS, rather than either of the two bright sources. The burst redshift of $z$ = 1.64 was determined from afterglow spectroscopy. Because this galaxy is therefore confirmed as the host, we reject the nearby bright objects and reclassify this host as a Pan-STARRS non-detection. \citet{080605} provide estimated parameters for the host galaxy, which we employ. 

\smallskip
\noindent{\it GRBs\,100816A and 110918A:}\ For these sources we use the physical parameters reported by \citet{Perez} and \citet{Elliott} respectively. In both cases, the IR bright source is aligned with the reported host galaxy.

\smallskip
\noindent{\it GRB\,120224A:}\ This object is only marginally detected in our X-shooter spectroscopy, however \citet{GRBXS} found a 2\,${\sigma}$ emission line in similar X-shooter data. If the line is H${\alpha}$, this corresponds to a redshift of 1.1. We tentatively assume this to be the case going forward, treating the source as an optically undetected galaxy at $z$ = 1.1.

\smallskip
\noindent{\it LGRBs\,130527A, 150323C and 161007A:}\ The positions of these bursts were observed in the optical. In each case, an extended object was seen, allowing galaxy classifications \citep{130527A,150323C,161007A}. There is insufficient photometry for SED fitting, so we continue to treat these as non-detections, but note the IR flux likely originates from faint galaxies (rather than stellar contaminants).  

\smallskip
\noindent{\it LGRB\,130528A:}\ While undetected in Pan-STARRS, this source is revealed to be in a crowded region in deeper imaging, with multiple objects in a 10\,arcsec region \citep{Jeong}. Because we cannot assign the IR flux to a single object with any certainty, we classify this burst as a potential chance alignment and do not consider it any further.

\subsection{Lacking Coverage}
\smallskip
\noindent{\it GRBs\,050219A, 070429B, 100316D:}\ We use the host galaxy physical parameters reported by \citet{Rossi}, \citet{Cenko} and \citet{100316D} respectively. In these cases, we compare the reported host coordinates to the \WISE\ positions. For all three, the IR flux is aligned with the galaxies identified as a potential hosts. Because these are good quality IR sources which satisfy our FAP cut, we include the reported host parameters in our analysis.  

\smallskip
\noindent{\it LGRB\,130515A:}\ \citet{130515A} observed the brightest source in the X-ray error circle with VLT/FORS2, finding it to be an M star. The position of the star is consistent with the \WISE\ source, so we discount this association.

\section{SED Fitting}\label{sec:sedfitting}
\subsection{Multiple Candidate Hosts}\label{sec:ambiguity}
In our sample, there are cases where multiple optical sources lie inside the search radius used to match optical imaging to \WISE. Visual inspection reveals 3 GRBs (141212A, 140331A and 150120A) where the candidate host IR flux is not uniquely associated with a single optical source. In these cases, the optical and IR images were aligned to check for astrometric offsets.

For LGRB\,140331A, the IR emission originates from the fainter of two optical sources within the XRT error circle.  Inspection of Chandra X-ray imaging with sub-arcsecond afterglow localisation suggests that the burst may not, in fact, be associated with either optical galaxy, but we cannot rule this out (Chrimes et al., in prep). The optical source aligned with the IR flux matched to LGRB\,140331A does not reach the threshold for cataloging in Pan-STARRS. Instead, and in addition to the WHT photometry (see section \ref{sec:wht}), we measure magnitudes from Pan-STARRS cutouts. It is this photometry which appears in the appendix, and is used for SED fitting.

In the case of SGRB\,141212A, the IR source centroid lies closest to the host identified by \citet{2014GCN..17170...1M}, although there may be some blending with neighbouring objects. We download and perform aperture photometry on the Pan-STARRS images in addition to the WHT images previously discussed. The photometry from both is in good agreement, and the Pan-STARRS measurements are again given in the appendix.

The IR source associated with SGRB\,150120A may have either of two optical counterparts, both catalogued in Pan-STARRS DR1. We fit their SEDs separately, in each case assuming that the entire IR flux is associated with the galaxy under consideration. The better of the two fits us used in the subsequent analysis.

\subsection{Galaxy SED fitting with MAGPHYS}
\begin{figure*}
\centering
\includegraphics[width=0.75\textwidth]{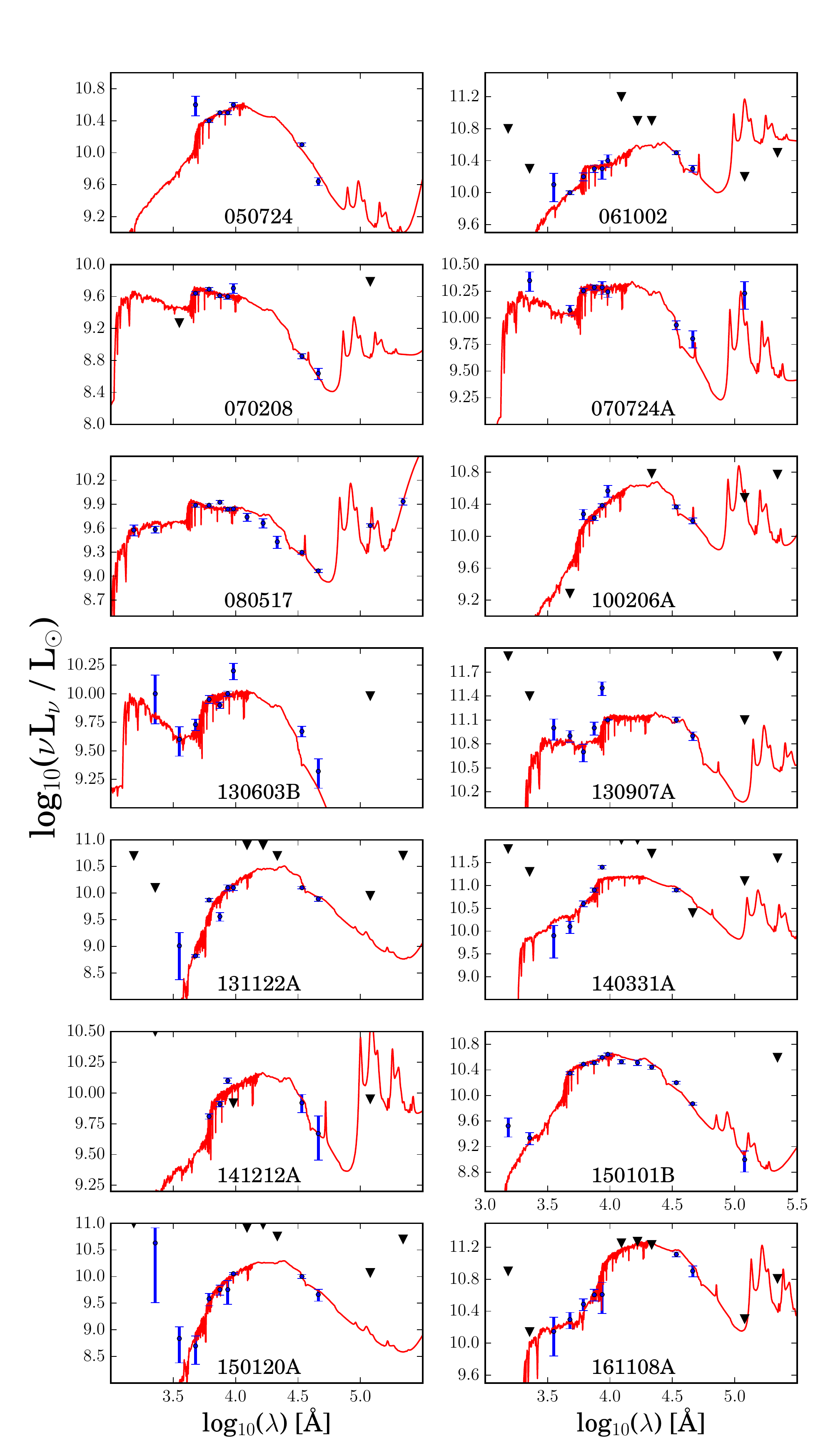}
\caption{SEDs for the objects which were best-fitting (or else confirmed) as galaxies. Some upper limits are too high on this scale to be visible.}
\label{fig:bestfitgalaxies}
\end{figure*}

\begin{figure}
\centering
\includegraphics[width=1.0\columnwidth]{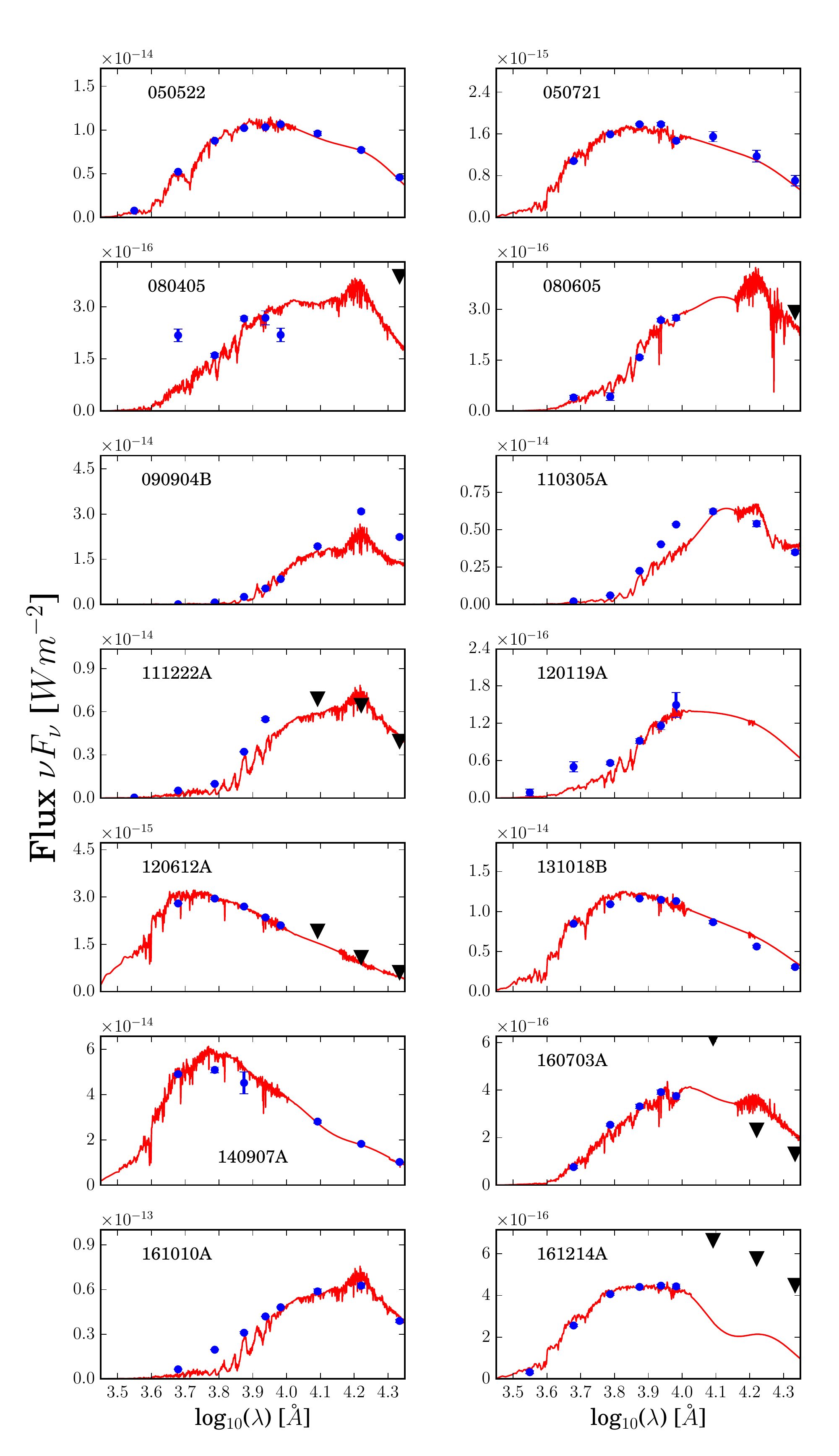}
\caption{SDSS stellar templates for sources which were best-fitting (or otherwise confirmed) as stars. Some upper limits are too high on this scale to be visible.}
\label{fig:bestfitstars}
\end{figure}

Spectral Energy Distribution (SED) fitting on this sample was performed by ${\chi}^{2}$ minimisation using templates derived from {\sc MAGPHYS} and its high redshift update \citep{Magphys08,Magphys15}. {\sc MAGPHYS} was chosen for fitting our sample due to its careful implementation of dust extinction and re-emission. Given that our sources are detected in the infrared, we expect dust to be an important influence on their spectra. The population synthesis models of \citet{Bruzal} are employed, with the dust absorption and re-emission model of \citet{CharlotFall}. The stellar populations are built up by assuming a range of ages distributed evenly from 0.1\,Gyr to the maximum age permitted at a given redshift (i.e. the age of the Universe). The star formation rate is modelled as declining proportional to e$^{-{\gamma}t}$, where ${\gamma}$ is the star formation timescale and $t$ is the time elapsed since the onset of star formation. Random constant SFR bursts are overlaid, with durations evenly distribution between 3${\times}$10$^{7}$\,yr and 3${\times}$10$^{8}$\,yr. The probability of a burst is such that 50 per cent of the model galaxies have experienced a starburst phase in the last 2\,Gyr. The amplitude of these bursts $A$ is defined as the ratio of stellar mass formed in the burst and all stellar mass assembled since the galaxy was formed at time t$_{\mathrm{form}}$; this parameter is distributed logarithmically from 0.03 to 4.00. 

Many of the candidates already have confirmed or likely redshifts, either from afterglow absorption lines or from the host. These were used to fix the redshift where possible. We also derive a photometric redshift for each galaxy. At each step in a grid of trial redshifts, we make use of the internal Bayesian fitting code built into {\sc MAGPHYS} to determine the redshift interval for which $\chi^2 \leq \chi_{\mathrm{min}}^2 + \Delta\chi^2$ \citep{AvniStats}. We did this at  redshift intervals of 0.05 over the range 0 ${\leq}$ z ${\leq}$ 1, or up to $z$ = 3 if no acceptable solution is found at lower redshifts. The distribution of ${\chi}^2$ over redshift is then minimized. In this way, {\sc MAGPHYS} is effectively used a photometric redshift code \citep{Magphys15}, with the redshift treated as an additional free parameter. Our photometric redshifts are generally in agreement with spectroscopic redshifts where available, as shown in table \ref{table:results}. Uncertainties on SED fitting parameters include the effect of photometric redshift uncertainty where a spectroscopic (or SDSS) redshift is not used. 



{\sc MAGPHYS} fits for stellar mass, current star formation rate, star formation history (age, timescale and burst amplitude), metallicity, and dust extinction, amongst other parameters. We caution that the number of free parameters is greater than the number of data points available for a given galaxy and that the fits may be overconstrained.  {\sc MAGPHYS} does not routinely report uncertainties on metallicity, as this is often a poorly constrained parameter, so we simply state the best-fitting metallicity given by {\sc MAGPHYS}.

\subsection{Galaxy SED fitting Summary}
Of the 55 GRBs with candidate \WISE\ counterparts, 29 have optical-NIR photometry (from Pan-STARRS, SDSS, APASS, 2MASS, WHT observations or some combination of the above) in addition to at least \WISE\ band 1. We correct this observed photometry for the Galactic extinction of \citet{Schlafly} and a Fitzpatrick reddening law with R$_{V}$ = 3.1, using the IRSA dust reddening and extinction service\footnote{http://irsa.ipac.caltech.edu/applications/DUST/} and the York Extinction Solver \citep[YES,][]{YES}. We fit 28 SEDs (excluding LGRB\,060429B, as this has been rejected as probable chance alignment). For the host of SGRB\,050120A, we fit the two optical components separately, assigning the entire \WISE\ flux each time. 14 of the 28 objects are best-fitting, or otherwise confirmed, as galaxies. Objects identified as galaxies or stars solely through the quality of fitting to stellar or galaxy templates are classified as photometric galaxies or stars ('PG' or 'Ps') in table \ref{tab:samp}. The limited number of stellar templates available makes it difficult to decide whether an object is best-fitting as a star or galaxy based on the reduced ${\chi}^{2}$ alone. Therefore, we note the reduced ${\chi}^{2}$ values for galaxy fitting, but make the assumption that a visually good star fit indicates that the object is a star, when the corresponding galaxy ${\chi}^{2}$ value is poor. 

The data for the objects best-fitting as galaxies, together with the best-fitting galaxy templates, are shown in figure \ref{fig:bestfitgalaxies} and table \ref{table:results}. There are some instances where the FUV/NUV flux appears inconsistent with the SED. This might be because we have matched to a different object - the matching to \WISE\ is independent for the UV, optical and NIR. However, this unlikely as we would expect a nearby UV source to also be seen in the optical. They might be UV upturns, a phenomena seen in otherwise red elliptical galaxies, or the fitting may simply be failing to properly account for complex stellar populations.

Some fits produced no clear minimum in ${\chi}^{2}$, and therefore lack a robust photometric redshift. In these cases, we use a spectroscopic or SDSS photometric redshift where available. For LRGB\,061002, we obtain a photometric redshift of 0.10$^{+0.45}_{-0.10}$. However, the SDSS photometric redshift, which is calculated using machine learning techniques, is much more precise at $z$ = 0.564. Therefore we fix our redshift to the SDSS value, which has sufficiently small uncertainties that the fit parameters are unaffected by variation with this redshift range. Similarly, the photometric redshift for LGRB\,131122A is unconstrained, likely because there is no clear Balmer or Lyman break evident from the photometric points. However, this object also has an SDSS photometric redshift, at $z$ = 0.399. This produces a visually acceptable fit, so we employ this redshift.

Another object, the candidate host of LGRB\,070208, is correctly identified as a galaxy in our analysis but our photometric redshift of $z$ = 0.16$^{+0.37}_{-0.16}$ is inconsistent with a previous spectroscopic redshift of $z$ = 1.165 \citep{2007GCN..6083....1C}. The foreground object is not detected in the spectroscopy reported by \citet{2007GCN..6083....1C}, but there is a marginally detected object seen offset from the foreground galaxy, nearer to the afterglow centroid. The {\it HST} imaging of \cite{GRBHST} confirms this interpretation. Given this information, we do not include LGRB\,070208 in the discussion of our sample properties.

\subsection{Stellar Fitting}
We also perform fitting to a library of 131 stellar spectra \citep{Pickles}, with the expectation that a significant fraction of our objects will be best-fitting as stars. The spectra span the range 1150-25000\AA, allowing fitting from the FUV to {\it K} bands. In most of the 14 cases where {\sc MAGPHYS} does not provide a good fit, stellar fitting does. The best-fitting stellar templates for these sources are shown in figure \ref{fig:bestfitstars}. Here, we show the SEDs for those objects with proper motion or other star diagnostics, in addition to sources where a stellar SED is a better fit than a galaxy SED. We note that the sources which fit best to stellar templates include all objects for which a spectroscopic or proper motion confirmation as a star is available.

\section{Results and Discussion}\label{sec:results}
\subsection{SED Fitting Results}
In table \ref{table:results} we present the results of our SED fitting for the 13 candidate or confirmed GRB hosts that match best to galaxy templates (the 14 shown in figure \ref{fig:bestfitgalaxies}, excluding the likely interloper LGRB\,070208), and also a compilation of information from the literature for seven of the GRB hosts discussed in section \ref{sec:prevreport}.

In comparing the SED fit results to independent measures of the host properties, potential discrepancies arises between them. The SED derived SFR, $\sim$ 0.15\,M$_{\odot}$yr$^{-1}$, for the host of SGRB\,150101B disagrees with the NVSS 1.4\,GHz radio SFR, $\sim$ 300\,M$_{\odot}$yr$^{-1}$ and the detection of this source at 5.8$\mu$m in the W3 band. This is due to the presence of an AGN \citep{Fong}. Unlike LGRB\,080307 which was rejected due to coincidence with a nearby AGN, the association of SGRB\,150101B with this galaxy has been secured with spectroscopic observations. Therefore, we keep it in the analysis. Only two other sources in our sample are detected in the W3 band. While this is sometimes used as a SFR indicator or AGN discriminator, the two cannot be unambiguously differentiated without a reliable (and low) redshift \citep{Davies}. We do not consider this data any further at this point.

\begin{table*}
\centering
\caption{SED fitting results for the IR-bright sources with photometry best-fitting to galaxy templates. Additionally, where we have non-detections or no coverage, we list well studied host galaxies identified in the literature in the last 7 rows. References are given below the table and provide spectroscopic redshifts in most cases, or additionally, parameter values for galaxies lacking the photometry needed for SED fitting.}
\scalebox{1.0}{
\hspace*{-1cm}
\begin{tabular}{p{0.5in} c c c c c c c c c}
\hline\hline 
GRB & z$_{phot}$ & z$_{spec}$ & SFR (M$_{\odot}$yr$^{-1}$) & M$_{\star}$ (10$^{10}$M$_{\odot}$) & sSFR (10$^{-10}$yr$^{-1}$) & A$_{V}$ & Z/Z$_{\odot}$ & ${\chi}^{2}$/dof & Ref. \\ [0.25ex] 
\hline 
050724$\ddagger$ & $0.05^{+0.60}_{-0.05}$ & 0.258 & $21.4^{+6.8}_{-20.7}$ & $8.1^{+0.8}_{-3.6}$ & $2.7^{+15.1}_{-2.6}$ & 2.45$\pm$0.13 & 0.038 & 1.81 & [1], [2]\\
061002$\ast$ & 0.564 & - & 1.6${\pm}$0.6 & $2^{+9}_{-1}$ & $7.08^{+8.23}_{-2.28}$ & $4.47^{+0.22}_{-0.22}$ & 1.46 & 1.90 & [3] \\
070724A$\dagger$ & 0.50$\pm$0.23 & 0.457 & $8.4^{+0.6}_{-6.1}$ & $1.8^{+2.8}_{-1.1}$ & $4.8^{+11.2}_{-3.0}$ & $0.43^{+5.81}_{-0.06}$ & 1.093  & 0.85 & [4], [5] \\
080517 & $0.01^{+0.20}_{-0.01}$ & 0.089 & $9.8^{+1.4}_{-2.7}$  & $0.26^{+0.10}_{-0.04}$ & $37^{+7}_{-2}$ & 3.3$\pm$0.1 & 1  & 2.51 & [6], [7] \\
100206A$\dagger$ & - & 0.4068 & $14^{+2}_{-2}$  & $8.8^{+0.1}_{-0.1}$ & $3.2^{+0.8}_{-0.8}$ & $2^{+1}_{-1}$ & 1.15  & 1.34 & [8] \\
130603B$\dagger$ & $0.36^{+0.12}_{-0.25}$ & 0.356 & $1.070^{+0.708}_{-0.001}$ & $1.1^{+0.7}_{-0.4}$ & $1.0^{+1.2}_{-0.6}$ & $0.03^{+0.85}_{-0.03}$ & 0.038  & 2.78 & [9], [10] \\
130907A & $1.00^{+0.04}_{-0.15}$ & 1.238 & 1.45$\pm$0.8 & $4.5^{+6.8}_{-2.7}$ & 71$\pm$64 & ${\sim}$1.9 & 1.9 & 2.54 & [11], [12] \\
131122A$\ast$ & 0.399 & - & $0.22^{+0.06}_{-0.04}$ & $8.9^{+1.7}_{-1.8}$ & $0.022^{+0.006}_{-0.004}$ & 2.2$\pm$0.1 & 0.038 & 1.25 & - \\
140331A & $1.00^{+0.11}_{-0.04}$ & - & $5.3^{+4.3}_{-2.4}$ & $16.5^{+4.1}_{-6.4}$ & $0.47^{+0.06}_{-0.35}$ & $1.4^{+0.9}_{-1.0}$ & $0.09^{+0.17}_{-0.06}$ & 6.49 & [13] \\
141212A$\dagger$ & -- & 0.596 & 0.65$\pm$0.4 & $1.4^{+0.8}_{-0.5}$ & 2.2$\pm$0.13 & $1.4^{+1.3}_{-1.2}$ & 1 & 2.42 & [14] \\
150101B$\dagger$ & $0.15^{+0.05}_{-0.04}$ & 0.134 & 0.15$\pm$0.03 & $6.3^{+0.8}_{-1.8}$ & 0.023$\pm$0.005 & ${\sim}$3.6 & 0.4 & 2.59 & [15] \\
150120A$\dagger$ & $0.1^{+0.2}_{-0.1}$ & 0.46 & $0.71^{+2.11}_{-0.08}$ & $5.6^{+8.5}_{-2.1}$ & $0.11^{+0.17}_{-0.07}$ & 1.7$\pm$0.3 & 0.89 & 0.90 & [16] \\
161108A & <1.3 & 1.159 & 0.25$\pm$0.11 & $11.2^{+0.2}_{-0.4}$ & $0.22^{+1.19}_{-0.05}$ & $0.3^{+0.3}_{-0.2}$ & 1.6 & 1.81 & [17] \\
\hline 
050219A & - & 0.2115 & ${\sim}$0.06 & ${\sim}$1 & ${\sim}$0.06 & <0.1 & - & - & [18] \\
070429B$\dagger$ & - & $\sim$0.9 & ${\gtrsim}$1.1 & ${\gtrsim}$440 & ${\gtrsim}$0.0025 & - & - & - & [19] \\
080207 & - & 2.086 & $\sim$119 & 32$\pm$8 & ${\sim}$4 & ${\sim}$1.9 & ${\sim}$1  & - & [20],[21],[22],[23] \\
080605 & - & 1.64 & $49^{+26}_{-13}$ & $0.80^{+0.13}_{-0.16}$ & $60^{+60}_{-20}$ & $0.22^{+0.40}_{-0.22}$ & 0.6$\pm$0.2 & - & [24] \\
100316D & - & 0.0591 & 1.20$\pm$0.08 & ${\sim}$0.0895 & ${\sim}$13 & 0.86 & 0.3 & - & [25],[26] \\
100816A & - & 0.804 & - & - & - & ${\sim}$0.2 & - & - & [27] \\
110918A & - & 0.984 & ${\sim}$40 & 10.68$\pm$0.16 & ${\sim}$3.7 & 0.10$\pm$0.16 & - & - & [28], [29] \\ 
\hline 
\end{tabular}}
\newline
{$\dagger$ These objects are short bursts. $\ddagger$ This object has extended emission but is likely a disguised short burst. $\ast$ Uses an SDSS photometric redshift. \\ References: [1] \citet{050724z}, [2] \citet{Berger05}, [3] \citet{SDSS12}, [4] \citet{Kocevski}, [5] \citet{2009ApJ...704..877B}, [6] \citet{Stanway}, [7] \citet{Stanway2}, [8] \citet{100206A}, [9] \citet{130603z}, [10] \citet{130603B_iso}, [11] \citet{130907z}, [12] \citet{130907A_iso}, [13] \citet{140331A_photz}, [14] \citet{Chornock}, [15] \citet{Fong}, [16] \citet{150120z}, [17] \citet{161108Az}, [18] \citet{Rossi}, [19] \citet{Cenko}, [20] \citet{2017arXiv170900424A}, [21] \citet{080207a}, [22] \citet{080207b}, [23] \citet{080207}, [24] \citet{080605}, [25] \citet{100316D}, [26] \citet{HIgrbhosts}, [27] \citet{Perez}, [28] \citet{Elliott}, [29] \citet{110918A_Iso}}
\label{table:results}
\end{table*}

\subsection{Short GRBs}\label{sec:svl}
Before considering the redshift distribution and other properties of this sample, it is important to consider the selection effects that will shape any comparison we make. The first issue to consider is whether we may be particularly biased towards long or short bursts. Since their progenitor mechanisms differ, we expect their host properties to also differ. As such it is essential to consider the short vs long divide. Of the 55 GRB locations identified as having a \WISE\ counterpart, 11 of the associated bursts had an observed $T_{90}<2$\,s. Of these, 8 are reported in table \ref{table:results} as galaxies. Two are identified as a Galactic stars.

The lack of a clear divide between the two populations can lead to ambiguity \citep[e.g.][and references therein]{2014ARA&A..52...43B}. Two of our targets have a $2<T_{90}<5$\,s and might be classified as short by some proposed criteria. Of these, one is identified in our galaxy sample, while one has insufficient photometry. In the following, we treat these "intermediate" sources as long bursts.


If we consider the 20 GRBs for which galaxy properties are assembled in table \ref{table:results}, the fraction of SGRBs is 40 per cent (8/20) and would be as high as 50 per cent if the intermediate bursts were included. This compares to the short burst fraction in the entire GRB catalog of only 6 per cent \citep{2014ARA&A..52...43B}, suggesting that we may be preferentially selecting short bursts. This may reflect the difference in the underlying redshift distribution of these sources, exemplified by the mean redshift of $z$ = 0.45 for the SGRBs in our sample. Short GRBs are typically of lower isotropic-equivalent luminosity and their distribution is biased towards low redshifts ($\langle z \rangle < 0.8$) relative to long GRBs ($\langle z \rangle \sim 2$), due to the differences in both their progenitors and detection probabilities \citep{2014ARA&A..52...43B}. Given the relatively shallow depth of the W1 band imaging, we might expect a low redshift, and therefore SGRB, excess in our sample.

\subsection{Redshift Distribution}\label{ref:zdist}

\begin{figure}
\centering
\includegraphics[width=0.47\textwidth]{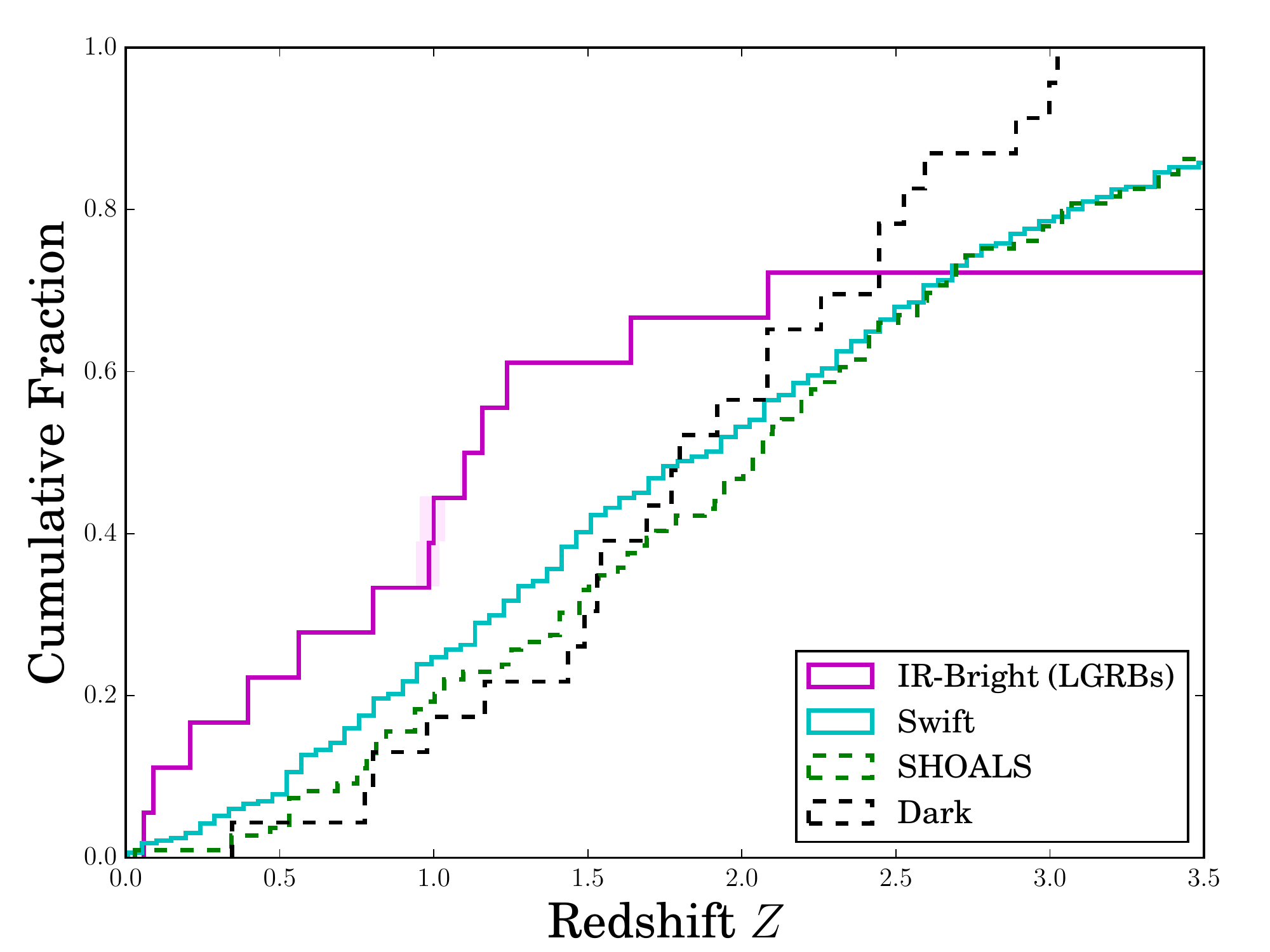}
\caption{The redshift cumulative distributions of the IR-bright LGRB host sample, all \textit{Swift} LGRBs with known redshift, the unbiased LGRB SHOALS host galaxy sample and the dark LGRB sample of \citet{PerleyDust}. We make the worst case assumption that the 5 optical-non detections that have not been ruled out as stars are higher redshift galaxies.}
\label{fig:zdist}
\end{figure}

We now consider the detailed properties of the IR-bright LGRB host population. Figure \ref{fig:zdist} shows the cumulative redshift distribution for the 13 candidate host galaxies that form our IR-bright LGRB host population (the 12 in table \ref{table:results} plus LGRB\,120224A at $z$ = 1.1). We compared these to the SHOALS sample of LGRB hosts \citep{Perley1,Perley2}, all \textit{Swift} LGRBs with a known redshift, and a sample of dark LGRB hosts \citep{PerleyDust}. In each case we indicate uncertainty on the cumulative distribution by performing an analysis in which each value is permitted to vary by addition of its associated random error, drawn from a skewed Gaussian distribution, in order to account for asymmetric errorbars. The scale and alpha parameters are chosen in each case such that the asymmetric distributions are reproduced. We show the standard deviation of 1000 realisations of the perturbed cumulative distribution as a shaded region.

In our sample, 7 objects have survey coverage but are undetected, and lack extensive study in the literature. It is possible that some of these may be M, L or T dwarfs, which can be a few magnitudes brighter in the \WISE\ bands than the optical \citep[e.g., ][]{browndwarf}. Indeed, the optically undetected object associated with LGRB\,050716 has significant \WISE\ proper motion. Late L and T dwarfs have the reddest colours of these stars and would be most able to satisfy the criteria of W1 detection and optically non-detection, however these are also the rarest classes of these objects.

Alternatively, the optical non-detections could be intrinsically faint or higher redshift galaxies. This possibility has been demonstrated by several such examples in section \ref{sec:prevreport}. Three of the optically undetected sources have been identified as extended sources in deeper imaging, as described in section \ref{sec:prevreport}. One of the non-detections, LGRB\,120224A, may be at $z$ = 1.1 \citep{GRBXS}. We can see in figure \ref{fig:zdist} that our sample is biased towards low redshifts when we do not included the non-detections. Therefore, the most extreme scenario is that all 5 of the remaining non-detections are in fact galaxies at higher redshift than the highest confirmed LGRB in our sample. Setting these 5 non-detections to an arbitrarily high redshift, we are still unable to match the slower rise of the SHOALS and dark samples, demonstrating that there is a low $z$ bias even in the 'worst case' scenario. In reality, their redshifts could well be lower, given that the optical undetected hosts of LGRBs\,080605, 100816A, 110918A and 120224A all lie at $z$ < 2. 

We must also consider the possibility that chance alignments remain in our sample. LGRB\,140331A is at high risk in this regard, since the XRT error circle position favours a different optical source, unaligned with the \WISE\ flux. In general, foreground chance alignments preferentially select lower redshift or foreground objects, and this remains a possible explanation for the difference in cumulative distribution between our sample and others. However, all but three of these sources have spectroscopic redshifts, and it is unlikely that a spectroscopic redshift measurement would be unaffected by a foreground interloper. We do, however, note that chance alignment is the most likely scenario for LGRB\,070208.

We further quantify the effect of the small number statistics in this sample by bootstrap resampling of the SHOALS sample.  We extract a subsample of  sources from one of the reference samples matching the IR bright sample in size, and calculate its redshift distribution. This is done 100,000 times to explore the frequency with which the subsample realisation matched the observed distribution. We define a match as a scenario in which an appropriate fraction of the sample lies at $z$ = 1.24 (our  highest optically detected galaxy redshift) or lower. We also consider the Kolmogorov-Smirnov (KS) statistic for the same distribution. As the results in table \ref{tab:LGRBproperties} demonstrate, the KS-test is unable to the reject the null hypothesis that IR-bright sample is drawn from the same population as SHOALS, while the bootstrapping estimate gives a ${\sim}$0.06 per cent chance of drawing this redshift distribution from SHOALS. Assuming instead that the undetected sources are dusty, lower redshift galaxies, or that they are stars, only increases the disparity between the samples (see later). 

The differences between the bootstrap and KS-test results are pronounced. The bootstrapping method supports a much stronger identification of IR-bright LGRB hosts being biased in redshift, compared to the KS-test. Fundamentally the two tests are exploring different aspects of the data. The KS-test is primarily sensitive to the {\it shape} of the distribution, and at $z$ < 1.24 these are similar. However, the entire distribution for the IR-bright galaxies is shifted towards lower redshifts, producing the clear bootstrap results indicated in table \ref{tab:LGRBproperties}. The IR-bright hosts possess a  biased distribution in redshift, but since the distribution shape is similar, this does not necessarily imply a distinct underlying population.



\begin{table}
\centering 
\caption{Bootstrap and KS-test results for the LGRB redshift, stellar mass, {\it V}-band attenuation and star formation rate distributions, compared to unbiased samples over the same redshift range.} 
\begin{tabular}{l p{3.4cm} p{1.5cm} p{1.cm}}
\hline 
Property & Boostrap Target & Bootstrap & KS-test   \\
 & & \%age prob. &  {p-value}$^1$\\
\hline 
z & 0.61 by $z$ = 1.24 & 0.056 & 0.45 \\ 
M$_{\star}$ & 0.50 by log$_{10}(\mathrm{M_{*}})=10.7$ & 0.001 & 0.004 \\ 
A$_{V}$ & 0.50 by $A_{V}$ = 0.84 & 20.2 & 0.07 \\ 
A$_{V}$ & 0.75 by $A_{V}$ = 1.89 & 0.09 & 0.07 \\ 
SFR & 0.50 by log$_{10}(\mathrm{SFR})$ = 0.17 & 50.1 & 0.62 \\ 
\hline 
\end{tabular}
$^1$The KS test requires a p-value of less than 0.05 to pass the widely used threshold of 2\,${\sigma}$ significance.
\label{tab:LGRBproperties}
\end{table}

\subsection{Masses, Dust Extinction and SFR}
\begin{figure}
\centering
\includegraphics[width=0.47\textwidth]{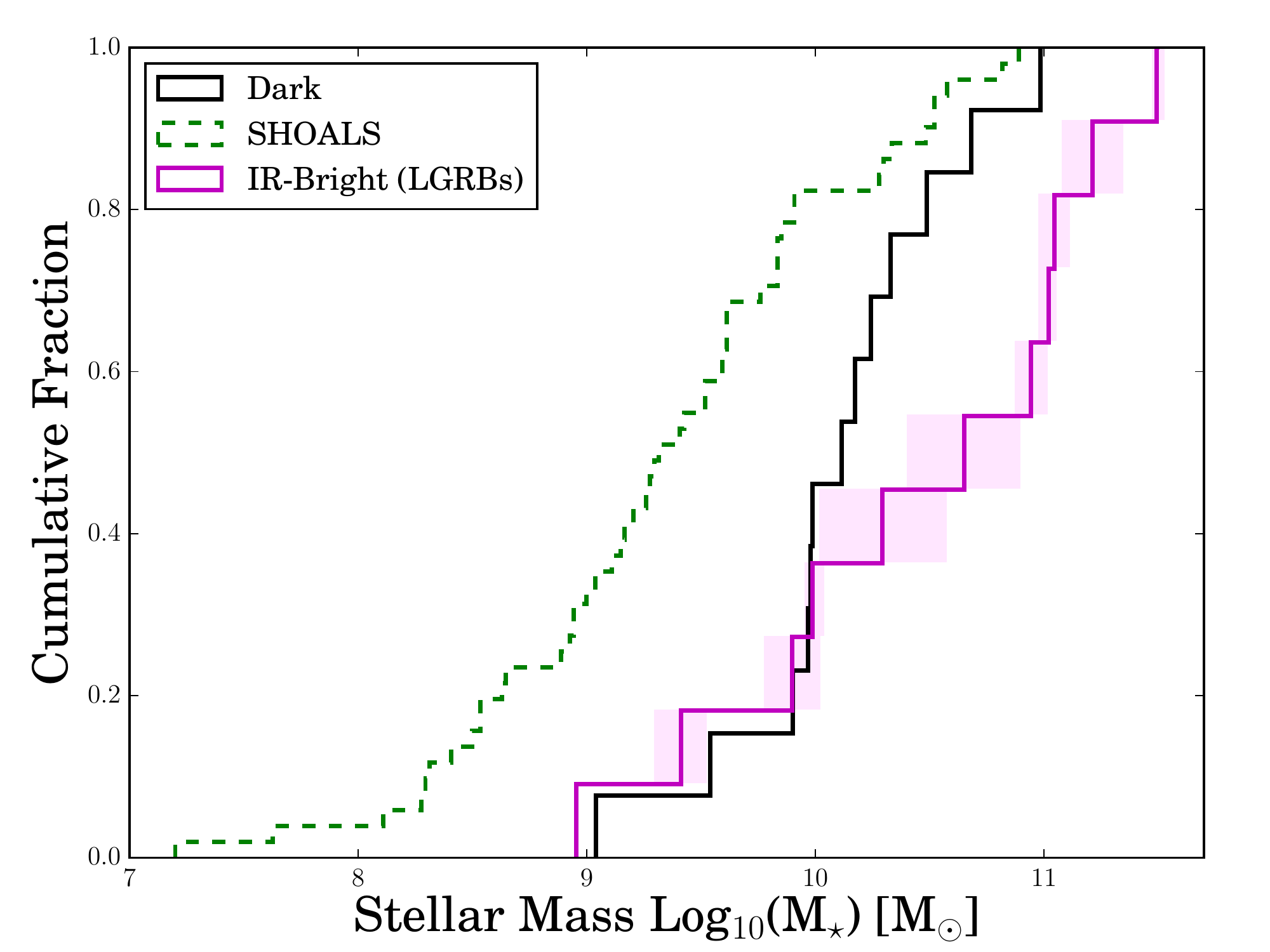}
\caption{The cumulative distribution of stellar masses in the IR-bright host population, the SHOALS sample, and the dark burst sample. The IR-bright hosts appear to be more massive than even the dark burst population.}
\label{fig:massdist}
\end{figure}

We compare the stellar masses derived for our IR-bright LGRB host population (11 objects - 6 from our SED fitting and 5 literature values), to existing samples over the same redshift range, in figure \ref{fig:massdist}. The IR-bright and SHOALS distributions yield a KS-test p-value that passes the 2$\sigma$ threshold for significance. This, and the corresponding bootstrap result, are given in table \ref{tab:LGRBproperties}. In comparing the mass distributions, we assume that the optically-faint subsample is not significantly biased in mass.

\begin{figure}
\centering
\includegraphics[width=0.47\textwidth]{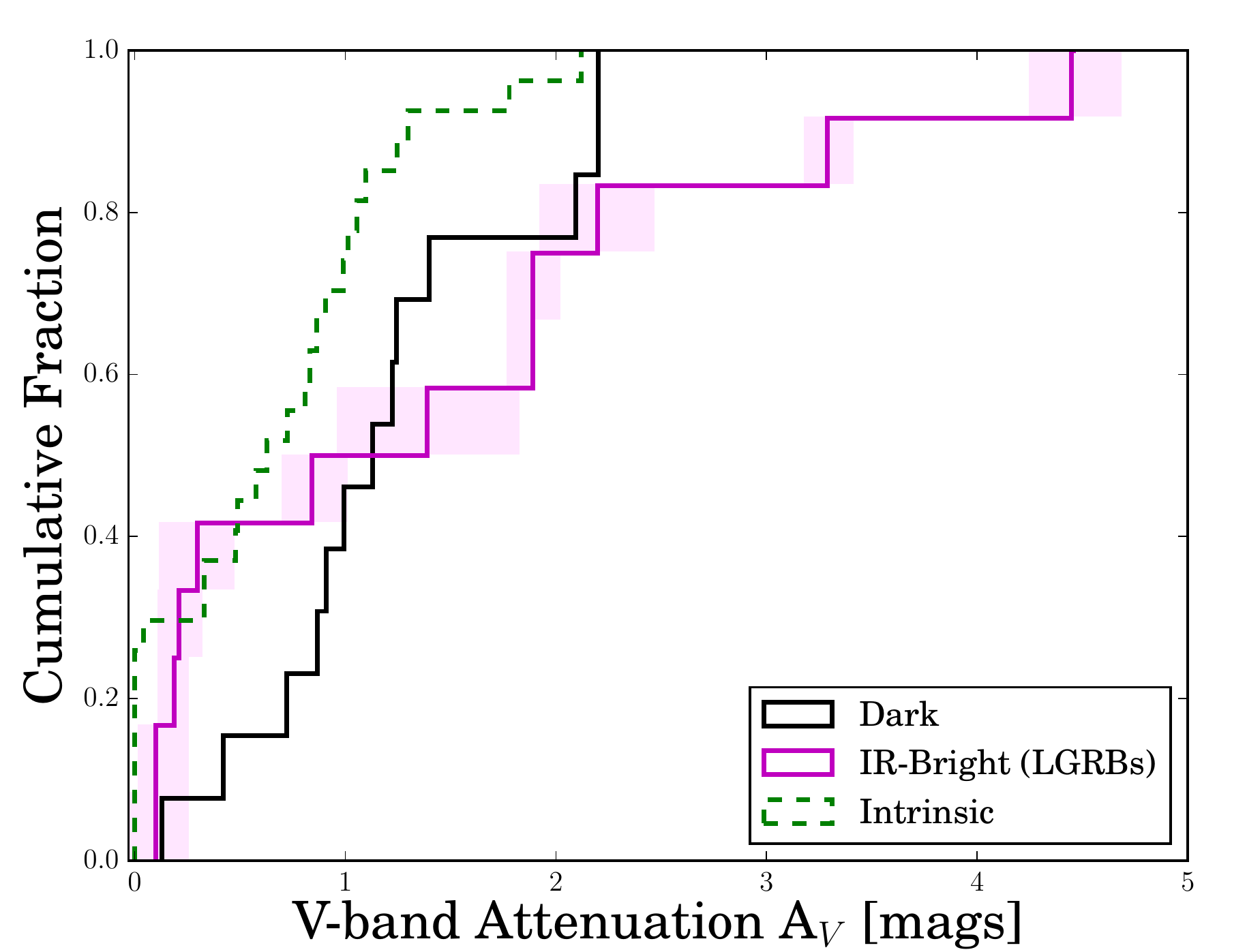}
\caption{The cumulative distribution of A$_{V}$ in the IR-bright host population, the dark sample and the derived `intrinsic' distribution of \citet{PerleyDust}.}
\label{fig:dustdist}
\end{figure}

Given that we are selecting in the infrared, we expect the dust extinction (and hence re-emission at long wavelengths) to be a parameter of interest. A constraint on this is obtained from the SED fitting, parameterised by the {\it V}-band attenuation A$_{V}$. The cumulative distribution of A$_{V}$ in our LGRB sample is shown in figure \ref{fig:dustdist}. This is compared to the  LGRB host distribution of A$_{V}$ (where we have restricted the sample to $z$ < 2), as determined by \citet{PerleyDust} by correcting previous optically biased studies. 


The bootstrap and KS-test results are again given in table \ref{tab:LGRBproperties}. 50 and 75 per cent bootstrap targets are provided to demonstrate the significance of the divergence of the distributions around A$_{V}$ = 1. We make no {\it a priori} assumption about the extinction in optical non-detections. As in the case of the redshift distribution, it is appropriate to consider the possibility that the 5 optically undetected sources may, in fact, be biased and have very high dust extinction values. This would only strengthen the conclusion that IR bright sources are dustier than the typical host galaxy, while also making them more extreme outliers in redshift. 


\begin{figure}
\centering
\includegraphics[width=0.47\textwidth]{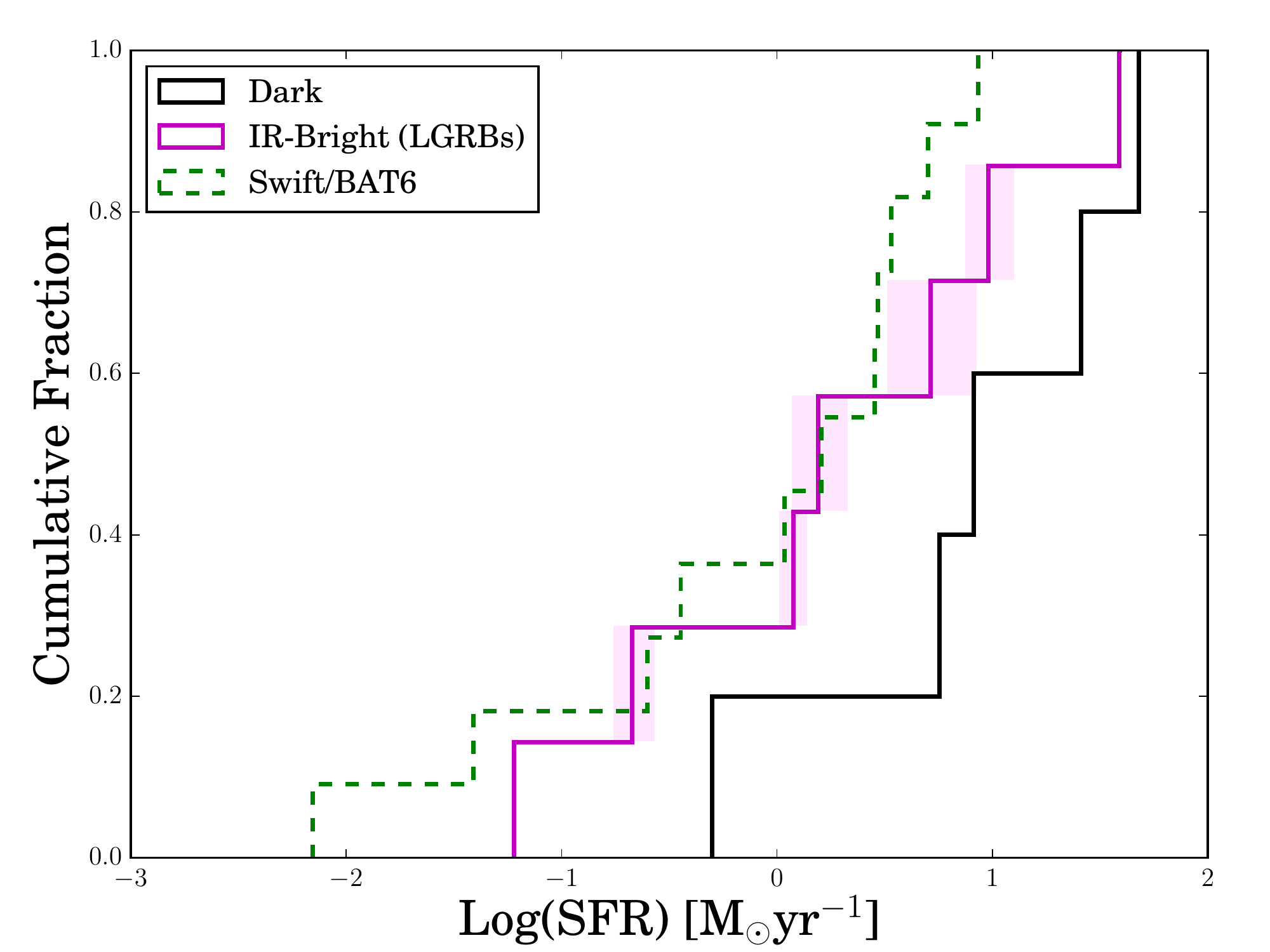}
\caption{The cumulative distribution of SFR in the IR-bright LGRB host population, dark sample and the $z<1$ unbiased distribution of \citet{Japelj}.}
\label{fig:sfrdist}
\end{figure}

Finally, we compare our population distribution in terms of the star formation rate in LGRB hosts. We compare the 7 IR-bright LGRB sources at $z$ < 1 for which we have SFR constraints (4 from SEDs, 3 from literature values) against the $z$ < 1 distribution reported by \citet{Salvaterra} and \citet{Japelj} for the BAT6 LGRB subsample in figure \ref{fig:sfrdist}. The KS-test p-value and boostrap result in table \ref{tab:LGRBproperties} fail to reject the null hypothesis that the IR bright sources are typical examples drawn from the underlying LGRB population.

\subsection{Host Luminosity}
\begin{figure}
\centering
\includegraphics[width=0.47\textwidth]{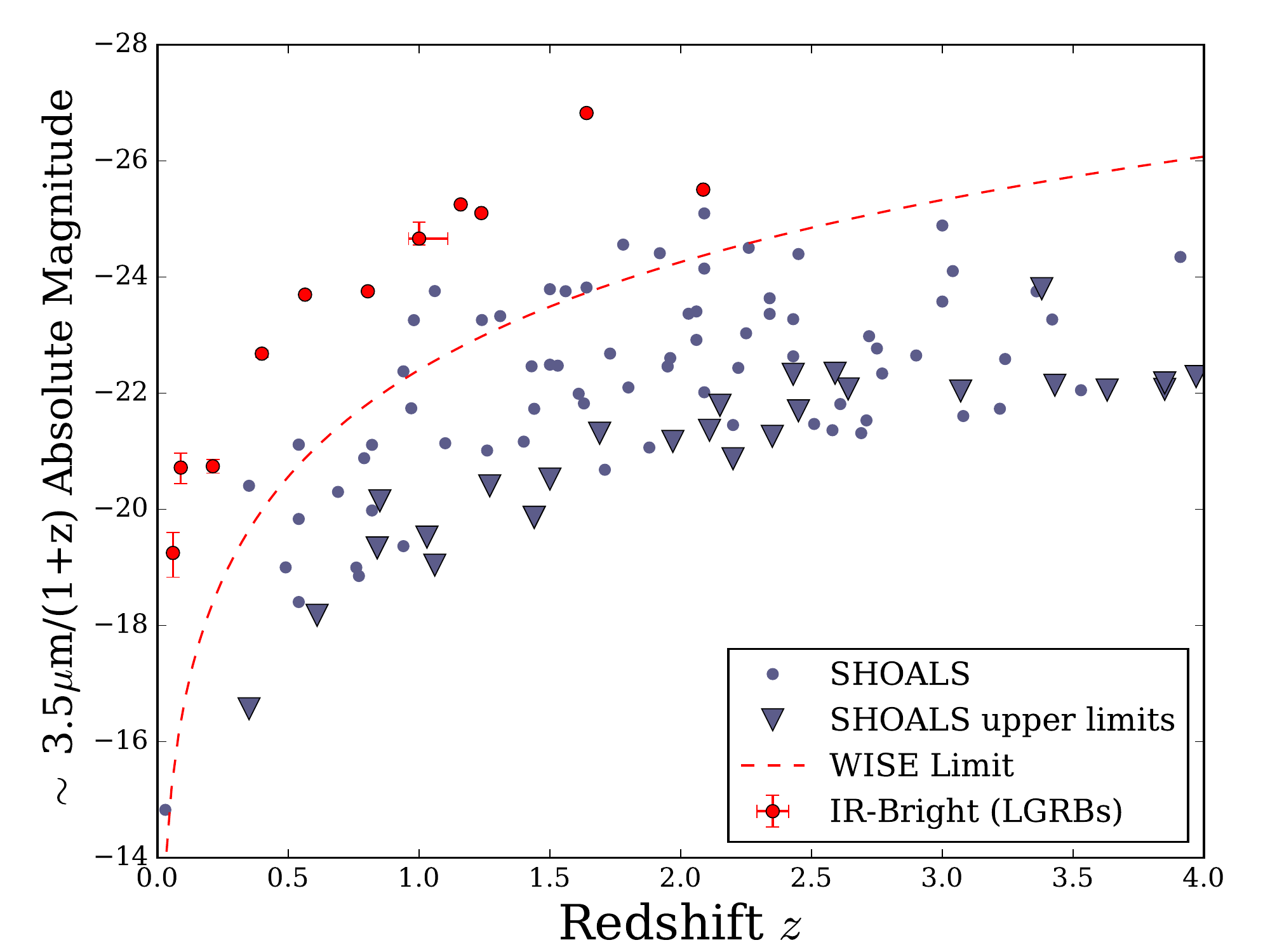}
\caption{The absolute, rest frame magnitude versus redshift for the IR selected hosts (red) and SHOALS. No K-correction is made to either sample. The absolute magnitude corresponding to the \WISE\ apparent magnitude 2\,$\sigma$ limit is given by the dashed curve.}
\label{fig:w1red}
\end{figure}

Given a large range of source redshifts, any single photometric band samples a range of rest frame wavelengths, complicating an analysis of their rest frame magnitudes. \citet{Perley2} obtained {\it Spitzer} 3.6\,${\mu}$m photometry for each of their targets in the unbiased SHOALS catalogue. We transform the 3.4$\mu$m W1 and {\it Spitzer} 3.6\,${\mu}$m apparent magnitudes to absolute magnitudes, without K-correction, while cautioning that these correspond to a rest-frame wavelength of ${\sim}$3.5\,$\mu$m/($1+z$) in each case. The absolute W1/($1+z$) magnitudes of the LGRB sample are shown in figure \ref{fig:w1red}, and are compared to the \citet{Perley2} distribution. Since this is a direct comparison with {\em Spitzer} data, the distribution in absolute magnitude at a given redshift is indepent of the K-correction uncertainties. At low redshift ($z\la$0.6), the W1 band is probing a fall off in the stellar flux and a rise in dust and PAH emission, rather than the stellar continuum. Above $z\sim$ 0.6 and below $z\sim$ 3, the W1 magnitude is probing a fairly flat region of stellar emission and so is a good mass indicator. Again, the comparison to literature work is valid because the SHOALS magnitudes also suffer from this effect. 

The faintest objects detected in \WISE\ band 1 have an apparent magnitude m$_{W1}$ ${\sim}$ 20. Placing such a source at $z$ = 2 corresponds to a 3.4${\mu}$m/(1+z) absolute magnitude of $\sim$-24.5, which is at the very high end for GRB hosts as figure \ref{fig:w1red} shows. The dashed line in figure \ref{fig:w1red} indicates the 2\,$\sigma$ detection threshold for \WISE\ as a function of redshift. It is clear that most SHOALS host galaxies, even at low redshift, fail to satisfy this threshold. 

We note that we are only matching against very bright sources, satisfying the cut-off for inclusion in the ALLWISE catalog. \citet{Perley2} obtained their 3.6${\mu}$m host galaxy fluxes by subtracting the flux from nearby bright sources to reveal an underlying host. This introduces the possibility that we may be overestimating the IR flux. However, in most cases we do not see another optical source that might be the true origin of the \WISE\ flux. In other words, if the IR sources we match to are simply chance alignments, then many of the GRB hosts would have to be too optically faint for detection in Pan-STARRS. In addition, we know that IR-bright GRB hosts exist from previous work \citep{Stanway,Stanway2}. Therefore, we have confidence that the differences between the 3.6${\mu}$m magnitudes in this sample, and those of the SHOALS sample, are real. 


The contrast between our sample and SHOALS is rather unsurprising, since our IR-bright sample is selected to be extreme in the W1 band. However, it does present the possibility that the bias towards low redshifts comes from sampling deeper into the host galaxy luminosity function. While our failure to identify higher redshift hosts may be a simple \WISE\ data selection effect, our most luminous host galaxy would, theoretically, have been detectable out to $z\sim$ 3. At these redshifts, the W1 band probes the rest-frame near-infrared and is unlikely to be strongly affected by dust, but the optical is probing the rest-frame ultraviolet, so dust extinction may account for observed optical non-detections. 

\begin{figure}
\centering
\includegraphics[width=0.47\textwidth]{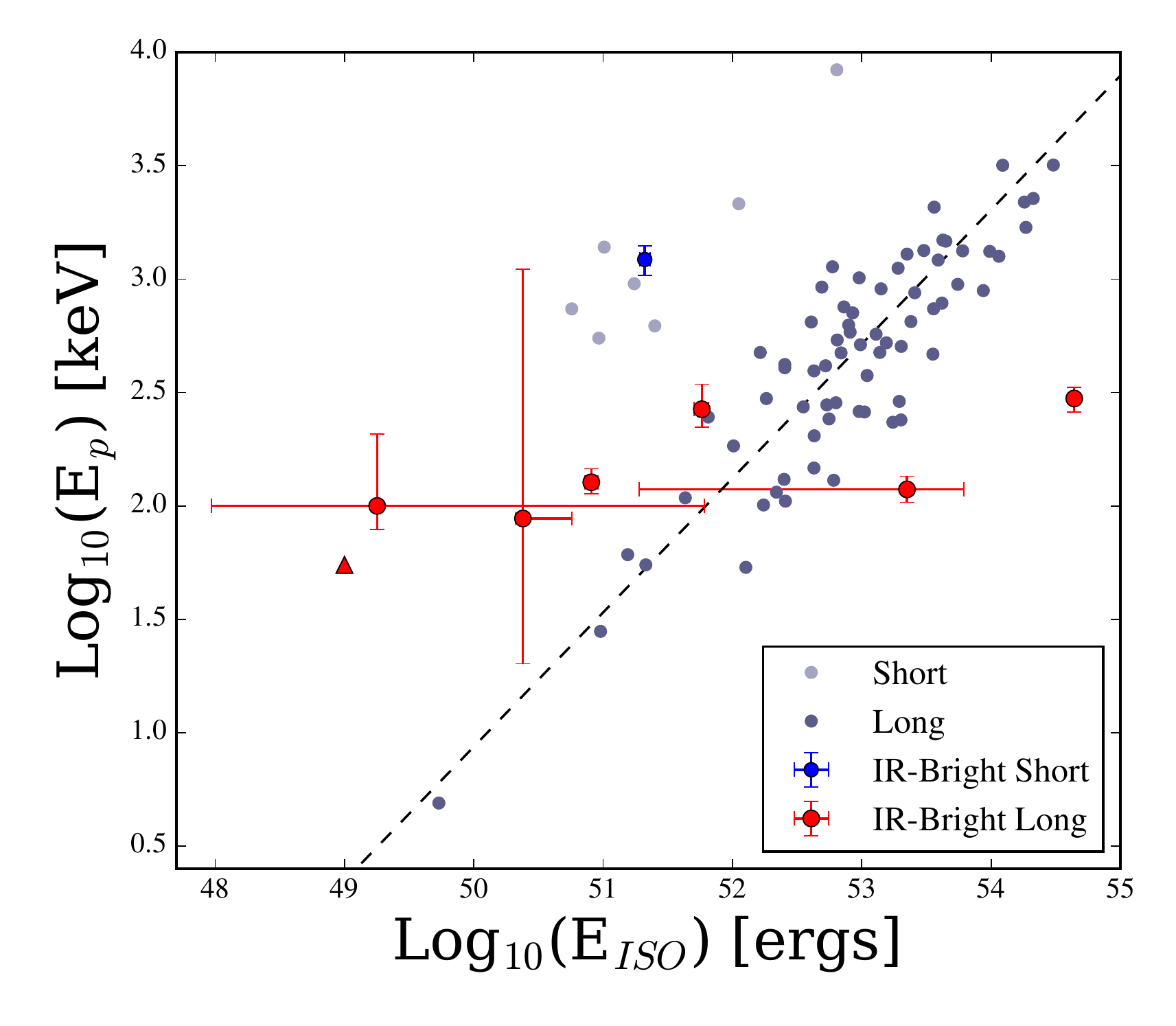}
\caption{Placement of the GRBs on the Amati relation. The darker grey points are LGRBs and paler SGRBs. The long bursts with IR-bright hosts have red errorbars, the single short burst (SGRB\,130603B) has blue. LGRB\,080517 is indicated by the lower limit. We have calculated E$_{\mathrm{iso}}$ for LGRB\,050219A and LGRB\,061002, based on the E$_{p}$ and S$_{\mathrm{bol}}$ values estimated by \citet{Butler}.} 
\label{fig:amati}
\end{figure}

\subsection{Burst Luminosity}
GRB studies have found a correlation between the rest-frame photon energy E$_{p,i}$ at peak prompt emission, and the isotropic equivalent energy E$_\mathrm{iso}$, known as the Amati relation \citep{Amati1,Amati2}. One member of our IR-bright host population, LGRB\,080517, has already been identified as a sub-luminous LLGRB on this relation \citep{Stanway} due to its low E$_\mathrm{iso}$ and lower limit on E$_{p,i}$. We thus consider whether other members of this population are low-luminosity bursts.

A number of targets in our sample have literature constraints on E$_{\mathrm{iso}}$ and E$_{p,i}$. Six bursts had E$_{\mathrm{iso}}$ and E$_{p,i}$ estimates in the literature (see the references in table \ref{table:results} for details). For LGRBs\,061002 and 050219A, we calculate E$_{\mathrm{iso}}$ and the rest frame peak energy E$_{p,i}$ using the SDSS photometric redshift given in section \ref{sec:sedfitting}, the spectroscopic redshift presented by \citet{Rossi} and the bolometric fluence and observed peak energy provided by \citet{Butler}. We use a cut-off power law to extrapolate from the gamma-ray band and infer E$_{\mathrm{iso}}$. 

These eight GRBs with IR-bright hosts are placed on the Amati relation in figure \ref{fig:amati}. Additional (non-IR bright) bursts are shown as grey circles in the background for reference, while the bursts from our sample are shown in red (long bursts) and blue (short) with error bars. While LGRB\,061002 lies in a similar region to LGRB\,080517, towards the low-luminosity region of parameter space, the uncertainties cannot rule out consistency with the Amati relation. Targets selected in the W1 band appear to cluster at the low $E_p$ end of the distribution, consistent with expectations given the low redshift bias in our sample.

\section{Interpretation}\label{sec:interp}

\begin{figure*}
\centering
\includegraphics[width=0.95\textwidth]{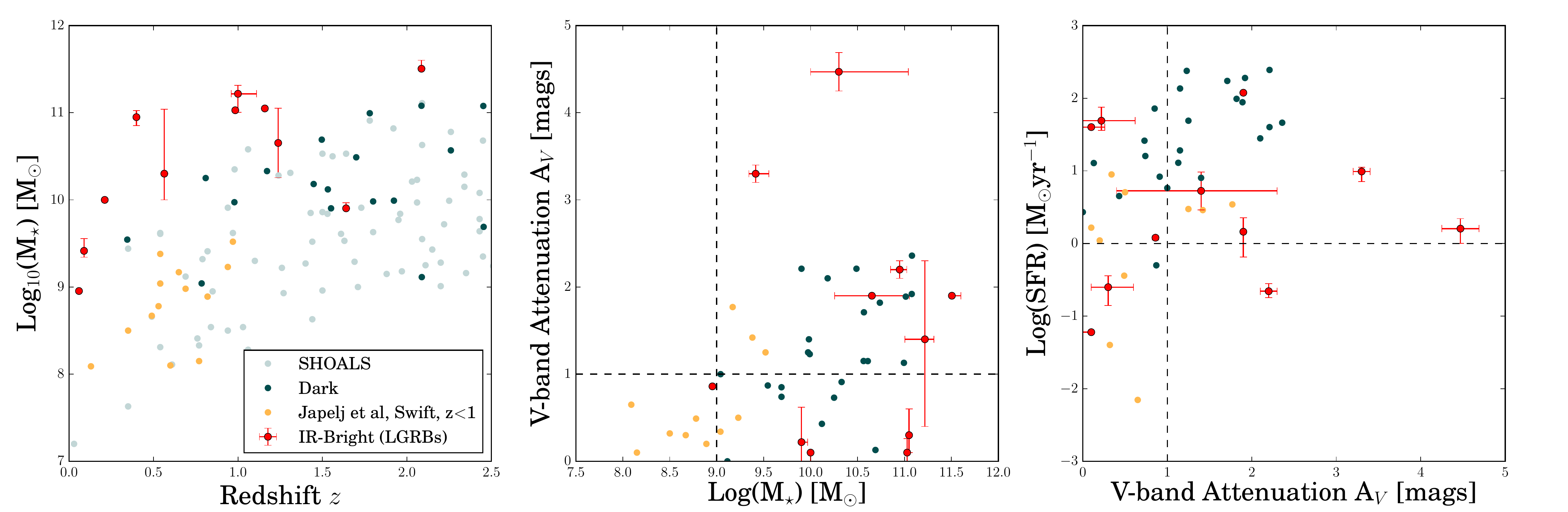}
\caption{The distribution in physical property parameter space (mass, redshift, attenuation and SFR) of the IR-bright LGRB hosts with redshift estimates, as compared to the comparison samples discussed above. In each case, the IR-bright hosts are shown as red points with error bars. The dashed lines indicate regions of parameter space discussed in section \ref{sec:interp}.}
\label{fig:multi}
\end{figure*}

Our targets were selected to be bright in W1 imaging. At low redshifts, W1 lies above the bulk of stellar photospheric emission, while at intermediate redshift $0.6<z<3$ it probes the bulk of the stellar mass.  As figure \ref{fig:w1red} makes clear, we are strongly limited by the shallow depth of W1 imaging in \WISE. This introduces a Malmquist bias \citep{1922MeLuF.100....1M} which limits us to comparatively low redshift. Given that the star forming galaxy luminosity function favours less luminous systems, we are able to sample a larger fraction of the extant population at the lowest redshifts, and likely miss many of the galaxies in the more distant Universe.  

However, many of the sources for which optical data is available are also extremely red (see appendix and figures \ref{fig:bestfitgalaxies} and \ref{fig:bestfitstars}). Red colours can arise from either old stellar populations or dusty populations. The Balmer break, which appears in a galaxy spectrum after the death of hot young O stars, is an age indicator and so may allow us to distinguish betwen these two cases. In 2 of the 14 galaxy SEDs presented in figure \ref{fig:bestfitgalaxies}, there is no clear constraint on the Balmer break. In 5 cases there is a clear indication of a Balmer break. Even those galaxies with Balmer breaks have moderate star formation rates. Thus for the sources which are detected in the optical, we can rule out IR-bright, old galaxies as being dominant in this sample. 

The population for which an IR-detection exists, but which remain undetected in optical surveys presents a slightly different challenge. It is possible that some of these may show a prominent 4000\AA\ break, if the galaxy in question lies at $z$>0.8. However, given the attenuation distribution presented in figure \ref{fig:dustdist}, our failure to detect galaxies in the range $1<z<3$ instead suggests that the rest-frame UV being probed at these redshifts has been attenuated by dust, resulting in observer-frame optical non-detections.


In figure \ref{fig:multi} we present the multivariate distribution of properties for our LGRB hosts. While our W1 band magnitudes are bright by construction, the emission may arise for different reasons. The sources could be extremely dusty, or, if lying at $0.6<z<3$, may instead be very massive. Figure \ref{fig:multi} indicates that our sources are amongst the most massive LGRB hosts known at their redshifts (figure \ref{fig:multi}, left). In terms of attenuation and mass, mass appears to be the dominant factor (figure \ref{fig:multi}, centre). Around half of the IR-bright sample are dusty for their star formation rate (figure \ref{fig:multi}, right). In turn, this implies they have high mass for their SFR \citep{Whitaker}, or relatively low sSFRs for LGRB hosts \citep[i.e. sit below what is known as the galaxy 'main sequence',][]{2007ApJ...660L..43N}. This is also seen in the sSFRs given in table \ref{table:results}. While some hosts have high sSFRs of $>10^{-9}\,$yr$^{-1}$, as is expected from LGRB hosts generally, others have low values (e.g. $6{\times}10^{-12}\,$yr$^{-1}$ for LGRB\,050219A and $3.8{\times}10^{-11}\,$yr$^{-1}$ for LGRB\,080517). 

This has implications for the supply and depletion of molecular gas in these systems. If our sample is preferentially selecting molecular gas (and therefore star formation) poor systems, this implies either that the hydrogen gas in these systems is in a different phase (i.e. heated by shocks, interactions or a large scale environment effect) or that there is very little gas, as might be seen in an old underlying system which accretes a small star forming satellite. This latter scenario has already been demonstrated in the most luminous example in this sample of IR-bright systems, the host of LGRB\,080517 \citep[][ and in prep]{Stanway2}. However a molecular gas detection has recently been reported for a second GRB host in our IR-bright selection, the dark burst LGRB\,080207 \citep{2017arXiv170900424A}, and two galaxies in our LGRB sample (080517 and 100316D) have detections of IR molecular Hydrogen emission lines (Wiersema et al., in prep). \citet{2017arXiv170900424A} find that the host of LGRB\,080207 appears to follow the normal scaling relations for star forming galaxies and is not noticeably poor in molecular gas. It is however a dark burst, in a host which has already been noted for being relatively massive and dusty - criteria which bring it into our overall sample. This implies that we are not preferentially selecting molecular gas rich systems.

Figure \ref{fig:multi} shows that the IR-bright LGRB host sample may not have a single set of consistent properties. If we divide the sample into dusty (A$_V$ > 1) vs non-dusty, and massive (log$_{10}$(M$_{\star}$/M$_\odot$) > 9) vs less massive, we find that of 11 LGRB host galaxies in our sample that have M$_{\star}$ and A$_{V}$ information, 6 would be classed as both massive and dusty, while 4 are simply massive, and one is on the border of being low in mass and not dusty.

A question thus arises: is this a distinct population, or is it sampling the tails of existing and known populations? The diversity of host galaxy properties in our sample favours the latter possibility. There is no clear separation between the IR luminous host galaxies and the bulk of the LGRB population in any projection of parameter space. While our sources are dustier than the bulk of the population over the same redshift range, they do not lie significantly outside of the range of typical LGRB host properties, with both A$_{V}$ and another extreme property (e.g. high stellar mass, low redshift) required for selection. This suggests that we are sampling the tails of previously established host distributions. As such, it is unlikely that implications for LGRB progenitors can be derived from this sample. None the less, selection in these relatively shallow infrared bands is useful for constraining the tails of existing distributions and therefore the full range of LGRB host properties.

This analysis has identified 2 new candidate GRB hosts. These are LGRB\,080517, which has already been extensively studied \citep[][]{Stanway,Stanway2} following selection with this methodology, and LGRB\,061002. This burst has a false alarm probability for association with the IR-bright galaxy of only 0.0003. We have also newly identified existing host candidates as being bright in the \WISE\ bands. Generally, the low number of IR-bright GRB hosts we have found (only 20 from 1001 bursts), is consistent with expectations for the following reasons. First and foremost, a shallow survey such as \WISE\ only allows us to sample low redshifts, covering a small cosmological volume. A secondary effect might be arising because the LGRB rate appears to be suppressed in the local Universe \citep[e.g. ][]{Perley1}, tracing the star formation rate density but also the increase in metallicity over cosmic time. Local and/or massive and dusty galaxies are typically detected in \WISE, however massive and dusty galaxies are typically higher in metallicity and are less often seen to host LGRBs. SGRBs could be hosted by these classes of galaxy, however short bursts are intrinsically less luminous and have a much lower observed redshift distribution. Finally, by the time we get to high enough redshifts that the rest frame bands LGRB hosts are bright in (e.g. UV) are redshifted to W1, the emission is too faint for detection by shallow surveys such as \WISE. These factors conspire to produce only small sample of GRB hosts that can be detected in \WISE. 

In theory, using these data, it is possible to calculate the fraction of low redshift and massive, dusty GRB hosts that have been missed in previous follow up studies. This is primarily true for low redshift galaxies. As figure \ref{fig:w1red} shows, the WISE 2\,${\sigma}$ limit corresponds to absolute magnitudes of ${\sim}$-18 at $z$=0.1. This is well below the knee of the galaxy luminosity function, probing galaxies of $\approx$0.1\,L$^\ast$ and above \citep{2006MNRAS.370.1159B}. We would therefore naively expect to see the majority of galaxies at $z<0.1$ in $WISE$. 
However, our cross matching procedure fails to capture bursts at very large apparent offsets from their hosts, and this primarily affects short bursts and the lowest redshift galaxies in the GRB population. In fact, of the three known long bursts at $z<0.05$, we identify zero as IR-bright. There are four long bursts known to be at $0.05<z<0.1$, of these we detect two and one of those was selected as low $z$ purely on the basis of its WISE detection. The remaining two GRBs constitute one source at large offset from a very extended galaxy, and one source in a highly sub-luminous host. This demonstrates that 3.6\,micron selection is not a highly efficient method for identifying low redshift bursts in all cases, but nonetheless that for typical galaxies in its redshift sweet spot, we would expect to have identified any source present in the archival data. Thus, while we cannot rule out the continued presence of $z<0.05$ galaxies in the archival data set, we can be reasonably confident that the fraction of GRB hosts without redshift identifications which lie at $0.05<z<0.10$ is very low. With the exception of GRB 080517, which was selected as part of this sample, we identify no further candidates in this range and would have expected to detect at least half of those present, unless the GRB host luminosity function has a steep faint end slope. This suggests that the total number of $0.05<z<0.1$ hosts still remaining unidentified in the archival sample is of order a few, at most.

We note that deeper IR data, for example that obtained with {\em Spitzer} or {\em JWST}, would detect galaxies further down the luminosity function, but would not overcome the issues of large projected offsets or galaxy sizes. It is therefore possible that future deeper infrared surveys will allow this analysis to be extended to slightly higher redshifts, and, for example, characterise the overlooked GRB hosts (if any) at $0.1<z<0.3$.


\section{Conclusions}\label{sec:conclusions}
A population of infrared-bright GRB host galaxies has been identified by cross-matching X-ray afterglow positions to the ALLWISE catalogue. Selective cuts in apparent magnitude and catalogue quality flags, in addition to a false alarm probability analysis, yield 55 IR-sources that are convincingly associated with a GRB X-ray position. Compiling photometry from surveys and our own observations, we perform SED fitting, finding that 14 sources fit well to galaxy templates and 14 are best-fittung to stellar SEDs. The remainder are either cut from the sample, optically undetected, lacking survey coverage or have been previously studied. Spectroscopy of 6 targets supplements the photometric data and allows us to rule out 3 stellar interlopers. Our methodology has identified the host of LGRB\,080517, and potentially LGRB\,061002. The former has already undergone extensive study, while the candidate host of LGRB\,061002 is newly reported here. Focusing on LGRB hosts in particular, we find that the population is biased towards massive, dusty and low redshift galaxies, with respect to unbiased samples of LGRB hosts. The low redshift bias appears to be due to the depth of \WISE, and within this low redshift population, galaxies with high stellar mass and dust content are selected. Dusty and local galaxies are the most frequently discovered hosts of dark and low-luminosity LGRBs respectively, classes of burst which are crucial to understand if we are to develop a full picture of the collapsar GRB phenomenon. We propose that the spatial association of IR-bright galaxies with LGRBs is therefore a useful technique for the identification of unusual host systems.


\section*{Acknowledgements}
We acknowledge PhD studentship 1763016 (AAC) from the UK Science and Technology Facilities Council (STFC). AAC also thanks the William Edwards Educational Charity. We acknowledge support from the European Research Council (AJL). CRA acknowledges receipt of a studentship from the Midlands Physics Alliance. SMLG is funded by a research studentship from the UK STFC. 

This paper contains original results based on observations taken at WHT and ATCA.
The Australia Telescope Compact Array is part of the Australia Telescope National Facility which is funded by the Australian Government for operation as a National Facility managed by CSIRO. The William Herschel Telescope is operated on the island of La Palma by the Isaac Newton Group of Telescopes in the Spanish Observatorio del Roque de los Muchachos of the Instituto de Astrofisica de Canarias.

This paper makes use of archival data. The Pan-STARRS Surveys and the PS1 public science archive have been made possible through contributions by the Institute for Astronomy, the University of Hawaii, the Pan-STARRS Project Office, the Max-Planck Society and its participating institutes. Funding for SDSS-III has been provided by the Alfred P. Sloan Foundation, the Participating Institutions, the National Science Foundation, and the U.S. Department of Energy Office of Science. The SDSS-III web site is http://www.sdss3.org/. SDSS-III is managed by the Astrophysical Research Consortium for the Participating Institutions of the SDSS-III Collaboration.

We use observations made with the NASA Galaxy Evolution Explorer. GALEX is operated for NASA by the California Institute of Technology under NASA contract NAS5-98034. This publication makes use of data products from the Wide-field Infrared Survey Explorer, which is a joint project of the University of California, Los Angeles, and the Jet Propulsion Laboratory/California Institute of Technology, funded by the National Aeronautics and Space Administration.
We acknowledge the use of public data from the \textit{Swift} data archive.
This research has made use of the APASS database, located at the AAVSO web site. Funding for APASS has been provided by the Robert Martin Ayers Sciences Fund.
This publication makes use of data products from the Two Micron All Sky Survey, which is a joint project of the University of Massachusetts and the Infrared Processing and Analysis Center/California Institute of Technology, funded by the National Aeronautics and Space Administration and the National Science Foundation. This research has made use of the NASA/ IPAC Infrared Science Archive, which is operated by the Jet Propulsion Laboratory, California Institute of Technology, under contract with the National Aeronautics and Space Administration.

We make use of a number of astronomical tools. {\sc IRAF} is distributed by the National Optical Astronomy Observatory, which is operated by the Association of Universities for Research in Astronomy (AURA) under a cooperative agreement with the National Science Foundation. We acknowledge the use of Ned Wright's cosmology calculator \citep{cosmocalc}. Finally, we thank the anonymous referee for their valuable feedback.





\begin{thebibliography}{}
\makeatletter
\relax
\def\mn@urlcharsother{\let\do\@makeother \do\$\do\&\do\#\do\^\do\_\do\%\do\~}
\def\mn@doi{\begingroup\mn@urlcharsother \@ifnextchar [ {\mn@doi@}
  {\mn@doi@[]}}
\def\mn@doi@[#1]#2{\def\@tempa{#1}\ifx\@tempa\@empty \href
  {http://dx.doi.org/#2} {doi:#2}\else \href {http://dx.doi.org/#2} {#1}\fi
  \endgroup}
\def\mn@eprint#1#2{\mn@eprint@#1:#2::\@nil}
\def\mn@eprint@arXiv#1{\href {http://arxiv.org/abs/#1} {{\tt arXiv:#1}}}
\def\mn@eprint@dblp#1{\href {http://dblp.uni-trier.de/rec/bibtex/#1.xml}
  {dblp:#1}}
\def\mn@eprint@#1:#2:#3:#4\@nil{\def\@tempa {#1}\def\@tempb {#2}\def\@tempc
  {#3}\ifx \@tempc \@empty \let \@tempc \@tempb \let \@tempb \@tempa \fi \ifx
  \@tempb \@empty \def\@tempb {arXiv}\fi \@ifundefined
  {mn@eprint@\@tempb}{\@tempb:\@tempc}{\expandafter \expandafter \csname
  mn@eprint@\@tempb\endcsname \expandafter{\@tempc}}}

\bibitem[\protect\citeauthoryear{{Alam} et~al.,}{{Alam} et~al.}{2015}]{SDSS12}
{Alam} S.,  et~al., 2015, \mn@doi [\apjs] {10.1088/0067-0049/219/1/12}, \href
  {http://adsabs.harvard.edu/abs/2015ApJS..219...12A} {219, 12}

\bibitem[\protect\citeauthoryear{{Altmann}, {Roeser}, {Demleitner}, {Bastian}
  \& {Schilbach}}{{Altmann} et~al.}{2017}]{HSOY}
{Altmann} M.,  {Roeser} S.,  {Demleitner} M.,  {Bastian} U.,   {Schilbach} E.,
  2017, \mn@doi [\aap] {10.1051/0004-6361/201730393}, \href
  {http://adsabs.harvard.edu/abs/2017A%26A...600L...4A} {600, L4}

\bibitem[\protect\citeauthoryear{{Amati}}{{Amati}}{2006}]{Amati2}
{Amati} L.,  2006, \mn@doi [\mnras] {10.1111/j.1365-2966.2006.10840.x}, \href
  {http://adsabs.harvard.edu/abs/2006MNRAS.372..233A} {372, 233}

\bibitem[\protect\citeauthoryear{{Amati} et~al.,}{{Amati}
  et~al.}{2002}]{Amati1}
{Amati} L.,  et~al., 2002, \mn@doi [\aap] {10.1051/0004-6361:20020722}, \href
  {http://adsabs.harvard.edu/abs/2002A%26A...390...81A} {390, 81}

\bibitem[\protect\citeauthoryear{{Arabsalmani} et~al.,}{{Arabsalmani}
  et~al.}{2017}]{2017arXiv170900424A}
{Arabsalmani} M.,  et~al., 2017, preprint, \href
  {http://adsabs.harvard.edu/abs/2017arXiv170900424A} {} (\mn@eprint {arXiv}
  {1709.00424})

\bibitem[\protect\citeauthoryear{{Avni}}{{Avni}}{1976}]{AvniStats}
{Avni} Y.,  1976, \mn@doi [\apj] {10.1086/154870}, \href
  {http://adsabs.harvard.edu/abs/1976ApJ...210..642A} {210, 642}

\bibitem[\protect\citeauthoryear{{Babbedge} et~al.,}{{Babbedge}
  et~al.}{2006}]{2006MNRAS.370.1159B}
{Babbedge} T.~S.~R.,  et~al., 2006, \mn@doi [\mnras]
  {10.1111/j.1365-2966.2006.10547.x}, \href
  {http://adsabs.harvard.edu/abs/2006MNRAS.370.1159B} {370, 1159}

\bibitem[\protect\citeauthoryear{{Barthelmy} et~al.,}{{Barthelmy}
  et~al.}{2005}]{BAT}
{Barthelmy} S.~D.,  et~al., 2005, \mn@doi [\ssr] {10.1007/s11214-005-5096-3},
  \href {http://adsabs.harvard.edu/abs/2005SSRv..120..143B} {120, 143}

\bibitem[\protect\citeauthoryear{{Becker}, {White}  \& {Helfand}}{{Becker}
  et~al.}{1994}]{FIRST}
{Becker} R.~H.,  {White} R.~L.,   {Helfand} D.~J.,  1994, in {Crabtree} D.~R.,
  {Hanisch} R.~J.,   {Barnes} J.,  eds,  ASP Conference Series Vol. 61,
  Astronomical Data Analysis Software and Systems III. p.~165

\bibitem[\protect\citeauthoryear{{Berger}}{{Berger}}{2014}]{2014ARA&A..52...43B}
{Berger} E.,  2014, \mn@doi [\araa] {10.1146/annurev-astro-081913-035926},
  \href {http://adsabs.harvard.edu/abs/2014ARA%26A..52...43B} {52, 43}

\bibitem[\protect\citeauthoryear{{Berger} et~al.,}{{Berger}
  et~al.}{2005}]{Berger05}
{Berger} E.,  et~al., 2005, \mn@doi [\nat] {10.1038/nature04238}, \href
  {http://adsabs.harvard.edu/abs/2005Natur.438..988B} {438, 988}

\bibitem[\protect\citeauthoryear{{Berger}, {Cenko}, {Fox}  \&
  {Cucchiara}}{{Berger} et~al.}{2009}]{2009ApJ...704..877B}
{Berger} E.,  {Cenko} S.~B.,  {Fox} D.~B.,   {Cucchiara} A.,  2009, \mn@doi
  [\apj] {10.1088/0004-637X/704/1/877}, \href
  {http://adsabs.harvard.edu/abs/2009ApJ...704..877B} {704, 877}

\bibitem[\protect\citeauthoryear{{Best} et~al.,}{{Best}
  et~al.}{2013}]{browndwarf}
{Best} W.~M.~J.,  et~al., 2013, \mn@doi [\apj] {10.1088/0004-637X/777/2/84},
  \href {http://adsabs.harvard.edu/abs/2013ApJ...777...84B} {777, 84}

\bibitem[\protect\citeauthoryear{{Blanchard}, {Berger}  \& {Fong}}{{Blanchard}
  et~al.}{2016}]{GRBHST}
{Blanchard} P.~K.,  {Berger} E.,   {Fong} W.-f.,  2016, \mn@doi [\apj]
  {10.3847/0004-637X/817/2/144}, \href
  {http://adsabs.harvard.edu/abs/2016ApJ...817..144B} {817, 144}

\bibitem[\protect\citeauthoryear{{Blandford} \& {Znajek}}{{Blandford} \&
  {Znajek}}{1977}]{BZmech}
{Blandford} R.~D.,  {Znajek} R.~L.,  1977, \mn@doi [\mnras]
  {10.1093/mnras/179.3.433}, \href
  {http://adsabs.harvard.edu/abs/1977MNRAS.179..433B} {179, 433}

\bibitem[\protect\citeauthoryear{{Bolmer}, {Greiner}, {Kr{\"u}hler}, {Schady},
  {Ledoux}, {Tanvir}  \& {Levan}}{{Bolmer} et~al.}{2017}]{Bolmer}
{Bolmer} J.,  {Greiner} J.,  {Kr{\"u}hler} T.,  {Schady} P.,  {Ledoux} C.,
  {Tanvir} N.~R.,   {Levan} A.~J.,  2017, preprint, \href
  {http://adsabs.harvard.edu/abs/2017arXiv170906867B} {} (\mn@eprint {arXiv}
  {1709.06867})

\bibitem[\protect\citeauthoryear{{Bruzual} \& {Charlot}}{{Bruzual} \&
  {Charlot}}{2003}]{Bruzal}
{Bruzual} G.,  {Charlot} S.,  2003, \mn@doi [\mnras]
  {10.1046/j.1365-8711.2003.06897.x}, \href
  {http://adsabs.harvard.edu/abs/2003MNRAS.344.1000B} {344, 1000}

\bibitem[\protect\citeauthoryear{{Burrows} et~al.,}{{Burrows}
  et~al.}{2004}]{XRT}
{Burrows} D.~N.,  et~al., 2004, in {Flanagan} K.~A.,  {Siegmund} O.~H.~W.,
  eds,  \procspie Vol. 5165, X-Ray and Gamma-Ray Instrumentation for Astronomy
  XIII. pp 201--216, \mn@doi{10.1117/12.504868}

\bibitem[\protect\citeauthoryear{{Butler}, {Kocevski}, {Bloom}  \&
  {Curtis}}{{Butler} et~al.}{2007}]{Butler}
{Butler} N.~R.,  {Kocevski} D.,  {Bloom} J.~S.,   {Curtis} J.~L.,  2007,
  \mn@doi [\apj] {10.1086/522492}, \href
  {http://adsabs.harvard.edu/abs/2007ApJ...671..656B} {671, 656}

\bibitem[\protect\citeauthoryear{{Cano}, {Malesani}, {Tanvir}, {Xu}, {Kangas}
  \& {Kajava}}{{Cano} et~al.}{2013}]{130527A}
{Cano} Z.,  {Malesani} D.,  {Tanvir} N.~R.,  {Xu} D.,  {Kangas} T.,   {Kajava}
  J.,  2013, GRB Coordinates Network, Circular Service, No.~14710, \#1 (2013),
  \href {http://adsabs.harvard.edu/abs/2013GCN.14710....1C} {14710}

\bibitem[\protect\citeauthoryear{{Cenko} et~al.,}{{Cenko} et~al.}{2008}]{Cenko}
{Cenko} S.~B.,  et~al., 2008, preprint, \href
  {http://adsabs.harvard.edu/abs/2008arXiv0802.0874C} {} (\mn@eprint {arXiv}
  {0802.0874})

\bibitem[\protect\citeauthoryear{{Chambers} et~al.,}{{Chambers}
  et~al.}{2016}]{PSDR1}
{Chambers} K.~C.,  et~al., 2016, preprint, \href
  {http://adsabs.harvard.edu/abs/2016arXiv161205560C} {} (\mn@eprint {arXiv}
  {1612.05560})

\bibitem[\protect\citeauthoryear{{Charlot} \& {Fall}}{{Charlot} \&
  {Fall}}{2000}]{CharlotFall}
{Charlot} S.,  {Fall} S.~M.,  2000, \mn@doi [\apj] {10.1086/309250}, \href
  {http://adsabs.harvard.edu/abs/2000ApJ...539..718C} {539, 718}

\bibitem[\protect\citeauthoryear{{Chornock} \& {Fong}}{{Chornock} \&
  {Fong}}{2015}]{150120z}
{Chornock} R.,  {Fong} W.,  2015, GRB Coordinates Network, \href
  {http://adsabs.harvard.edu/abs/2015GCN..17358...1C} {17358}

\bibitem[\protect\citeauthoryear{{Chornock}, {Fong}  \& {Fox}}{{Chornock}
  et~al.}{2014}]{Chornock}
{Chornock} R.,  {Fong} W.,   {Fox} D.~B.,  2014, GRB Coordinates Network, \href
  {http://adsabs.harvard.edu/abs/2014GCN..17177...1C} {17177}

\bibitem[\protect\citeauthoryear{{Christensen}, {Vreeswijk}, {Sollerman},
  {Th{\"o}ne}, {Le Floc'h}  \& {Wiersema}}{{Christensen}
  et~al.}{2008}]{IFU1998}
{Christensen} L.,  {Vreeswijk} P.~M.,  {Sollerman} J.,  {Th{\"o}ne} C.~C.,  {Le
  Floc'h} E.,   {Wiersema} K.,  2008, \mn@doi [\aap]
  {10.1051/0004-6361:200809896}, \href
  {http://adsabs.harvard.edu/abs/2008A%26A...490...45C} {490, 45}

\bibitem[\protect\citeauthoryear{{Condon}, {Cotton}, {Greisen}, {Yin},
  {Perley}, {Taylor}  \& {Broderick}}{{Condon} et~al.}{1998}]{NVSS}
{Condon} J.~J.,  {Cotton} W.~D.,  {Greisen} E.~W.,  {Yin} Q.~F.,  {Perley}
  R.~A.,  {Taylor} G.~B.,   {Broderick} J.~J.,  1998, \mn@doi [\aj]
  {10.1086/300337}, \href {http://adsabs.harvard.edu/abs/1998AJ....115.1693C}
  {115, 1693}

\bibitem[\protect\citeauthoryear{{Cucchiara}, {Fox}, {Cenko}  \&
  {Price}}{{Cucchiara} et~al.}{2007}]{2007GCN..6083....1C}
{Cucchiara} A.,  {Fox} D.~B.,  {Cenko} S.~B.,   {Price} P.~A.,  2007, GRB
  Coordinates Network, \href
  {http://adsabs.harvard.edu/abs/2007GCN..6083....1C} {6083}

\bibitem[\protect\citeauthoryear{{Cucchiara} et~al.,}{{Cucchiara}
  et~al.}{2011}]{Cucchiara}
{Cucchiara} A.,  et~al., 2011, \mn@doi [\apj] {10.1088/0004-637X/736/1/7},
  \href {http://adsabs.harvard.edu/abs/2011ApJ...736....7C} {736, 7}

\bibitem[\protect\citeauthoryear{{Dainotti}, {Petrosian}, {Willingale},
  {O'Brien}, {Ostrowski}  \& {Nagataki}}{{Dainotti} et~al.}{2015}]{Dainotti}
{Dainotti} M.,  {Petrosian} V.,  {Willingale} R.,  {O'Brien} P.,  {Ostrowski}
  M.,   {Nagataki} S.,  2015, \mn@doi [\mnras] {10.1093/mnras/stv1229}, \href
  {http://adsabs.harvard.edu/abs/2015MNRAS.451.3898D} {451, 3898}

\bibitem[\protect\citeauthoryear{{Davies} et~al.,}{{Davies}
  et~al.}{2017}]{Davies}
{Davies} L.~J.~M.,  et~al., 2017, \mn@doi [\mnras] {10.1093/mnras/stw3080},
  \href {http://adsabs.harvard.edu/abs/2017MNRAS.466.2312D} {466, 2312}

\bibitem[\protect\citeauthoryear{{Eckart}, {McGreer}, {Stern}, {Harrison}  \&
  {Helfand}}{{Eckart} et~al.}{2010}]{Eckart}
{Eckart} M.~E.,  {McGreer} I.~D.,  {Stern} D.,  {Harrison} F.~A.,   {Helfand}
  D.~J.,  2010, \mn@doi [\apj] {10.1088/0004-637X/708/1/584}, \href
  {http://adsabs.harvard.edu/abs/2010ApJ...708..584E} {708, 584}

\bibitem[\protect\citeauthoryear{{Elliott} et~al.,}{{Elliott}
  et~al.}{2013}]{Elliott}
{Elliott} J.,  et~al., 2013, preprint, \href
  {http://adsabs.harvard.edu/abs/2013arXiv1308.5520E} {} (\mn@eprint {arXiv}
  {1308.5520})

\bibitem[\protect\citeauthoryear{{Farrow} et~al.,}{{Farrow}
  et~al.}{2014}]{Farrow}
{Farrow} D.~J.,  et~al., 2014, \mn@doi [\mnras] {10.1093/mnras/stt1933}, \href
  {http://adsabs.harvard.edu/abs/2014MNRAS.437..748F} {437, 748}

\bibitem[\protect\citeauthoryear{{Fong} et~al.,}{{Fong} et~al.}{2016}]{Fong}
{Fong} W.,  et~al., 2016, \mn@doi [\apj] {10.3847/1538-4357/833/2/151}, \href
  {http://adsabs.harvard.edu/abs/2016ApJ...833..151F} {833, 151}

\bibitem[\protect\citeauthoryear{{Frederiks}}{{Frederiks}}{2013}]{130603B_iso}
{Frederiks} D.,  2013, GRB Coordinates Network, \href
  {http://adsabs.harvard.edu/abs/2013GCN..14772...1F} {14772}

\bibitem[\protect\citeauthoryear{{Frederiks} \& {Pal'Shin}}{{Frederiks} \&
  {Pal'Shin}}{2011}]{110918A_Iso}
{Frederiks} D.,  {Pal'Shin} V.,  2011, GRB Coordinates Network, \href
  {http://adsabs.harvard.edu/abs/2011GCN..12370...1F} {12370}

\bibitem[\protect\citeauthoryear{{Fruchter} et~al.,}{{Fruchter}
  et~al.}{2006}]{Fruchter}
{Fruchter} A.~S.,  et~al., 2006, \mn@doi [\nat] {10.1038/nature04787}, \href
  {http://adsabs.harvard.edu/abs/2006Natur.441..463F} {441, 463}

\bibitem[\protect\citeauthoryear{{Galama} et~al.,}{{Galama}
  et~al.}{1998}]{1998discovery}
{Galama} T.~J.,  et~al., 1998, \mn@doi [\nat] {10.1038/27150}, \href
  {http://adsabs.harvard.edu/abs/1998Natur.395..670G} {395, 670}

\bibitem[\protect\citeauthoryear{{Gehrels} et~al.,}{{Gehrels}
  et~al.}{2004}]{Swift}
{Gehrels} N.,  et~al., 2004, \mn@doi [\apj] {10.1086/422091}, \href
  {http://adsabs.harvard.edu/abs/2004ApJ...611.1005G} {611, 1005}

\bibitem[\protect\citeauthoryear{{Goldstein} et~al.,}{{Goldstein}
  et~al.}{2017}]{2017ApJ...848L..14G}
{Goldstein} A.,  et~al., 2017, \mn@doi [\apjl] {10.3847/2041-8213/aa8f41},
  \href {http://adsabs.harvard.edu/abs/2017ApJ...848L..14G} {848, L14}

\bibitem[\protect\citeauthoryear{{Golenetskii}, {Aptekar}, {Frederiks},
  {Pal'Shin}, {Oleynik}, {Ulanov}, {Svinkin}  \& {Cline}}{{Golenetskii}
  et~al.}{2013}]{130907A_iso}
{Golenetskii} S.,  {Aptekar} R.,  {Frederiks} D.,  {Pal'Shin} V.,  {Oleynik}
  P.,  {Ulanov} M.,  {Svinkin} D.,   {Cline} T.,  2013, GRB Coordinates
  Network, \href {http://adsabs.harvard.edu/abs/2013GCN..15203...1G} {15203}

\bibitem[\protect\citeauthoryear{{Greiner} et~al.,}{{Greiner}
  et~al.}{2009}]{Greiner_hi-z}
{Greiner} J.,  et~al., 2009, \mn@doi [\apj] {10.1088/0004-637X/693/2/1610},
  \href {http://adsabs.harvard.edu/abs/2009ApJ...693.1610G} {693, 1610}

\bibitem[\protect\citeauthoryear{{Heintz}, {Malesani}, {de Ugarte Postigo},
  {Pursimo}, {Kruehler}, {Telting}  \& {Fynbo}}{{Heintz}
  et~al.}{2016}]{161007A}
{Heintz} K.~E.,  {Malesani} D.,  {de Ugarte Postigo} A.,  {Pursimo} T.,
  {Kruehler} T.,  {Telting} J.,   {Fynbo} J.~P.~U.,  2016, GRB Coordinates
  Network, Circular Service, No.~20020, \#1 (2016), \href
  {http://adsabs.harvard.edu/abs/2016GCN.20020....1H} {20020}

\bibitem[\protect\citeauthoryear{{Henden} \& {Munari}}{{Henden} \&
  {Munari}}{2014}]{APASS}
{Henden} A.,  {Munari} U.,  2014, Contributions of the Astronomical Observatory
  Skalnate Pleso, \href {http://adsabs.harvard.edu/abs/2014CoSka..43..518H}
  {43, 518}

\bibitem[\protect\citeauthoryear{{Hjorth}}{{Hjorth}}{2013}]{Hjorth}
{Hjorth} J.,  2013, \mn@doi [Philosophical Transactions of the Royal Society of
  London Series A] {10.1098/rsta.2012.0275}, \href
  {http://adsabs.harvard.edu/abs/2013RSPTA.37120275H} {371, 20120275}

\bibitem[\protect\citeauthoryear{{Hjorth} et~al.,}{{Hjorth}
  et~al.}{2012}]{TOUGH}
{Hjorth} J.,  et~al., 2012, \mn@doi [\apj] {10.1088/0004-637X/756/2/187}, \href
  {http://adsabs.harvard.edu/abs/2012ApJ...756..187H} {756, 187}

\bibitem[\protect\citeauthoryear{{Hopkins} \& {Beacom}}{{Hopkins} \&
  {Beacom}}{2006}]{HopkinsBeacom}
{Hopkins} A.~M.,  {Beacom} J.~F.,  2006, \mn@doi [\apj] {10.1086/506610}, \href
  {http://adsabs.harvard.edu/abs/2006ApJ...651..142H} {651, 142}

\bibitem[\protect\citeauthoryear{{Hunt}, {Palazzi}, {Rossi}, {Savaglio},
  {Cresci}, {Klose}, {Micha{\l}owski}  \& {Pian}}{{Hunt}
  et~al.}{2011}]{080207a}
{Hunt} L.,  {Palazzi} E.,  {Rossi} A.,  {Savaglio} S.,  {Cresci} G.,  {Klose}
  S.,  {Micha{\l}owski} M.,   {Pian} E.,  2011, \mn@doi [\apjl]
  {10.1088/2041-8205/736/2/L36}, \href
  {http://adsabs.harvard.edu/abs/2011ApJ...736L..36H} {736, L36}

\bibitem[\protect\citeauthoryear{{Jakobsson}, {Hjorth}, {Fynbo}, {Watson},
  {Pedersen}, {Bj{\"o}rnsson}  \& {Gorosabel}}{{Jakobsson}
  et~al.}{2004}]{Jakobsson}
{Jakobsson} P.,  {Hjorth} J.,  {Fynbo} J.~P.~U.,  {Watson} D.,  {Pedersen} K.,
  {Bj{\"o}rnsson} G.,   {Gorosabel} J.,  2004, \mn@doi [\apjl]
  {10.1086/427089}, \href {http://adsabs.harvard.edu/abs/2004ApJ...617L..21J}
  {617, L21}

\bibitem[\protect\citeauthoryear{{Japelj} et~al.,}{{Japelj}
  et~al.}{2016}]{Japelj}
{Japelj} J.,  et~al., 2016, \mn@doi [\aap] {10.1051/0004-6361/201628314}, \href
  {http://adsabs.harvard.edu/abs/2016A%26A...590A.129J} {590, A129}

\bibitem[\protect\citeauthoryear{{Jeong} et~al.,}{{Jeong} et~al.}{2014}]{Jeong}
{Jeong} S.,  et~al., 2014, \mn@doi [\aap] {10.1051/0004-6361/201423979}, \href
  {http://adsabs.harvard.edu/abs/2014A%26A...569A..93J} {569, A93}

\bibitem[\protect\citeauthoryear{{Kennicutt} \& {Evans}}{{Kennicutt} \&
  {Evans}}{2012}]{2012ARA&A..50..531K}
{Kennicutt} R.~C.,  {Evans} N.~J.,  2012, \mn@doi [\araa]
  {10.1146/annurev-astro-081811-125610}, \href
  {http://adsabs.harvard.edu/abs/2012ARA%26A..50..531K} {50, 531}

\bibitem[\protect\citeauthoryear{{Kocevski} et~al.,}{{Kocevski}
  et~al.}{2010}]{Kocevski}
{Kocevski} D.,  et~al., 2010, \mn@doi [\mnras]
  {10.1111/j.1365-2966.2010.16327.x}, \href
  {http://adsabs.harvard.edu/abs/2010MNRAS.404..963K} {404, 963}

\bibitem[\protect\citeauthoryear{{Kov{\'a}cs} \& {Szapudi}}{{Kov{\'a}cs} \&
  {Szapudi}}{2015}]{Kovacs}
{Kov{\'a}cs} A.,  {Szapudi} I.,  2015, \mn@doi [\mnras] {10.1093/mnras/stv063},
  \href {http://adsabs.harvard.edu/abs/2015MNRAS.448.1305K} {448, 1305}

\bibitem[\protect\citeauthoryear{{Kr{\"u}hler} et~al.,}{{Kr{\"u}hler}
  et~al.}{2011}]{Kruhler2011}
{Kr{\"u}hler} T.,  et~al., 2011, \mn@doi [\aap] {10.1051/0004-6361/201117428},
  \href {http://adsabs.harvard.edu/abs/2011A%26A...534A.108K} {534, A108}

\bibitem[\protect\citeauthoryear{{Kr{\"u}hler} et~al.,}{{Kr{\"u}hler}
  et~al.}{2012a}]{080605}
{Kr{\"u}hler} T.,  et~al., 2012a, \mn@doi [\aap] {10.1051/0004-6361/201118670},
  \href {http://adsabs.harvard.edu/abs/2012A%26A...546A...8K} {546, A8}

\bibitem[\protect\citeauthoryear{{Kr{\"u}hler} et~al.,}{{Kr{\"u}hler}
  et~al.}{2012b}]{TOUGH2}
{Kr{\"u}hler} T.,  et~al., 2012b, \mn@doi [\apj] {10.1088/0004-637X/758/1/46},
  \href {http://adsabs.harvard.edu/abs/2012ApJ...758...46K} {758, 46}

\bibitem[\protect\citeauthoryear{{Kr{\"u}hler} et~al.,}{{Kr{\"u}hler}
  et~al.}{2015}]{Kruhler}
{Kr{\"u}hler} T.,  et~al., 2015, \mn@doi [\aap] {10.1051/0004-6361/201425561},
  \href {http://adsabs.harvard.edu/abs/2015A%26A...581A.125K} {581, A125}

\bibitem[\protect\citeauthoryear{{LIGO Scientific Collaboration} et~al.,}{{LIGO
  Scientific Collaboration} et~al.}{2017}]{NSNSmulti1}
{LIGO Scientific Collaboration} et~al., 2017, preprint, \href
  {http://adsabs.harvard.edu/abs/2017arXiv171005833L} {} (\mn@eprint {arXiv}
  {1710.05833})

\bibitem[\protect\citeauthoryear{{Levan} \& {Tanvir}}{{Levan} \&
  {Tanvir}}{2013}]{130515A}
{Levan} A.~J.,  {Tanvir} N.~R.,  2013, GRB Coordinates Network, Circular
  Service, No.~14667, \#1 (2013), \href
  {http://adsabs.harvard.edu/abs/2013GCN.14667....1L} {14667}

\bibitem[\protect\citeauthoryear{{Levan}, {Crowther}, {de Grijs}, {Langer},
  {Xu}  \& {Yoon}}{{Levan} et~al.}{2016}]{Levan}
{Levan} A.,  {Crowther} P.,  {de Grijs} R.,  {Langer} N.,  {Xu} D.,   {Yoon}
  S.-C.,  2016, \mn@doi [\ssr] {10.1007/s11214-016-0312-x}, \href
  {http://adsabs.harvard.edu/abs/2016SSRv..202...33L} {202, 33}

\bibitem[\protect\citeauthoryear{{Liang}, {Zhang}, {Virgili}  \& {Dai}}{{Liang}
  et~al.}{2007}]{LLGRBs}
{Liang} E.,  {Zhang} B.,  {Virgili} F.,   {Dai} Z.~G.,  2007, \mn@doi [\apj]
  {10.1086/517959}, \href {http://adsabs.harvard.edu/abs/2007ApJ...662.1111L}
  {662, 1111}

\bibitem[\protect\citeauthoryear{{Littlejohns} et~al.,}{{Littlejohns}
  et~al.}{2014}]{140331A_photz}
{Littlejohns} O.,  et~al., 2014, GRB Coordinates Network, Circular Service,
  No.~16050, \#1 (2014), \href
  {http://adsabs.harvard.edu/abs/2014GCN.16050....1L} {16050}

\bibitem[\protect\citeauthoryear{{Lyman} et~al.,}{{Lyman} et~al.}{2017}]{Lyman}
{Lyman} J.~D.,  et~al., 2017, \mn@doi [\mnras] {10.1093/mnras/stx220}, \href
  {http://adsabs.harvard.edu/abs/2017MNRAS.467.1795L} {467, 1795}

\bibitem[\protect\citeauthoryear{{Madau} \& {Dickinson}}{{Madau} \&
  {Dickinson}}{2014}]{Madau}
{Madau} P.,  {Dickinson} M.,  2014, \mn@doi [\araa]
  {10.1146/annurev-astro-081811-125615}, \href
  {http://adsabs.harvard.edu/abs/2014ARA%26A..52..415M} {52, 415}

\bibitem[\protect\citeauthoryear{{Malesani}, {D'Avanzo}, {D'Elia}, {Vergani},
  {Andreuzzi}, {Garcia}, {Escudero}  \& {Bonomo}}{{Malesani}
  et~al.}{2014}]{2014GCN..17170...1M}
{Malesani} D.,  {D'Avanzo} P.,  {D'Elia} V.,  {Vergani} S.~D.,  {Andreuzzi} G.,
   {Garcia} A.,  {Escudero} G.,   {Bonomo} A.,  2014, GRB Coordinates Network,
  \href {http://adsabs.harvard.edu/abs/2014GCN..17170...1M} {17170}

\bibitem[\protect\citeauthoryear{{Malesani}, {Xu}, {Jakobsson}  \&
  {Harmanen}}{{Malesani} et~al.}{2015}]{150323C}
{Malesani} D.,  {Xu} D.,  {Jakobsson} P.,   {Harmanen} J.,  2015, GRB
  Coordinates Network, Circular Service, No.~17655, \#1 (2015), \href
  {http://adsabs.harvard.edu/abs/2015GCN.17655....1M} {17655}

\bibitem[\protect\citeauthoryear{{Malesani} et~al.,}{{Malesani}
  et~al.}{2016}]{161214A}
{Malesani} D.,  et~al., 2016, GRB Coordinates Network, Circular Service,
  No.~20319, \#1 (2016), \href
  {http://adsabs.harvard.edu/abs/2016GCN.20319....1M} {20319}

\bibitem[\protect\citeauthoryear{{Malmquist}}{{Malmquist}}{1922}]{1922MeLuF.100....1M}
{Malmquist} K.~G.,  1922, Meddelanden fran Lunds Astronomiska Observatorium
  Serie I, \href {http://adsabs.harvard.edu/abs/1922MeLuF.100....1M} {100, 1}

\bibitem[\protect\citeauthoryear{{Martin} \& {GALEX Science Team}}{{Martin} \&
  {GALEX Science Team}}{2005}]{Martin}
{Martin} D.~C.,  {GALEX Science Team} 2005, in American Astronomical Society
  Meeting Abstracts. p.~1235

\bibitem[\protect\citeauthoryear{{Martin} et~al.,}{{Martin}
  et~al.}{2003}]{GALEX}
{Martin} C.,  et~al., 2003, in {Blades} J.~C.,  {Siegmund} O.~H.~W.,  eds,
  \procspie Vol. 4854, Future EUV/UV and Visible Space Astrophysics Missions
  and Instrumentation.. pp 336--350, \mn@doi{10.1117/12.460034}

\bibitem[\protect\citeauthoryear{{McCall}}{{McCall}}{2004}]{YES}
{McCall} M.~L.,  2004, \mn@doi [\aj] {10.1086/424933}, \href
  {http://adsabs.harvard.edu/abs/2004AJ....128.2144M} {128, 2144}

\bibitem[\protect\citeauthoryear{{McGuire} et~al.,}{{McGuire}
  et~al.}{2016}]{ManyHiZ}
{McGuire} J.~T.~W.,  et~al., 2016, \mn@doi [\apj]
  {10.3847/0004-637X/825/2/135}, \href
  {http://adsabs.harvard.edu/abs/2016ApJ...825..135M} {825, 135}

\bibitem[\protect\citeauthoryear{{Micha{\l}owski} et~al.,}{{Micha{\l}owski}
  et~al.}{2015}]{HIgrbhosts}
{Micha{\l}owski} M.~J.,  et~al., 2015, \mn@doi [\aap]
  {10.1051/0004-6361/201526542}, \href
  {http://adsabs.harvard.edu/abs/2015A%26A...582A..78M} {582, A78}

\bibitem[\protect\citeauthoryear{{Micha{\l}owski} et~al.,}{{Micha{\l}owski}
  et~al.}{2016}]{Michalowski}
{Micha{\l}owski} M.~J.,  et~al., 2016, preprint, \href
  {http://adsabs.harvard.edu/abs/2016arXiv161006928M} {} (\mn@eprint {arXiv}
  {1610.06928})

\bibitem[\protect\citeauthoryear{{Noeske} et~al.,}{{Noeske}
  et~al.}{2007}]{2007ApJ...660L..43N}
{Noeske} K.~G.,  et~al., 2007, \mn@doi [\apjl] {10.1086/517926}, \href
  {http://adsabs.harvard.edu/abs/2007ApJ...660L..43N} {660, L43}

\bibitem[\protect\citeauthoryear{{Oke} \& {Gunn}}{{Oke} \& {Gunn}}{1983}]{Oke}
{Oke} J.~B.,  {Gunn} J.~E.,  1983, \mn@doi [\apj] {10.1086/160817}, \href
  {http://adsabs.harvard.edu/abs/1983ApJ...266..713O} {266, 713}

\bibitem[\protect\citeauthoryear{{Page} et~al.,}{{Page} et~al.}{2009}]{Page}
{Page} K.~L.,  et~al., 2009, \mn@doi [\mnras]
  {10.1111/j.1365-2966.2009.14509.x}, \href
  {http://adsabs.harvard.edu/abs/2009MNRAS.395..328P} {395, 328}

\bibitem[\protect\citeauthoryear{{P{\'e}rez-Ram{\'{\i}}rez}
  et~al.,}{{P{\'e}rez-Ram{\'{\i}}rez} et~al.}{2013}]{Perez}
{P{\'e}rez-Ram{\'{\i}}rez} D.,  et~al., 2013, in {Castro-Tirado} A.~J.,
  {Gorosabel} J.,   {Park} I.~H.,  eds,  EAS Publications Series Vol. 61, EAS
  Publications Series. pp 345--349, \mn@doi{10.1051/eas/1361055}

\bibitem[\protect\citeauthoryear{{Perley}, {Foley}, {Bloom}  \&
  {Butler}}{{Perley} et~al.}{2006}]{051109B}
{Perley} D.~A.,  {Foley} R.~J.,  {Bloom} J.~S.,   {Butler} N.~R.,  2006, GRB
  Coordinates Network, \href
  {http://adsabs.harvard.edu/abs/2006GCN..5387....1P} {5387}

\bibitem[\protect\citeauthoryear{{Perley}, {Bloom}, {Butler}, {Li}  \&
  {Chen}}{{Perley} et~al.}{2007}]{lensedgrb}
{Perley} D.~A.,  {Bloom} J.~S.,  {Butler} N.~R.,  {Li} W.,   {Chen} H.-W.,
  2007, in {Immler} S.,  {Weiler} K.,   {McCray} R.,  eds,  American Institute
  of Physics Conference Series Vol. 937, Supernova 1987A: 20 Years After:
  Supernovae and Gamma-Ray Bursters. pp 526--529, \mn@doi{10.1063/1.3682955}

\bibitem[\protect\citeauthoryear{{Perley}, {Modjaz}, {Morgan}, {Cenko},
  {Bloom}, {Butler}, {Filippenko}  \& {Miller}}{{Perley}
  et~al.}{2012}]{100206A}
{Perley} D.~A.,  {Modjaz} M.,  {Morgan} A.~N.,  {Cenko} S.~B.,  {Bloom} J.~S.,
  {Butler} N.~R.,  {Filippenko} A.~V.,   {Miller} A.~A.,  2012, \mn@doi [\apj]
  {10.1088/0004-637X/758/2/122}, \href
  {http://adsabs.harvard.edu/abs/2012ApJ...758..122P} {758, 122}

\bibitem[\protect\citeauthoryear{{Perley} et~al.,}{{Perley}
  et~al.}{2013}]{PerleyDust}
{Perley} D.~A.,  et~al., 2013, \mn@doi [\apj] {10.1088/0004-637X/778/2/128},
  \href {http://adsabs.harvard.edu/abs/2013ApJ...778..128P} {778, 128}

\bibitem[\protect\citeauthoryear{{Perley} et~al.,}{{Perley}
  et~al.}{2016a}]{Perley1}
{Perley} D.~A.,  et~al., 2016a, \mn@doi [\apj] {10.3847/0004-637X/817/1/7},
  \href {http://adsabs.harvard.edu/abs/2016ApJ...817....7P} {817, 7}

\bibitem[\protect\citeauthoryear{{Perley} et~al.,}{{Perley}
  et~al.}{2016b}]{Perley2}
{Perley} D.~A.,  et~al., 2016b, \mn@doi [\apj] {10.3847/0004-637X/817/1/8},
  \href {http://adsabs.harvard.edu/abs/2016ApJ...817....8P} {817, 8}

\bibitem[\protect\citeauthoryear{{Pickles}}{{Pickles}}{1998}]{Pickles}
{Pickles} A.~J.,  1998, \mn@doi [\pasp] {10.1086/316197}, \href
  {http://adsabs.harvard.edu/abs/1998PASP..110..863P} {110, 863}

\bibitem[\protect\citeauthoryear{{Prochaska}, {Chen}, {Bloom}  \&
  {Stephens}}{{Prochaska} et~al.}{2005}]{050724z}
{Prochaska} J.~X.,  {Chen} H.-W.,  {Bloom} J.~S.,   {Stephens} A.,  2005, GRB
  Coordinates Network, \href
  {http://adsabs.harvard.edu/abs/2005GCN..3679....1P} {3679}

\bibitem[\protect\citeauthoryear{{Qi} et~al.,}{{Qi} et~al.}{2015}]{APOP}
{Qi} Z.,  et~al., 2015, \mn@doi [\apj] {10.1088/0004-6256/150/4/137}, \href
  {http://adsabs.harvard.edu/abs/2015AJ....150..137Q} {150, 137}

\bibitem[\protect\citeauthoryear{{Rol}, {Wijers}, {Kouveliotou}, {Kaper}  \&
  {Kaneko}}{{Rol} et~al.}{2005}]{Rol_darks}
{Rol} E.,  {Wijers} R.~A.~M.~J.,  {Kouveliotou} C.,  {Kaper} L.,   {Kaneko} Y.,
   2005, \mn@doi [\apj] {10.1086/429082}, \href
  {http://adsabs.harvard.edu/abs/2005ApJ...624..868R} {624, 868}

\bibitem[\protect\citeauthoryear{{Rol} et~al.,}{{Rol}
  et~al.}{2007}]{050716_rol}
{Rol} E.,  et~al., 2007, \mn@doi [\mnras] {10.1111/j.1365-2966.2006.11224.x},
  \href {http://adsabs.harvard.edu/abs/2007MNRAS.374.1078R} {374, 1078}

\bibitem[\protect\citeauthoryear{{Roming} et~al.,}{{Roming}
  et~al.}{2005}]{UVOT}
{Roming} P.~W.~A.,  et~al., 2005, \mn@doi [\ssr] {10.1007/s11214-005-5095-4},
  \href {http://adsabs.harvard.edu/abs/2005SSRv..120...95R} {120, 95}

\bibitem[\protect\citeauthoryear{{Rossi} et~al.,}{{Rossi} et~al.}{2014}]{Rossi}
{Rossi} A.,  et~al., 2014, \mn@doi [\aap] {10.1051/0004-6361/201423865}, \href
  {http://adsabs.harvard.edu/abs/2014A%26A...572A..47R} {572, A47}

\bibitem[\protect\citeauthoryear{{Ruiz-Velasco} et~al.,}{{Ruiz-Velasco}
  et~al.}{2007}]{Ruiz-Velasco}
{Ruiz-Velasco} A.~E.,  et~al., 2007, \mn@doi [\apj] {10.1086/521546}, \href
  {http://adsabs.harvard.edu/abs/2007ApJ...669....1R} {669, 1}

\bibitem[\protect\citeauthoryear{{Salvaterra} et~al.,}{{Salvaterra}
  et~al.}{2009}]{Salvaterra}
{Salvaterra} R.,  et~al., 2009, \mn@doi [\nat] {10.1038/nature08445}, \href
  {http://adsabs.harvard.edu/abs/2009Natur.461.1258S} {461, 1258}

\bibitem[\protect\citeauthoryear{{Schady}}{{Schady}}{2017}]{Schady}
{Schady} P.,  2017, preprint, \href
  {http://adsabs.harvard.edu/abs/2017arXiv170705214S} {} (\mn@eprint {arXiv}
  {1707.05214})

\bibitem[\protect\citeauthoryear{{Schlafly} \& {Finkbeiner}}{{Schlafly} \&
  {Finkbeiner}}{2011}]{Schlafly}
{Schlafly} E.~F.,  {Finkbeiner} D.~P.,  2011, \mn@doi [\apj]
  {10.1088/0004-637X/737/2/103}, \href
  {http://adsabs.harvard.edu/abs/2011ApJ...737..103S} {737, 103}

\bibitem[\protect\citeauthoryear{{Selsing} et~al.,}{{Selsing}
  et~al.}{2018}]{GRBXS}
{Selsing} J.,  et~al., 2018, preprint, \href
  {http://adsabs.harvard.edu/abs/2018arXiv180207727S} {} (\mn@eprint {arXiv}
  {1802.07727})

\bibitem[\protect\citeauthoryear{{Shanks} et~al.,}{{Shanks}
  et~al.}{2015}]{ATLAS}
{Shanks} T.,  et~al., 2015, \mn@doi [\mnras] {10.1093/mnras/stv1130}, \href
  {http://adsabs.harvard.edu/abs/2015MNRAS.451.4238S} {451, 4238}

\bibitem[\protect\citeauthoryear{{Skrutskie} et~al.,}{{Skrutskie}
  et~al.}{2006}]{2MASS}
{Skrutskie} M.~F.,  et~al., 2006, \mn@doi [\aj] {10.1086/498708}, \href
  {http://adsabs.harvard.edu/abs/2006AJ....131.1163S} {131, 1163}

\bibitem[\protect\citeauthoryear{{Stanway}, {Levan}, {Tanvir}, {Wiersema}, {van
  der Horst}, {Mundell}, {Guidorzi}  \& et al.}{{Stanway}
  et~al.}{2015a}]{Stanway}
{Stanway} E.~R.,  {Levan} A.~J.,  {Tanvir} N.,  {Wiersema} K.,  {van der Horst}
  A.,  {Mundell} C.~G.,  {Guidorzi} C.,   et al. 2015a, \mn@doi [\mnras]
  {10.1093/mnras/stu2286}, \href
  {http://adsabs.harvard.edu/abs/2015MNRAS.446.3911S} {446, 3911}

\bibitem[\protect\citeauthoryear{{Stanway}, {Levan}, {Tanvir}, {Wiersema}  \&
  {van der Laan}}{{Stanway} et~al.}{2015b}]{Stanway2}
{Stanway} E.~R.,  {Levan} A.~J.,  {Tanvir} N.~R.,  {Wiersema} K.,   {van der
  Laan} T.~P.~R.,  2015b, \mn@doi [\apjl] {10.1088/2041-8205/798/1/L7}, \href
  {http://adsabs.harvard.edu/abs/2015ApJ...798L...7S} {798, L7}

\bibitem[\protect\citeauthoryear{{Starling} et~al.,}{{Starling}
  et~al.}{2011}]{100316D}
{Starling} R.~L.~C.,  et~al., 2011, \mn@doi [\mnras]
  {10.1111/j.1365-2966.2010.17879.x}, \href
  {http://adsabs.harvard.edu/abs/2011MNRAS.411.2792S} {411, 2792}

\bibitem[\protect\citeauthoryear{{Svensson}, {Levan}, {Tanvir}, {Fruchter}  \&
  {Strolger}}{{Svensson} et~al.}{2010}]{Svensson}
{Svensson} K.~M.,  {Levan} A.~J.,  {Tanvir} N.~R.,  {Fruchter} A.~S.,
  {Strolger} L.-G.,  2010, \mn@doi [\mnras] {10.1111/j.1365-2966.2010.16442.x},
  \href {http://adsabs.harvard.edu/abs/2010MNRAS.405...57S} {405, 57}

\bibitem[\protect\citeauthoryear{{Svensson} et~al.,}{{Svensson}
  et~al.}{2012a}]{080207}
{Svensson} K.~M.,  et~al., 2012a, \mn@doi [\mnras]
  {10.1111/j.1365-2966.2011.19811.x}, \href
  {http://adsabs.harvard.edu/abs/2012MNRAS.421...25S} {421, 25}

\bibitem[\protect\citeauthoryear{{Svensson} et~al.,}{{Svensson}
  et~al.}{2012b}]{080207b}
{Svensson} K.~M.,  et~al., 2012b, \mn@doi [\mnras]
  {10.1111/j.1365-2966.2011.19811.x}, \href
  {http://adsabs.harvard.edu/abs/2012MNRAS.421...25S} {421, 25}

\bibitem[\protect\citeauthoryear{{Tanvir} et~al.,}{{Tanvir}
  et~al.}{2009}]{Tanvir2}
{Tanvir} N.~R.,  et~al., 2009, \mn@doi [\nat] {10.1038/nature08459}, \href
  {http://adsabs.harvard.edu/abs/2009Natur.461.1254T} {461, 1254}

\bibitem[\protect\citeauthoryear{{Tanvir} et~al.,}{{Tanvir}
  et~al.}{2017a}]{Tanvir_hi-z}
{Tanvir} N.~R.,  et~al., 2017a, preprint, \href
  {http://adsabs.harvard.edu/abs/2017arXiv170309052T} {} (\mn@eprint {arXiv}
  {1703.09052})

\bibitem[\protect\citeauthoryear{{Tanvir} et~al.,}{{Tanvir}
  et~al.}{2017b}]{2017ApJ...848L..27T}
{Tanvir} N.~R.,  et~al., 2017b, \mn@doi [\apjl] {10.3847/2041-8213/aa90b6},
  \href {http://adsabs.harvard.edu/abs/2017ApJ...848L..27T} {848, L27}

\bibitem[\protect\citeauthoryear{{Taylor}}{{Taylor}}{2005}]{TOPCAT}
{Taylor} M.~B.,  2005, in {Shopbell} P.,  {Britton} M.,   {Ebert} R.,  eds,
  ASP Conference Series Vol. 347, Astronomical Data Analysis Software and
  Systems XIV. p.~29

\bibitem[\protect\citeauthoryear{{The LIGO Scientific Collaboration} \& {The
  Virgo Collaboration}}{{The LIGO Scientific Collaboration} \& {The Virgo
  Collaboration}}{2017}]{NSNSGW1}
{The LIGO Scientific Collaboration} {The Virgo Collaboration} 2017, preprint,
  \href {http://adsabs.harvard.edu/abs/2017arXiv171005832T} {} (\mn@eprint
  {arXiv} {1710.05832})

\bibitem[\protect\citeauthoryear{{Troja} et~al.,}{{Troja}
  et~al.}{2017}]{2017arXiv171005433T}
{Troja} E.,  et~al., 2017, preprint, \href
  {http://adsabs.harvard.edu/abs/2017arXiv171005433T} {} (\mn@eprint {arXiv}
  {1710.05433})

\bibitem[\protect\citeauthoryear{{Vernet} et~al.,}{{Vernet} et~al.}{2011}]{XS}
{Vernet} J.,  et~al., 2011, \mn@doi [\aap] {10.1051/0004-6361/201117752}, \href
  {http://adsabs.harvard.edu/abs/2011A%26A...536A.105V} {536, A105}

\bibitem[\protect\citeauthoryear{{Vink} \& {Harries}}{{Vink} \&
  {Harries}}{2017}]{VinkBH}
{Vink} J.~S.,  {Harries} T.~J.,  2017, preprint, \href
  {http://adsabs.harvard.edu/abs/2017arXiv170309857V} {} (\mn@eprint {arXiv}
  {1703.09857})

\bibitem[\protect\citeauthoryear{{Vink} \& {de Koter}}{{Vink} \& {de
  Koter}}{2005}]{Vink}
{Vink} J.~S.,  {de Koter} A.,  2005, \mn@doi [\aap]
  {10.1051/0004-6361:20052862}, \href
  {http://adsabs.harvard.edu/abs/2005A%26A...442..587V} {442, 587}

\bibitem[\protect\citeauthoryear{{Whitaker}, {Pope}, {Cybulski}, {Casey},
  {Popping}  \& {Yun}}{{Whitaker} et~al.}{2017}]{Whitaker}
{Whitaker} K.~E.,  {Pope} A.,  {Cybulski} R.,  {Casey} C.~M.,  {Popping} G.,
  {Yun} M.~S.,  2017, \mn@doi [\apj] {10.3847/1538-4357/aa94ce}, \href
  {http://adsabs.harvard.edu/abs/2017ApJ...850..208W} {850, 208}

\bibitem[\protect\citeauthoryear{{Wiersema} et~al.,}{{Wiersema}
  et~al.}{2012}]{120224A}
{Wiersema} K.,  et~al., 2012, GRB Coordinates Network, \href
  {http://adsabs.harvard.edu/abs/2012GCN..12991...1W} {12991}

\bibitem[\protect\citeauthoryear{{Williams}}{{Williams}}{1995}]{Penrose}
{Williams} R.~K.,  1995, \mn@doi [\prd] {10.1103/PhysRevD.51.5387}, \href
  {http://adsabs.harvard.edu/abs/1995PhRvD..51.5387W} {51, 5387}

\bibitem[\protect\citeauthoryear{{Woosley} \& {Bloom}}{{Woosley} \&
  {Bloom}}{2006}]{Woolsey}
{Woosley} S.~E.,  {Bloom} J.~S.,  2006, \mn@doi [\araa]
  {10.1146/annurev.astro.43.072103.150558}, \href
  {http://adsabs.harvard.edu/abs/2006ARA%26A..44..507W} {44, 507}

\bibitem[\protect\citeauthoryear{{Wright}}{{Wright}}{2006}]{cosmocalc}
{Wright} E.~L.,  2006, \mn@doi [\pasp] {10.1086/510102}, \href
  {http://adsabs.harvard.edu/abs/2006PASP..118.1711W} {118, 1711}

\bibitem[\protect\citeauthoryear{{Wright} et~al.,}{{Wright}
  et~al.}{2010}]{Wright}
{Wright} E.~L.,  et~al., 2010, \mn@doi [\aj] {10.1088/0004-6256/140/6/1868},
  \href {http://adsabs.harvard.edu/abs/2010AJ....140.1868W} {140, 1868}

\bibitem[\protect\citeauthoryear{{Zheng}, {Filippenko}  \& {Graham}}{{Zheng}
  et~al.}{2016}]{160703A}
{Zheng} W.,  {Filippenko} A.~V.,   {Graham} M.,  2016, GRB Coordinates Network,
  Circular Service, No.~19668, \#1 (2016), \href
  {http://adsabs.harvard.edu/abs/2016GCN.19668....1Z} {19668}

\bibitem[\protect\citeauthoryear{{da Cunha}, {Charlot}  \& {Elbaz}}{{da Cunha}
  et~al.}{2008}]{Magphys08}
{da Cunha} E.,  {Charlot} S.,   {Elbaz} D.,  2008, \mn@doi [\mnras]
  {10.1111/j.1365-2966.2008.13535.x}, \href
  {http://adsabs.harvard.edu/abs/2008MNRAS.388.1595D} {388, 1595}

\bibitem[\protect\citeauthoryear{{da Cunha} et~al.,}{{da Cunha}
  et~al.}{2015}]{Magphys15}
{da Cunha} E.,  et~al., 2015, \mn@doi [\apj] {10.1088/0004-637X/806/1/110},
  \href {http://adsabs.harvard.edu/abs/2015ApJ...806..110D} {806, 110}

\bibitem[\protect\citeauthoryear{{de Ugarte Postigo}, {Xu}, {Malesani},
  {Gorosabel}, {Jakobsson}  \& {Kajava}}{{de Ugarte Postigo}
  et~al.}{2013}]{130907z}
{de Ugarte Postigo} A.,  {Xu} D.,  {Malesani} D.,  {Gorosabel} J.,  {Jakobsson}
  P.,   {Kajava} J.,  2013, GRB Coordinates Network, \href
  {http://adsabs.harvard.edu/abs/2013GCN..15187...1D} {15187}

\bibitem[\protect\citeauthoryear{{de Ugarte Postigo} et~al.,}{{de Ugarte
  Postigo} et~al.}{2014}]{130603z}
{de Ugarte Postigo} A.,  et~al., 2014, \mn@doi [\aap]
  {10.1051/0004-6361/201322985}, \href
  {http://adsabs.harvard.edu/abs/2014A%26A...563A..62D} {563, A62}

\bibitem[\protect\citeauthoryear{{de Ugarte Postigo} et~al.,}{{de Ugarte
  Postigo} et~al.}{2016}]{161108Az}
{de Ugarte Postigo} A.,  et~al., 2016, GRB Coordinates Network, \href
  {http://adsabs.harvard.edu/abs/2016GCN..20150...1D} {20150}

\makeatother
\end{thebibliography}




\appendix
\section{Additional Information}
\subsection{Photometric Compilation}
In the  table \ref{tab:photresults}, we compile archival photometric data for the 30 GRBs with reliable \WISE\ detected counterparts and detections or limits in archival optical photometry. One target is listed twice, since two plausible optical counterparts are detected. All magnitudes, including those for WISE are given in the AB magnitude system.

\begin{table*}
\caption{The full table of archival photometry for 30 GRB locations with reliable \WISE\ detected counterparts, where optical data was available. No Galactic dust correction has been applied. If a magnitude is entered without an associated error, it is a 2\,${\sigma}$ upper limit. If no value is entered, then photometry in that band was unavailable. GRB\,150120A has two sets of photometry, one for each component (see section \ref{sec:ambiguity}).} 
\begin{sideways}
\scalebox{0.7}{
\begin{tabular}{l c c c c c c c c c c c c c c c c c c c c c c c c c c c c c c r}
\hline\hline 
GRB & FUV & ${\delta}$FUV & NUV & ${\delta}$NUV & u & ${\delta}$u & g & ${\delta}$g & r & ${\delta}$r & i & ${\delta}$i & z & ${\delta}$z & y & ${\delta}$y & J & ${\delta}$j & H & ${\delta}$h & K & ${\delta}$k & W1 & ${\delta}$W1 & W2 & ${\delta}$W2 & W3 & ${\delta}$W3 & W4 & ${\delta}$W4\\ [0.25ex] 
\hline 
050522 & 20.89 &  - & 21.79 &  - & 19.01 & 0.02 & 16.60 & 0.00 & 15.77 & 0.00 & 15.38 & 0.00 & 15.21 & 0.00 & 15.07 & 0.00 & 14.02 & 0.03 & 13.45 & 0.02 & 13.27 & 0.04 & 13.28 & 0.02 & 13.27 & 0.03 & 11.95 &  - & 9.07 &  - & \\
050721 & 20.89 &  - & 21.79 &  - &  - &  - & 18.31 & 0.01 & 17.62 & 0.00 & 17.28 & 0.01 & 17.12 & 0.02 & 17.22 & 0.02 & 16.00 & 0.07 & 15.50 & 0.11 & 15.30 & 0.17 & 15.31 & 0.06 & 16.04 & 0.26 & 11.48 &  - & 7.99 &  - & \\
050724 & 20.89 &  - & 21.79 &  - &  - &  - & 20.75 & 0.35 & 20.21 & 0.05 & 19.48 & 0.03 & 18.99 & 0.06 & 18.74 & 0.08 & 17.13 &  - & 16.51 &  - & 16.19 &  - & 15.35 & 0.05 & 15.42 & 0.13 & 12.37 &  - & 8.42 &  - & \\
060428B & 20.89 &  - & 21.79 &  - & 23.37 & 0.81 & 21.34 & 0.15 & 19.91 & 0.04 & 19.46 & 0.02 & 19.25 & 0.03 & 18.86 & 0.09 & 16.90 &  - & 16.00 &  - & 15.52 &  - & 15.90 & 0.04 & 15.76 & 0.08 & 12.81 &  - & 9.44 &  - & \\
061002 & 20.89 &  - & 21.79 &  - & 22.60 & 0.53 & 22.27 & 0.06 & 21.67 & 0.14 & 21.13 & 0.12 & 20.81 & 0.33 & 20.40 & 0.22 & 16.55 &  - & 16.43 &  - & 15.69 &  - & 16.20 & 0.05 & 15.76 & 0.11 & 12.21 &  - & 9.27 &  - & \\
070208 & 20.89 &  - & 21.79 &  - & 23.29 &  - & 20.00 & 0.02 & 19.60 & 0.04 & 19.57 & 0.03 & 19.43 & 0.08 & 19.06 & 0.16 & 17.10 &  - & 16.30 &  - & 17.20 &  - & 17.12 & 0.08 & 16.68 & 0.19 & 12.84 &  - & 9.54 &  - & \\
070724A & 20.89 &  - & 21.70 & 0.25 &  - &  - & 21.52 & 0.12 & 20.79 & 0.04 & 20.48 & 0.06 & 20.32 & 0.16 & 20.31 & 0.13 & 16.50 &  - & 16.00 &  - & 16.00 &  - & 17.03 & 0.11 & 16.37 & 0.22 & 12.34 & 0.37 & 9.13 &  - & \\
080307 & 20.89 &  - & 21.79 &  - & 23.29 &  - & 24.30 &  - & 20.88 & 0.16 & 21.65 & 0.19 & 23.90 &  - & 22.30 &  - & 16.90 &  - & 16.40 &  - & 15.80 &  - & 17.64 & 0.21 & 16.65 &  - & 12.36 &  - & 9.10 &  - & \\
080405 & 19.85 & 0.18 & 19.54 & 0.07 &  - &  - & 20.05 & 0.10 & 20.11 & 0.04 & 19.35 & 0.03 & 19.18 & 0.09 & 19.29 & 0.10 & 16.74 &  - & 16.14 &  - & 15.95 &  - & 14.82 & 0.03 & 13.89 & 0.04 & 11.66 & 0.24 & 8.67 & 0.36 & \\
080517 & 20.85 & 0.18 & 20.46 & 0.11 &  - &  - & 18.31 & 0.01 & 17.94 & 0.03 & 17.56 & 0.01 & 17.57 & 0.03 & 17.45 & 0.03 & 16.48 & 0.13 & 15.85 & 0.15 & 15.66 & 0.21 & 14.65 & 0.03 & 14.25 & 0.04 & 9.85 & 0.05 & 7.07 & 0.11 & \\
080605 & 20.89 &  - & 21.79 &  - &  - &  - & 21.90 & 0.18 & 21.57 & 0.34 & 19.91 & 0.03 & 19.18 & 0.02 & 19.04 & 0.03 & 16.93 &  - & 16.43 &  - & 16.26 &  - & 15.92 & 0.05 & 15.98 & 0.17 & 12.56 &  - & 8.52 &  - & \\
090904B & 20.89 &  - & 21.79 &  - &  - &  - & 21.04 & 0.08 & 18.56 & 0.02 & 16.90 & 0.01 & 15.94 & 0.01 & 15.32 & 0.01 & 13.26 & 0.02 & 11.95 & 0.02 & 11.54 & 0.02 & 11.41 & 0.04 & 11.89 & 0.05 & 11.97 & 0.44 & 8.89 &  - & \\
100206A & 20.89 &  - & 21.79 &  - &  - &  - & 24.30 &  - & 21.26 & 0.17 & 20.94 & 0.08 & 20.22 & 0.05 & 19.68 & 0.20 & 16.74 &  - & 16.14 &  - & 15.95 &  - & 15.71 & 0.05 & 15.15 & 0.09 & 11.31 & 0.18 & 8.56 &  - & \\
110305A & 20.89 &  - & 21.79 &  - &  - &  - & 20.10 & 0.03 & 18.67 & 0.02 & 17.03 & 0.01 & 16.24 & 0.01 & 15.82 & 0.01 & 14.49 & 0.03 & 13.84 & 0.04 & 13.56 & 0.04 & 13.30 & 0.05 & 13.16 & 0.05 & 11.41 &  - & 8.41 &  - & \\
111222A & 20.89 &  - & 21.79 &  - & 22.34 & 0.30 & 19.25 & 0.01 & 18.10 & 0.01 & 16.68 & 0.00 & 15.98 & 0.00 & 15.61 & 0.00 & 14.38 & 0.03 & 13.65 & 0.03 & 13.43 & 0.04 & 13.26 & 0.02 & 13.16 & 0.03 & 12.45 &  - & 8.91 &  - & \\
120119A & 20.89 &  - & 21.79 &  - & 23.85 & 1.00 & 21.65 & 0.19 & 21.25 & 0.06 & 20.50 & 0.05 & 20.09 & 0.06 & 19.70 & 0.15 & 16.75 &  - & 16.06 &  - & 15.41 &  - & 16.64 & 0.08 & 16.92 & 0.37 & 12.69 &  - & 8.97 &  - & \\
120612A & 20.89 &  - & 21.31 & 0.25 &  - &  - & 17.28 & 0.00 & 16.95 & 0.01 & 16.83 & 0.00 & 16.82 & 0.01 & 16.83 & 0.01 & 15.77 & 0.07 & 15.59 & 0.13 & 15.47 & 0.24 & 15.64 & 0.05 & 15.71 & 0.12 & 12.64 &  - & 8.78 &  - & \\
130603B & 20.89 &  - & 21.99 & 0.66 & 22.46 & 0.37 & 21.78 & 0.13 & 20.92 & 0.09 & 20.81 & 0.07 & 20.32 & 0.06 & 19.84 & 0.19 & 17.11 &  - & 16.50 &  - & 16.00 &  - & 17.05 & 0.12 & 16.95 & 0.37 & 12.53 &  - & 8.91 &  - & \\
130907A & 20.89 &  - & 21.79 &  - & 22.09 & 0.38 & 22.21 & 0.19 & 22.29 & 0.31 & 21.38 & 0.22 & 19.88 & 0.23 & 20.80 & 0.01 & 16.75 &  - & 16.06 &  - & 15.41 &  - & 16.89 & 0.09 & 16.18 & 0.14 & 12.69 &  - & 8.86 &  - & \\
131018B & 20.89 &  - & 21.79 &  - &  - &  - & 16.07 & 0.00 & 15.53 & 0.00 & 15.25 & 0.00 & 15.10 & 0.00 & 15.00 & 0.00 & 14.13 & 0.03 & 13.79 & 0.04 & 13.70 & 0.06 & 16.15 & 0.03 & 13.43 & 0.03 & 12.64 &  - & 8.98 &  - & \\
131122A & 20.89 &  - & 21.79 &  - & 24.13 & 1.59 & 24.30 &  - & 21.42 & 0.09 & 21.94 & 0.22 & 20.52 & 0.15 & 20.25 & 0.19 & 17.16 &  - & 16.19 &  - & 15.97 &  - & 16.32 & 0.06 & 15.81 & 0.12 & 12.65 &  - & 8.73 &  - & \\
140331A & 20.89 &  - & 21.79 &  - & 23.78 & 1.00 & 23.648 & 0.38 & 21.99 & 0.17 & 21.08 & 0.12 & 19.67 & 0.10 & 20.18 & - & 16.74 &  - & 16.14 &  - & 15.95 &  - & 16.76 & 0.10 & 16.88 &  - & 12.23 &  - & 9.06 &  - & \\
140927A & 20.89 &  - & 18.65 & 0.05 &  - &  - & 14.49 & 0.02 & 14.09 & 0.03 & 13.98 & 0.15 & 0.01 & 0.01 & 0.01 & 0.01 & 13.06 & 0.03 & 12.76 & 0.03 & 12.74 & 0.03 & 12.64 & 0.02 & 12.65 & 0.03 & 12.21 & 0.33 & 8.70 &  - & \\
141212A & 20.89 &  - & 21.79 &  - &  - &  - & 22.81 &  - & 22.80 & 0.02 & 22.26 & 0.01 & 21.70 & 0.05 & 20.85 &  - & 16.49 &  - & 16.06 &  - & 15.72 &  - & 17.76 & 0.20 & 17.40 & 0.54 & 12.78 &  - & 9.23 &  - & \\
150101B & 21.42 & 0.44 & 21.46 & 0.26 &  - &  - & 17.90 & 0.01 & 17.24 & 0.02 & 16.94 & 0.02 & 16.55 & 0.03 & 16.31 & 0.06 & 15.41 & 0.09 & 14.65 & 0.11 & 14.06 & 0.09 & 13.43 & 0.03 & 13.18 & 0.03 & 12.39 & 0.49 & 8.78 &  - & \\
150120A & 20.89 &  - & 21.76 & 2.82 & 25.34 & 1.14 & 25.25 & 0.86 & 22.68 & 0.33 & 21.97 & 0.25 & 21.76 & 0.70 & 21.21 & 0.01 & 17.16 &  - & 16.19 &  - & 15.97 &  - & 16.89 & 0.10 & 16.75 & 0.31 & 12.38 &  - & 8.71 &  - & \\
150120A & 20.89 &  - & 21.76 & 2.82 & 25.59 & 0.87 & 22.81 & 0.19 & 21.95 & 0.15 & 21.75 & 0.18 & 20.53 & 0.23 & 20.80 &  0.01 & 17.16 &  - & 16.19 &  - & 15.97 &  - & 16.89 & 0.10 & 16.75 & 0.31 & 12.38 &  - & 8.71 &  - & \\
160703A & 20.89 &  - & 21.79 &  - &  - &  - & 21.19 & 0.09 & 19.61 & 0.03 & 19.11 & 0.03 & 18.77 & 0.02 & 18.71 & 0.04 & 17.00 &  - & 17.26 &  - & 17.13 &  - & 16.81 & 0.07 & 17.71 & 0.50 & 13.04 &  - & 9.08 &  - & \\
161010A & 20.89 &  - & 21.79 &  - &  - &  - & 16.39 & 0.00 & 14.90 & 0.01 & 14.18 & 0.02 & 13.70 & 0.00 & 13.43 & 0.01 & 12.05 & 0.03 & 11.18 & 0.03 & 10.94 & 0.03 & 10.75 & 0.03 & 10.95 & 0.02 & 11.25 &  - & 8.85 &  - & \\
161108A & 20.89 &  - & 22.36 & 0.48 & 24.19 & 0.77 & 23.46 & 0.28 & 22.70 & 0.20 & 22.18 & 0.20 & 21.99 & 0.59 & 22.09 &  - & 17.19 &  - & 16.34 &  - & 16.64 &  - & 16.57 & 0.08 & 16.10 & 0.18 & 12.23 &  - & 8.95 &  - & \\
161214A & 20.89 &  - & 21.79 &  - & 22.48 & 0.29 & 19.88 & 0.02 & 19.10 & 0.01 & 18.79 & 0.01 & 18.62 & 0.01 & 18.52 & 0.03 & 16.92 &  - & 16.27 &  - & 15.79 &  - & 16.78 & 0.11 & 16.50 & 0.29 & 11.79 &  - & 8.88 &  - & \\
\hline 
\end{tabular}}
\end{sideways}
\label{tab:photresults}
\end{table*}


\subsection{Candidate Host Image Stamps}
\begin{figure*}
\begin{flushright}
\begin{minipage}[t]{0.423\linewidth}
\includegraphics[width=1\textwidth]{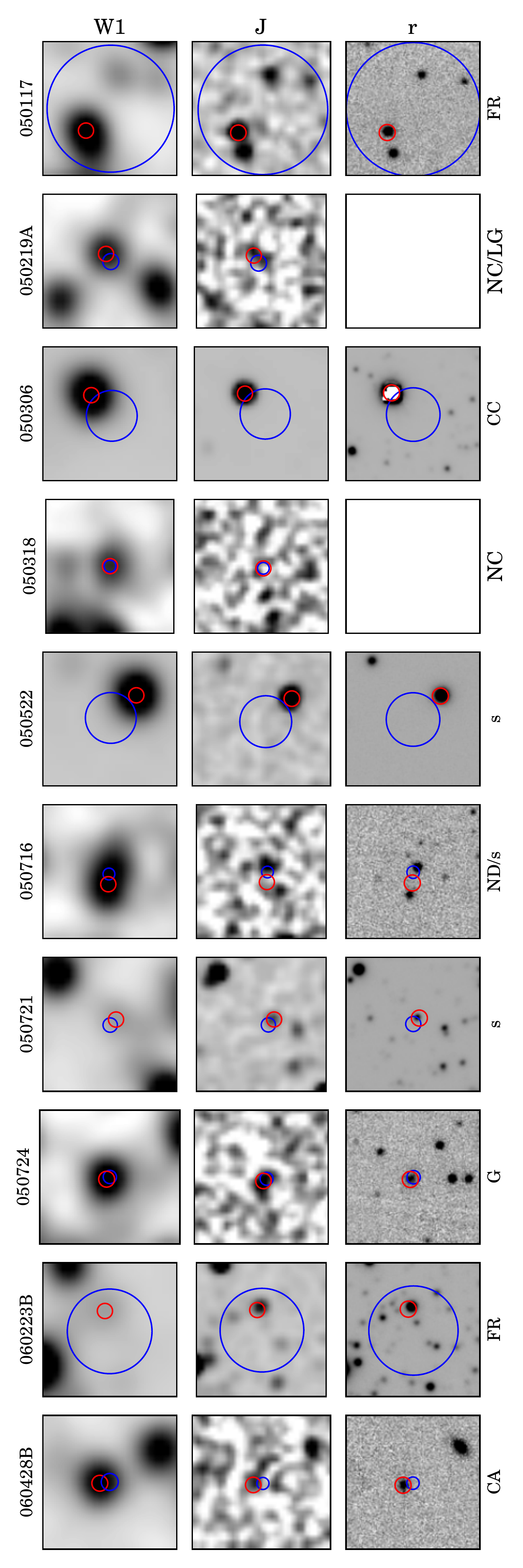}
\end{minipage}
\end{flushright}
\caption{Archival imaging in the W1, {\it J} and {\it r} bands (from top to bottom of each column). Overlaid are the X-ray R$_{90}$ error circles (blue, varying size) and the \WISE\ source centroids (red, fixed diameter). The images are 30\,arcsec on each side and are centered on the GRB location. The classification of each object, as defined in table \ref{tab:samp} is shown at the bottom of each column. This figure extends over seven panels. \label{fig:stamps}}
\end{figure*}

In figure \ref{fig:stamps} we provide archival imaging postage stamp cut-outs in the W1, {\it J} and {\it r} bands. Overlaid are the X-ray R$_{90}$ error circles (blue, varying size) and the \WISE\ source centroids (red, fixed diameter). The images are 30\,arcsec on each side and are centered on the GRB location. The figure extends over seven panels.

\begin{figure*}
\centering
\begin{minipage}[t]{0.423\linewidth}
\centering
\includegraphics[width=1\textwidth]{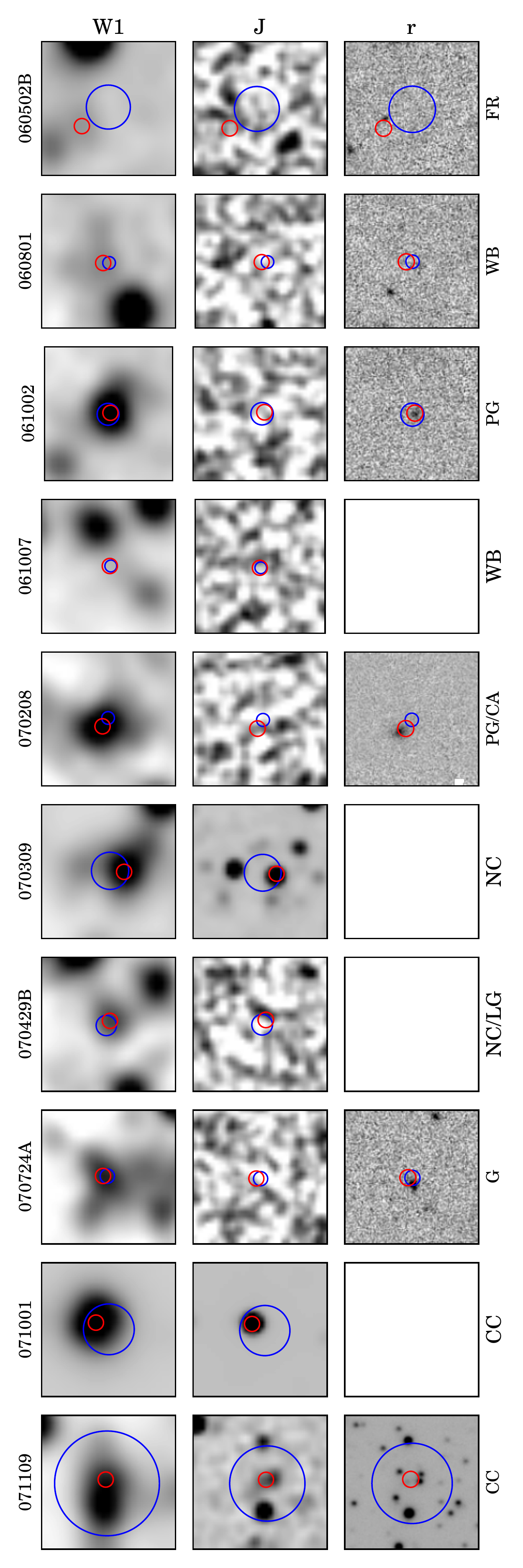}
\end{minipage}
\quad
\begin{minipage}[t]{0.423\linewidth}
\centering
\includegraphics[width=1\textwidth]{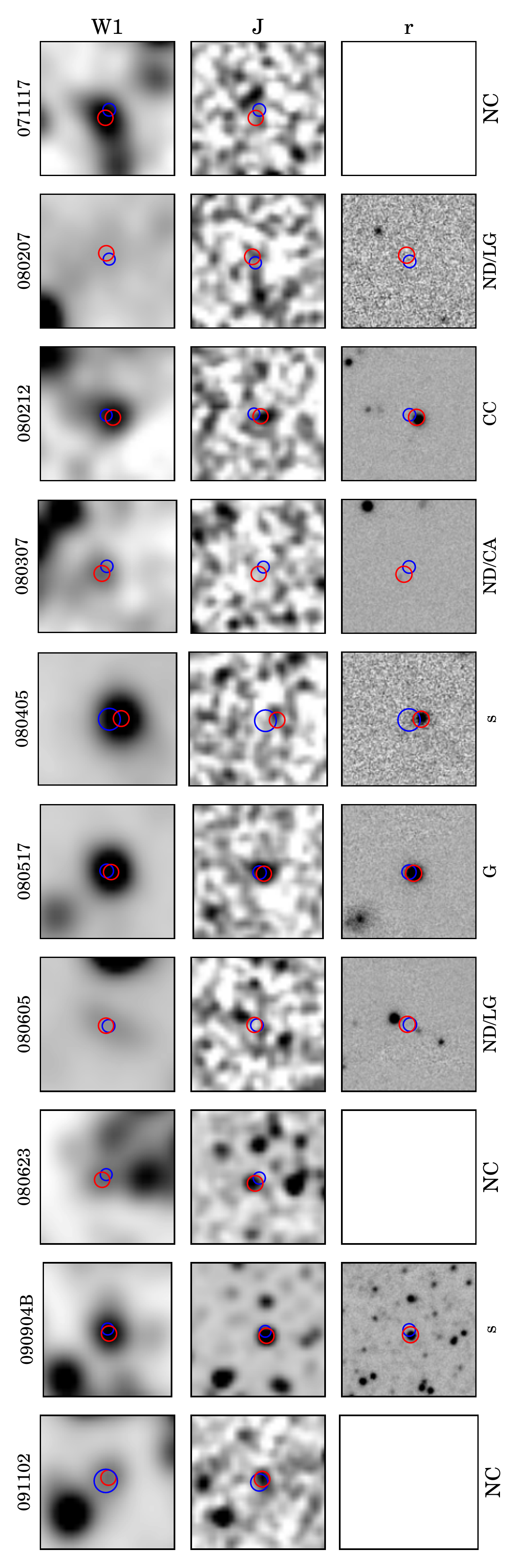}
\end{minipage}
\end{figure*} 

\begin{figure*}
\centering
\begin{minipage}[t]{0.423\linewidth}
\centering
\includegraphics[width=1\textwidth]{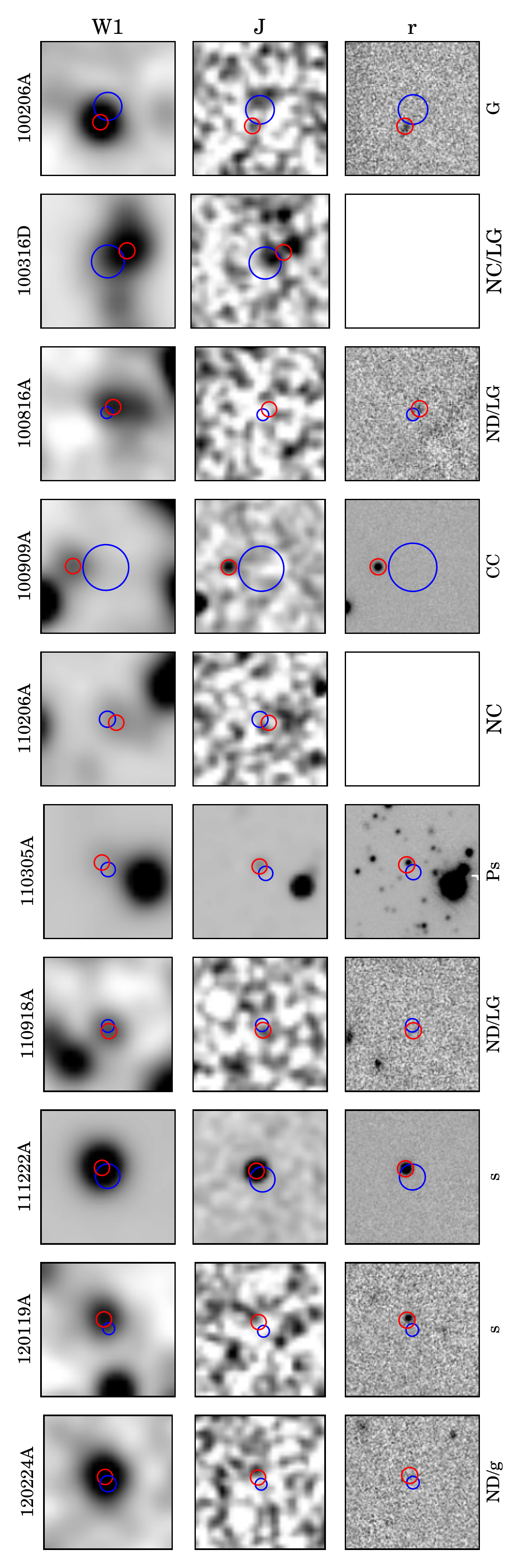}
\end{minipage}
\quad
\begin{minipage}[t]{0.423\linewidth}
\centering
\includegraphics[width=1\textwidth]{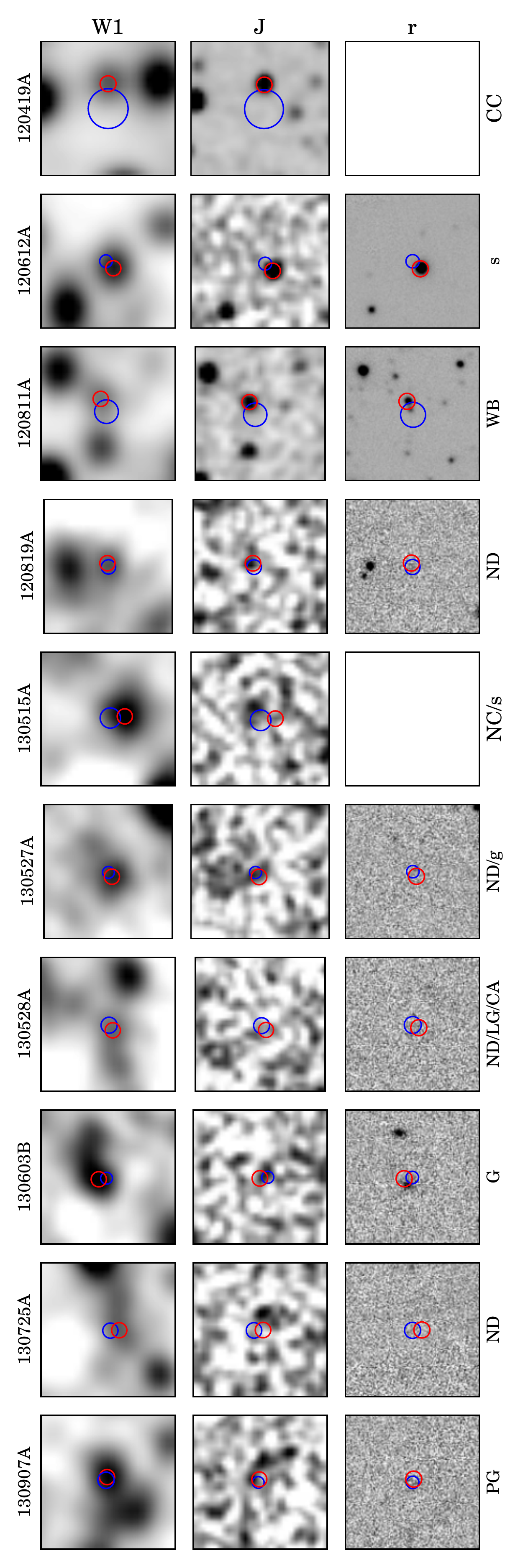}
\end{minipage}
\end{figure*} 

\begin{figure*}
\centering
\begin{minipage}[t]{0.423\linewidth}
\centering
\includegraphics[width=1\textwidth]{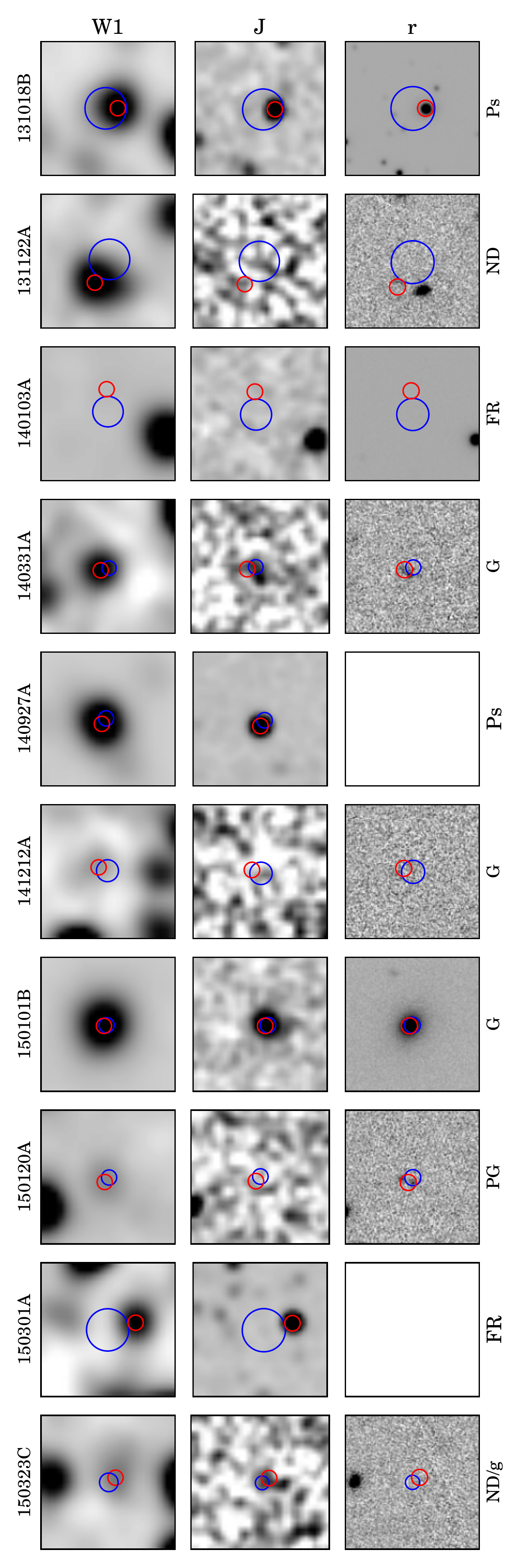}
\end{minipage}
\quad
\begin{minipage}[t]{0.423\linewidth}
\centering
\includegraphics[width=1\textwidth]{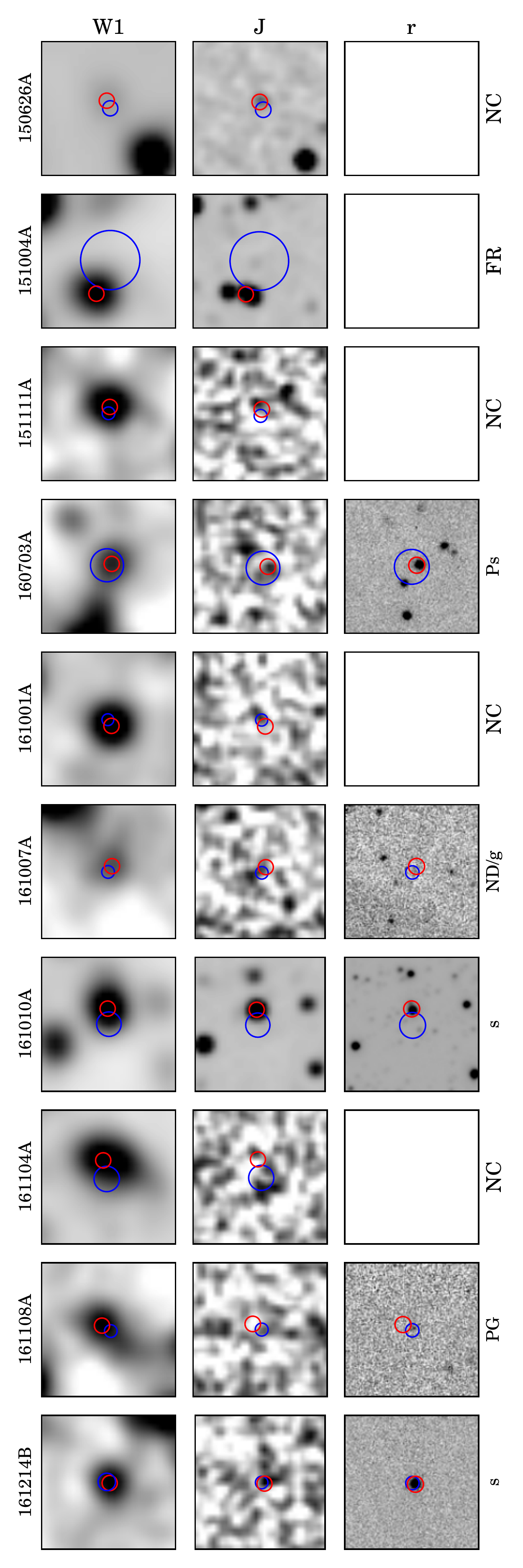}
\end{minipage}
\end{figure*}


\bsp	
\label{lastpage}
\end{document}